\documentclass[useAMS,usenatbib,usegraphics]{mn2e}
\usepackage{ifpdf} 
\pdfoutput=1
\usepackage{graphicx}
\usepackage{latexsym,amssymb}
\usepackage{amsmath}
\usepackage{gensymb}
\usepackage{capt-of}
\usepackage{textpos}
\usepackage{url}
\usepackage{multirow}
\usepackage{color}
\usepackage[T1]{fontenc}
\usepackage{float}
\usepackage{rotating}
\usepackage{tikz}

\usepackage{hyperref}
    \hypersetup{
        bookmarks=true,         
        unicode=false,          
        pdftoolbar=true,        
        pdfmenubar=true,        
        pdffitwindow=false,     
        pdfstartview={FitH},    
        pdftitle={My title},    
        pdfauthor={Author},     
        pdfsubject={Subject},   
        pdfcreator={Creator},   
        pdfproducer={Producer}, 
        pdfkeywords={keyword1} {key2} {key3}, 
        pdfnewwindow=true,      
        colorlinks=false,       
        linkcolor=black,          
        citecolor=blue,        
        filecolor=black,      
        urlcolor=black,           
        linkbordercolor={1 0 0},
        citebordercolor={0 1 0}
    }

\makeatletter
\Hy@AtBeginDocument{%
  \def\@pdfborder{0 0 1}
  \def\@pdfborderstyle{/S/U/W 0}
}

\makeatother

\newcommand{\kms}{km\,s$^{-1}$}

\newcommand{\Msun}{M$_\odot$}
\newcommand{\Lsun}{L$_\odot$}

\newcommand{\kmsM}{km\,s$^{-1}$\,Mpc$^{-1}$}

\newcommand{\Reff}{\ensuremath{R_e}}

\newcommand{\MLstar}{\ensuremath{\Upsilon_\star}}

\newcommand{\Mbh}{\ensuremath{M_\mathrm{\bullet}}}
\newcommand{\Mstar}{\ensuremath{M_{\star}}}

\newcommand{\sauron}{\textsc{SAURON}}      
\newcommand{\sersic}{S\'{e}rsic}		

\title[An orbit-based dynamical analysis of compact, high velocity dispersion galaxies]{MRK\,1216 \& NGC\,1277 - An orbit-based dynamical analysis of compact, high velocity dispersion galaxies}
\author[A. Y{\i}ld{\i}r{\i}m et al.]
{\parbox{\textwidth}{Ak{\i}n Y{\i}ld{\i}r{\i}m$^{1}$\thanks{E-mail: yildirim@mpia.de},
Remco C. E. van den Bosch$^{1}$,
Glenn van de Ven$^{1}$,
Bernd Husemann$^{2, 3}$,
Mariya Lyubenova$^{1, 4}$,
Jonelle L. Walsh$^{5, 6}$,
Karl Gebhardt$^{5}$
and Kayhan G\"ultekin$^{7}$}\vspace{0.4cm}\\
\parbox{\textwidth}{
$^{1}$Max Planck Institute for Astronomy, K\"onigstuhl 17, 69117 Heidelberg, Germany\\
$^{2}$European Southern Observatory, Karl-Schwarzschild-Str. 2, 85748 Garching, Germany\\
$^{3}$Leibniz Institute for Astrophysics Potsdam, An der Sternwarte 16, 14482 Potsdam, Germany\\
$^{4}$Kapteyn Astronomical Institute, University of Groningen, PO Box 800, 9700 AV Groningen, The Netherlands\\
$^{5}$Department of Astronomy, The University of Texas at Austin, Austin, TX 78712, USA\\
$^{6}$George P. and Cynthia Woods Mitchell Institute for Fundamental Physics and Astronomy, Department of Physics and Astronomy, Texas A\&M University, College Station, TX 77843, USA\\
$^{7}$Department of Astronomy, University of Michigan, Ann Arbor, MI 48109, USA}}

\begin{document}

\date{Accepted 2015 June 19. Received 2015 June 18; in original form 2014 September 10}


\def\LaTeX{L\kern-.36em\raise.3ex\hbox{a}\kern-.15em
    T\kern-.1667em\lower.7ex\hbox{E}\kern-.125emX}

\maketitle

\label{firstpage}

\begin{abstract}
We present a dynamical analysis to infer the structural parameters and properties of the two nearby, compact, high velocity dispersion galaxies MRK\,1216 and NGC\,1277. Combining deep \textit{Hubble Space Telescope} imaging, wide-field integral field unit stellar kinematics, and complementary long-slit spectroscopic data out to three effective radii, we construct orbit-based models to constrain their black hole masses, dark matter content and stellar mass-to-light ratios. We obtain a black hole mass of log(\Mbh/$M_{\scriptscriptstyle \odot}$) = 10.1$_{-0.2}^{+0.1}$ for NGC\,1277 and an upper limit of log(\Mbh/$M_{\scriptscriptstyle \odot}$) = 10.0 for MRK\,1216, within 99.7 per cent (3$\sigma$) confidence. The stellar mass-to-light ratios span a range of $\Upsilon_{V}$ = $6.5_{-1.5}^{+1.5}$ in NGC\,1277 and $\Upsilon_{H}$ = $1.8_{-0.8}^{+0.5}$ in MRK\,1216 and are in good agreement with single stellar population models of a single power-law Salpeter initial mass function. Even though our models do not place strong constraints on the dark halo parameters, they suggest that dark matter is a necessary ingredient in MRK\,1216, with a dark matter contribution of $22^{+30}_{-20}$ per cent to the total mass budget within one effective radius. NGC\,1277, on the other hand, can be reproduced without the need for a dark halo, and a maximal dark matter fraction of 13 per cent within the same radial extent. In addition, we investigate the orbital structures of both galaxies, which are rotationally supported and consistent with photometric multi-\sersic\ decompositions, indicating that these compact objects do not host classical, non-rotating bulges formed during recent ($z \le 2$) dissipative events or through violent relaxation. Finally, both MRK\,1216 and NGC\,1277 are anisotropic, with a global anisotropy parameter $\delta$ of 0.33 and 0.58, respectively. While MRK\,1216 follows the trend of fast-rotating, oblate galaxies with a flattened velocity dispersion tensor in the meridional plane of the order $\beta_z \sim \delta$, NGC\,1277 is highly tangentially anisotropic and seems to belong kinematically to a distinct class of objects.
\end{abstract}


\section{Introduction}
\label{sec:introduction}

The highest velocity dispersion galaxies ($\sigma \ge 300$ \kms) are typically large and massive galaxies. Examples are the central brightest cluster galaxies (BCGs) such as M\,87 and NGC\,4884. These object are very round, have  half-light radii larger than $R_e > 8$\,kpc and absolute magnitudes brighter than $M_{Ks} \ge -25.5$. Surprisingly, the Hobby-Eberly Telescope Massive Galaxy Survey \citep[\textsc{HETMGS}]{2015ApJS..218...10V} found several small galaxies with very high velocity dispersions. In \citet[][hereafter vdB12]{2012Natur.491..729V}, six such objects were highlighted with sizes smaller than \Reff\ $\le$ 3\,kpc and central stellar velocity dispersions higher than $\sigma_{c} \ge 300$ \kms.\\

These features indicate extremely high dynamical mass densities for which there are two possible explanations, assuming reasonable stellar densities: over-massive black holes that weigh a significant fraction of the total baryonic galaxy mass, or high stellar mass-to-light ratios which would increase the stellar dynamical mass considerably but imply a stellar initial mass function (IMF) much more bottom-heavy than a Salpeter IMF.\\

According to the orbit-based dynamical models of vdB12, NGC\,1277 hosts an over-massive SMBH and possesses a stellar IMF that is consistent with a Chabrier IMF, ruling out a Salpeter IMF at $3\sigma$. Interestingly,  \citet[][hereafter E13]{2013MNRAS.433.1862E} showed a hand-picked alternative model with a smaller black hole and a Salpeter-like IMF and no dark matter, which produces a reasonable fit. Furthermore, spatially resolved spectroscopic data along NGC\,1277's major axis have been obtained and investigated in \citet[][hereafter T14]{2014ApJ...780L..20T} and \cite{2015arXiv150501485M}, which indicate a uniformly old stellar population, high constant $\alpha$-abundance and bottom heavy IMF. Their reconstructed stellar mass-to-light ratio of \MLstar\ $\simeq$ 7 is consistent with the values reported in vdB12, but much lower than the \MLstar\ = 10 adopted by E13. Clearly, all these differences call for a re-examination of NGC\,1277's stellar and central dark component.

Dark matter is not expected to be an important contributor at kpc to sub-kpc scales, but is nonetheless a key ingredient in many early-type galaxies \citep{1997ApJ...488..702R,2006MNRAS.366.1126C,2007MNRAS.382..657T,2013MNRAS.432.1709C} that needs to be taken into account in any dynamical analysis due to its degeneracy with the stellar mass-to-light ratio and hence with the black hole mass (\Mbh) \citep{2009ApJ...700.1690G}. Most studies that aimed to constrain the halo contribution to the overall mass profile either used long-slit spectroscopic observations or spatially limited integral field unit (IFU) data that rarely go beyond $\sim$ 1-2 effective radii ($\Reff$). The effective radius is only a relative scale that neither guarantees nor excludes the coverage of a substantial amount of dark matter. However, the \textsc{SAURON} and \textsc{ATLAS$^{3D}$} survey found a mean dark matter contribution of about 30\% inside 1 \Reff, which corresponds to a mean absolute scale of only $\le$ 5\,kpc. The aforementioned \textsc{HETMGS}'s compact galaxy sub-sample, though, should provide more interesting constraints in this regard. Their apparent sizes are relatively small and thus allow us to obtain detailed two dimensional stellar kinematics out to several effective radii (which at the same time also corresponds to a larger absolute coverage of up to 10\,kpc, given their mean distance) to study their mass profiles and hence to probe the existence of dark matter, which is assumed to dominate the mass budget in these remote regions.\\

The aim of this paper is to set the stage for an investigation of compact, high-dispersion galaxies from the \textsc{HETMGS} by combining long-slit spectroscopy with the \textsc{HET}, high-spatial resolution imaging with the \textit{HST} and large-field, medium- and low-resolution spectroscopic observations with the \textit{PPAK} IFU at Calar Alto. In doing so, we want to tackle several issues.

\begin{enumerate}
\item
dynamically infer the black hole mass, mass-to-light ratio and dark matter content of each galaxy,
\item
identify the dynamically hot and cold components to see whether violent relaxation or dissipative events played an important role in the recent evolution of these objects,
\item
analyse the stellar populations to obtain and further constrain reliable stellar mass-to-light ratios and IMF slopes as well as to gain insight into their formation histories,
\item
compare our results with the current picture of how galaxies and their constituents scale and evolve.
\end{enumerate}

In this paper, we focus on orbit-based dynamical models of only two objects, namely MRK\,1216 and NGC\,1277, with effective radii smaller than 2.5\,kpc and exceptional central dispersion peaks of $\sigma_{c} \ge 300$ \kms\ (Table \hyperref[tab:photometric_props]{\ref{tab:photometric_props}}). The \textit{PPAK} observations of both were taken with the V1200 medium-resolution grating, covering a wavelength range of 3400-4840 \AA. This restricted range makes a stellar population analysis not very suitable for answering IMF related questions. In addition, the kinematics for these two galaxies are currently available to a radius of $\sim$ 15\arcsec\ only, due to a much shorter exposure strategy compared to the rest of the sample. Nevertheless, the wide-field IFU data presented here still cover these objects out to $\sim$ 3 \Reff\ (i.e. $\ge$ 5\,kpc) which should be sufficient for a dynamical examination.\\

The paper is organised as follows: In Section \hyperref[sec:photometry]{\ref{sec:photometry}} we present the photometry. Section \hyperref[sec:stellar_kinematics]{\ref{sec:stellar_kinematics}} covers the stellar kinematics. In Section \hyperref[sec:dynamical_analysis]{\ref{sec:dynamical_analysis}} we carry out a dynamical analysis to constrain the black hole mass, stellar mass-to-light ratio and dark matter content of both galaxies. Section \hyperref[sec:discussion]{\ref{sec:discussion}} rounds up and discusses the results. Section \hyperref[sec:uncertainties]{\ref{sec:uncertainties}} highlights uncertainties and potential error sources, followed by a brief summary in Section \hyperref[sec:summary]{\ref{sec:summary}}.\\

Throughout this paper, we adopt 5th year results of the Wilkinson Microwave Anisotropy Probe (\textit{WMAP}) \citep{2009ApJS..180..225H}, with a Hubble constant of $H_{0}$ = 70.5 \kmsM, a matter density of $\Omega_{M}$ = 0.27 and a cosmological constant of $\Omega_{\Lambda}$ = 0.73.

\begin{table}
	\caption{Photometric properties of MRK\,1216 and NGC\,1277. (1) Morphological classification according to the NED, (2) Hubble flow distance, (3) scale at this distance, (4) effective radius in arcsec or (5) kpc, measured by a circular aperture that contains half of the light , (6) extinction corrected total luminosity of the \textit{HST} F160W (MRK\,1216) and F814W (NGC\,1277) exposures, (7) peak and effective velocity dispersion in the \textit{PPAK} data, and (8) adopted inclination.}
	\begin{center}
	\begin{tabular}{ c  c  c }
		\hline
		 & MRK\,1216 & NGC\,1277 \\
		\hline
		Type & E & S0 \\
		Distance [Mpc]  &  94$\pm$2  &    71$\pm$1    \\
		Distance Scale [kpc/arcsec] & 0.45$\pm$0.01 & 0.34$\pm$0.01 \\
		$\Reff$ [arcsec]  &  5.1   &    3.5   \\
		$\Reff$ [kpc]  &  2.3   &  1.2   \\
		Luminosity [log(\Lsun)] &  11.1   &    10.3    \\
		$\sigma_c$ / $\sigma_e$ [\kms]  & 338$\pm$8 / 308$\pm$7  & 355$\pm$7 / 317$\pm$5 \\
		Inclination [deg] & 70 & 75 \\
		\hline
	\end{tabular}
	\vspace{2ex}
	\label{tab:photometric_props}
	\end{center}
\end{table}

\section{Photometry}
\label{sec:photometry}

In this section, we present the photometric data, consisting of high-spatial resolution imaging with the \textit{HST}. The first part of this paragraph covers the reduction and combination of dithered \textit{HST} exposures to a final, super-sampled image. The second part then describes the photometric analysis of MRK\,1216 and NGC\,1277.

\subsection{HST Imaging}
\label{sec:hst_img}

We obtained single orbit imaging of MRK\,1216 with the \textit{HST} WFC3 in \textit{I}- (F814W) and \textit{H}-band (F160W), as part of program GO: 13050 (PI: van den Bosch). The data set comprises three dithered images in \textit{I}-band with a total integration time of 500 seconds and seven images in \textit{H}-band with a total integration time of 1354 seconds. The \textit{H}-band images consist of three dithered full- and four sub-array exposures. The 16\arcsec$\times$16\arcsec\ sub-array images are short 1.7 second exposures, to mitigate possible saturation in the 450 seconds full-frames of the high surface brightness nucleus.

For the photometric as well as the dynamical analysis, we give preference to the deep \textit{HST} \textit{H}-band exposures, mainly due to less dust susceptibility in the near-infrared (NIR) and the fact that the inferred stellar mass-to-light ratios become a weaker function of the underlying stellar populations \citep{2001ApJ...550..212B,2001MNRAS.326..255C}. 

The reduction and combination of the individual F160W exposures is performed via \textsc{Astrodrizzle} \citep{2012drzp.book.....G} in two major steps. First, a bad pixel mask for each flat-field calibrated image is generated, which then again is used during combination of dithered exposures, while correcting for geometric and photometric distortions. Both the deep full- and sub-array exposures in F160W are dominated by galaxy light of the huge stellar halo and hence the standard sky subtraction routine in \textsc{Astrodrizzle} (consisting of iterative sigma-clipping of uniformly distributed pixels) overestimates the background flux. We therefore measure the background flux separately in all the frames, manually, before combining the images. For the full-frames, the background level is measured in the less contaminated corners of each image, while the background flux in the sub-frames is estimated by measuring the flux difference between the (already) sky subtracted full-frames and the different sub-frames. We don't reach the sky/noise dominated regions in the full-frames, but the surface brightness (SB) in the corners of each image (where the sky estimates have been performed) is more than 10 magnitudes below the central SB, and will thus have little impact on the accurate recovery of the central stellar light/mass.

\begin{figure}
	\begin{minipage}{1.\textwidth}
		\includegraphics[width=.49\textwidth]{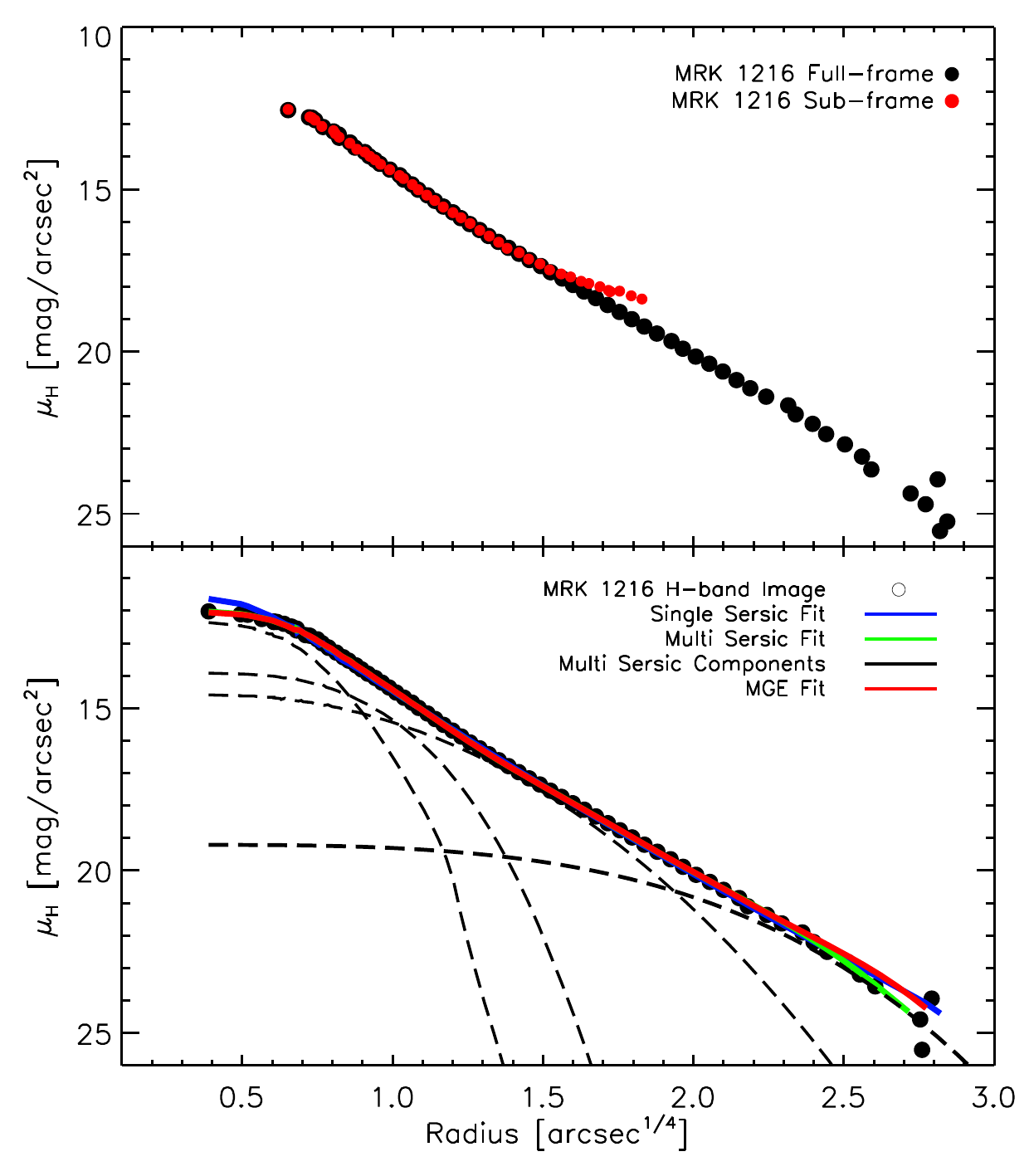}
	\end{minipage}
	\caption{\textit{Top:} Surface brightness comparison of the sky subtracted \textit{HST} (F160W) \textit{H}-band full- and sub-frames of MRK\,1216. The sky of the full-frames was calculated by iterative sigma-clipping of non-contaminated regions. The sky of the sub-frames was inferred by matching their non-sky subtracted SB profile with the sky subtracted SB profile of the full-exposures. At radii beyond 5\arcsec, the SB of the sub-exposures are noise dominated. \textit{Bottom:} Fits to the final, \textit{HST H}-band image, showing the match with a single \sersic, multi-\sersic\ and multi-Gaussian expansion. The single \sersic\ overestimates the SB measurement at both the inner- and outermost radii. The MGE accurately reproduces the SB in the inner parts but is more extended, whereas the multi-\sersic\ fit represents a fair match at all radii.}
	\label{fig:mrk1216_sb_plot}
\end{figure}

Fig. \hyperref[fig:mrk1216_sb_plot]{\ref{fig:mrk1216_sb_plot}} (top panel) shows the match of the surface brightness profiles of MRK\,1216 after subtraction of sky background in the full- and sub-array exposures. Once the sky values have been determined, we combine the frames via \textsc{Astrodrizzle} and obtain a super-sampled image with a resolution of 0.06\arcsec/pixel and a FOV of 1.5 arcmin$^{2}$ (Fig. \hyperref[fig:mrk1216]{\ref{fig:mrk1216}}).

For the photometric analysis, we adopt a \textsc{CANDELS} point-spread function (PSF) \citep{2012ApJS..203...24V}. In brief, the PSF has a size of 0.17\arcsec\ FWHM and has been generated with \textsc{TinyTim} \citep{1995ASPC...77..349K} for the F160W filter. The PSF is created in the centre of the WFC3 detector, to minimise distortion, and is 10 $\times$ sub-sampled. Resampling it back to the original \textit{HST} \textit{IR} scale of 0.13\arcsec/pixel, and applying a kernel to replicate the effects of inter-pixel capacitance, creates a synthetic star in the centre of each frame. The final PSF is then obtained by "drizzling" the images and thus the PSFs. In this way, we produce a point-spread function at the same scale as our final science image.\\

High-resolution imaging of NGC\,1277 is available in the Hubble Legacy Archive. Observations of this galaxy have been carried out in program GO: 10546 (PI: Fabian), resulting in three dithered exposures in \textit{R}- (F625W) and \textit{V}-band (F550M) with a total integration time of 1654s and 2439s respectively. In contrast to the \textit{I}- and \textit{H}-band images of MRK\,1216, the redder \textit{R}-band does not have a significant advantage over the \textit{V}-band. The leverage between the two filters is small and consequently both are equally subject to the effects of extinction (see Section \hyperref[sec:ngc1277_photometry]{\ref{sec:ngc1277_photometry}}). Here, we employ the \textit{V}-band photometry because of its longer exposure time and for the sake of consistency with the modelling results of vdB12.

The F550M flat-field calibrated images have been sky subtracted, cosmic ray rejected, corrected for photometric and geometric distortions via \textsc{Astrodrizzle} before being combined into a final image with a resolution of 0.05\arcsec/pixel. The PSF of these observations was recovered with \textsc{TinyTim}, created in each of the three individual, dithered exposures and drizzled to match the resolution of the corresponding science frame.

\subsection{MRK\,1216}
\label{sec:mrk1216_photometry}

MRK\,1216 is a sparsely investigated early-type galaxy (ETG) ($\rmn{R.A.}: 08^{\rmn{h}} 28^{\rmn{m}} 4\degr$, $\rmn{Decl.}: -06\degr 56\arcmin 22\arcsec$) with strong excess of UV radiation in its centre \citep{1963SoByu..34....3M}. A few redshift measurements have been carried out for this object \citep{2007ApJS..170...33P, 2009MNRAS.399..683J}, which translate to a Hubble flow distance of $94{\pm 2}$ Mpc. Given its distance, 1\arcsec\ is equivalent to $450\pm10$ pc/arcsec. The final, combined \textit{HST} image thus covers a field of view (FOV) of 65\,kpc$^2$.\\

To examine its photometric properties, structure and morphology we decompose the galaxy into multiple \sersic\ components using \textsc{Galfit} \citep{2002AJ....124..266P}. The analysis is done for 3 different scenarios: First, a single \sersic\ fit is carried out to obtain the single \sersic\ index and thus the overall steepness of the light profile. Second, we perform a bulge-disk decomposition, if possible. Although such a decomposition is a matter of debate, we do this for comparison with literature studies, where similar procedures have been carried out to relate central black hole masses to bulge luminosities. Third, we execute a fit with multiple \sersic\ components that best matches the light profile \footnote{All magnitudes presented throughout this section are corrected for Galactic extinction \citep{2011ApJ...737..103S}; 0.017 mag in \textit{H}-band and 0.431 mag in {V}-band. The sizes are semi-major axis radii, unless mentioned otherwise.}.

\begin{figure*}
		\begin{center}
		\includegraphics[width=.49\textwidth]{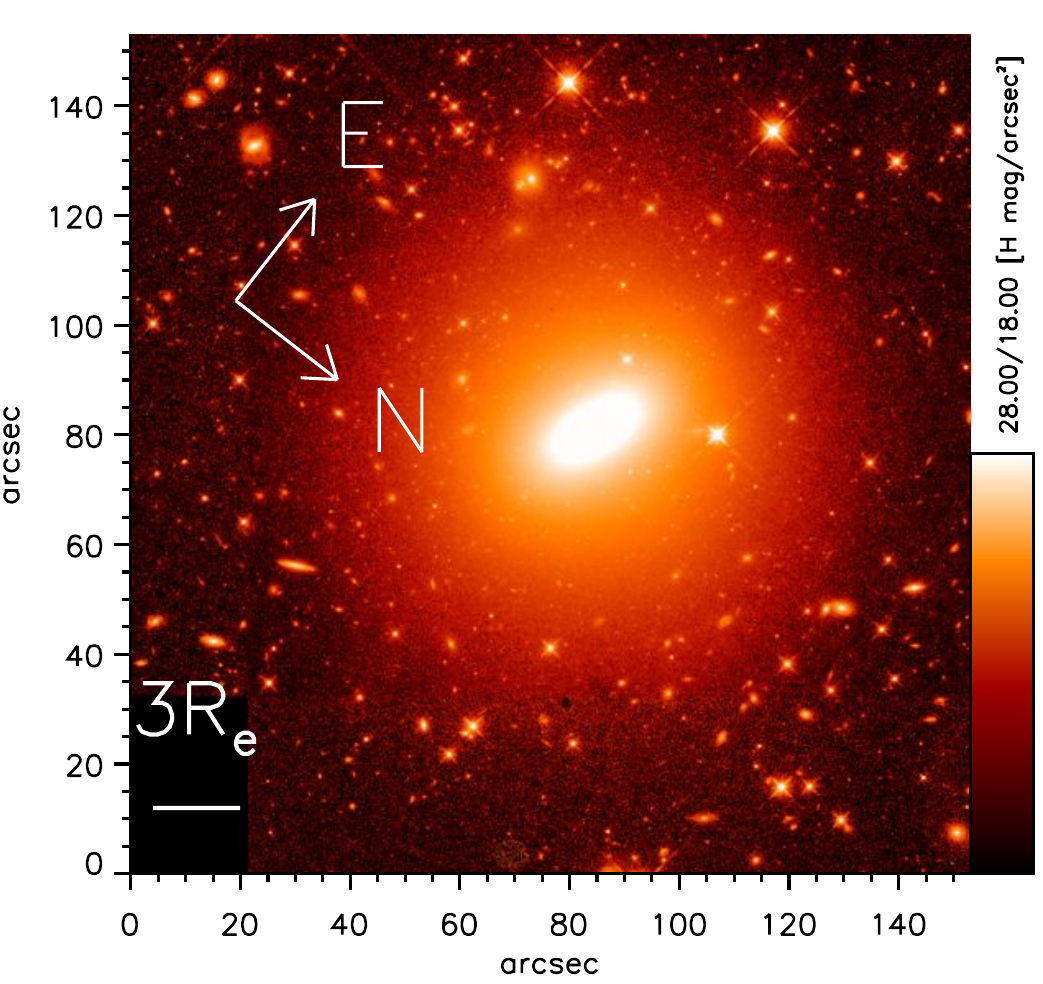}
		\includegraphics[width=.49\textwidth]{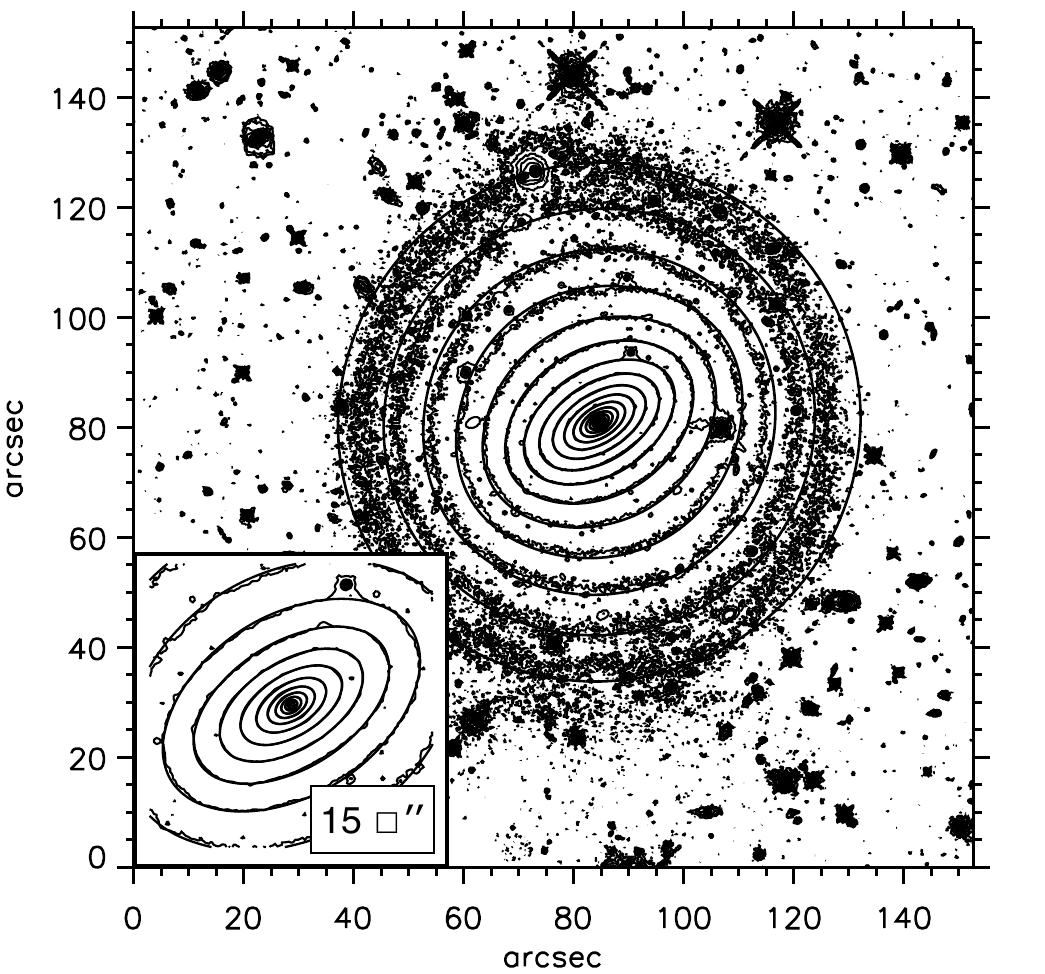}
		\end{center}
	\caption{\textit{Left:} \textit{HST} \textit{H}-band image of MRK\,1216, which covers a field of $\sim$ 150 $\square$\arcsec\ (i.e. $\sim$ 70\,kpc$^{2}$), with a final scale of 0.06\arcsec/pixel. \textit{Right:} Contour map of the same image.  The MGE contours are over-plotted in black. The bottom left plot shows the accurate reproduction of the surface brightness profile within the central 15  $\square$\arcsec\ ($\sim$ 3 \Reff).}
	\label{fig:mrk1216}
\end{figure*}

A single \sersic\ fit to MRK\,1216's \textit{H}-band image has an apparent magnitude of $m_{H, Vega} = 10.47$, an effective radius of \Reff\ = 6.34\arcsec, a projected axis ratio $(b/a) = 0.58$ and a single \sersic\ index of $n = 4.93$. Residuals of this fit are strong. In comparison to the SB measurement, the single \sersic\ fit shows an excess of light in the very centre and also tends to overpredict the light of the large outer halo (Fig. \hyperref[fig:mrk1216_sb_plot]{\ref{fig:mrk1216_sb_plot}}, bottom panel).

We further investigate the stellar structure by gradually increasing the number of \sersic\ components.
A two-component model yields a very centrally concentrated (\Reff\ = 3.42\arcsec) "bulge" with a \sersic\ index of $n = 3.61$ - although remarkably flat $(q = 0.56)$ - which is embedded in a (close to) exponential, round and very extended (\Reff = 17.22\arcsec) stellar "disk"/envelope with a \sersic\ index of $n = 0.96$. Pronounced residuals remain, hinting at a more complex stellar composition. Even so, the "bulge" of the two-component fit will serve as an upper limit to the bulge luminosity. According to this fitting scenario, we obtain a bulge-to-total luminosity ratio of $B/T = 0.69$.

A decent fit is obtained with at least four \sersic\ components, resulting in notably lower and less prominent residuals. In this case, the outer stellar "disk"/envelope persists, whereas indications of a "bulge"-like component totally disappear (Table \hyperref[tab:mrk1216_sersic]{\ref{tab:mrk1216_sersic}}). All components show rather low \sersic\ indices, which complicates any attempt of a morphological interpretation. We therefore do not present a unique classification but rather stick to the conclusion that MRK\,1216 is indeed a compact ETG, harbouring a complex, flat substructure, that is embedded in a round, extended stellar halo. As the innermost component is too small to be considered a bulge, we adopt the luminosity of the second innermost \sersic\ as a conservative lower limit of a bulge luminosity, which accounts for 12 per cent of the total light and extents to 1.34\arcsec\ (or roughly 0.3 $R_{e}$).\\

\begin{table}
	\caption{\sersic\ decomposition of MRK\,1216's \textit{HST} (F160W) \textit{H}-band image. The columns represent the number of \sersic s for a given fitting scenario (1), their apparent total magnitude (extinction corrected) (2), their effective semi major axis radius (3), the corresponding \sersic\ index (4) and their apparent flattening (5).}
	\begin{center}
	\begin{tabular}{ c  c  c  c  c }
		\hline
		\# of components & mag [H, Vega] & \Reff\ [arcsec] & $n$ & $q$ \\
		\hline
		1  &  10.49   &    6.34    &    4.93 & 0.58 \\
		\hline
		1  &  10.89   &    3.42    &    3.61 & 0.56 \\
		2 &  11.77   &    17.22    &    0.96 & 0.88 \\
		\hline
		1  &  13.19   &    0.40    &    1.27 & 0.81 \\
		2  &  12.77   &    1.34    &    0.99 & 0.57 \\
		3 &  11.32   &    5.59    &    1.61 & 0.52 \\
		4  &  11.60   &    19.31    &    1.05 & 0.99 \\
		\hline
	\end{tabular}
	\vspace{2ex}
	\label{tab:mrk1216_sersic}
	\end{center}
\end{table}

Our orbit-based dynamical models need a stellar mass model, from which we can infer the stellar gravitational potential.  This is accomplished by deprojecting the surface brightness distribution of a galaxy which however is a non-unique task, as has been convincingly illustrated by \cite{1987IAUS..127..397R}. Even the surface brightness distribution of an axisymmetric stellar system only provides information about its density outside a so-called "cone of ignorance". This means that in principle, and unless the galaxy is observed edge-on, there could be a family of "konus densities" \citep{1996MNRAS.279..993G} which alter the intrinsic mass distribution but are invisible to the observer as they project to zero surface brightness. Making use of physically and observationally motivated criteria for the luminosity profile of axisymmetric galaxies, \cite{1997MNRAS.287..543V} found that the addition of mass due to konus densities cannot be arbitrary and is most likely confined to be less than 10 per cent for (cusped) ellipticals, implying a marginal role in the dynamics of early-type galaxies. We employ a similar, empirically motivated approach by parameterising the surface brightness distribution of galaxies with a set of multiple, two-dimensional Gaussian functions (\textsc{MGE}: \citealt{1992A&A...253..366M, 1994A&A...285..723E}). Although a set of Gaussians does not form a complete set, the MGE method has been very successful in the recovery of the surface brightness profiles and features of realistic multi-component galaxies \citep{2002MNRAS.333..400C}. We obtain the intrinsic luminosity density by deprojecting the parameterised surface brightness distribution for a given/assumed set of viewing angles, adopting an absolute magnitude of 3.32 for the Sun in \textit{H}-band \citep{1998gaas.book.....B}. In the case of an MGE, the deprojection can be performed analytically while the gravitational potential is then obtained by means of a simple, one-dimensional integral.

Our final MGE contains 10 components with a fixed position angle (PA) of 70.2\degree (measured counter-clockwise from the y-axis to the galaxy major axis, with the image aligned N.-E., i.e. north is up and east is left) and a common centre, as listed in Tab. \hyperref[tab:mrk1216_mge]{\ref{tab:mrk1216_mge}}. The flattest Gaussian has an axis ratio of $q = 0.52$, which forces the lower boundary of possible inclinations to be greater than 59$\degree$ (with 90$\degree$ being edge-on), assuming oblate axial symmetry (see Section \hyperref[sec:uncertainties]{\ref{sec:uncertainties}}). A dust disc would be helpful in further constraining the inclination of the galaxy, although it would also pose a major concern for the modelling of the stellar mass, but is not evident in either of the \textit{H}- and \textit{I}-band images of MRK\,1216.

\begin{table}
	\caption{Multi-Gaussian-Expansion of MRK\,1216's \textit{HST} (F160W) \textit{H}-band image. The columns display the number of each Gaussian, beginning with the innermost one (1), their total \textit{H}-band magnitude (2), their effective radius (3) and their corresponding \sersic\ index (4) as well as their apparent flattening (5).}
	\begin{center}
	\begin{tabular}{ c  c  c  c  c }
		\hline
		\# of components & mag [H, Vega] & \Reff\ [arcsec] & $n$ & $q$ \\
		\hline
		1 & 15.92   &    0.094    & 0.5 &  0.72 \\
		2 & 14.25   &    0.230    & 0.5 &  0.77 \\
		3 & 13.75   &    0.458   &  0.5 &  0.73 \\
		4 & 13.19   &    0.919   &  0.5 &  0.59 \\
		5 & 12.99   &    1.711   &  0.5 &  0.55 \\
		6 & 12.73   &    3.208  &  0.5  & 0.52 \\
		7 & 12.41   &    5.905  &  0.5 & 0.54 \\
		8 & 12.52   &    9.848  &  0.5  & 0.67 \\
		9 & 12.50   &    14.72  &  0.5  & 0.96 \\
		10 & 12.37   &  29.80  &  0.5  & 0.99 \\
		\hline
	\end{tabular}
	\vspace{2ex}
	\label{tab:mrk1216_mge}
	\end{center}
\end{table}

Fig. \hyperref[fig:mrk1216]{\ref{fig:mrk1216}} shows the combined, final \textit{H}-band image of MRK\,1216 (left) and its contour map (right). Over-plotted are contours of the MGE (black) as well as an excerpt of the central 15 arcsec$^{2}$ ($\sim$ 3 \Reff). The MGE reproduces the SB profile within the central 40 arcsec$^{2}$, but tends to overpredict it at the largest radii. Note that the SB profile lacks any PA twists. The PA is almost constant within 30\arcsec\ from the centre ($\Delta$PA $\le$ 2\degree) and changes only at larger radii where it is virtually unconstrained as the round outer halo has close to zero ellipticity.

\subsection{NGC\,1277}
\label{sec:ngc1277_photometry}

Given several redshift measurements, NGC\,1277 is located at a Hubble flow distance of $71{\pm 1}$ Mpc and deeply embedded in the Perseus cluster. It is classified as a lenticular S0 galaxy \citep{2001ApJS..132..129M} without any noticeable substructures or prominent features besides the clearly visible central dust disk with a semi-major axis radius of 0.13\,kpc and a flattening of $q = 0.3$. The presence of dust complicates the recovery of the central stellar mass and hence the black hole mass in NGC\,1277 (see Section \hyperref[sec:uncertainties]{\ref{sec:uncertainties}}), but assuming that this nearly edge-on disk traces the PA of its host, we can pin down the inclination of the galaxy to 75\degree. As in the case of MRK\,1216, photometry shows that a superposition of galaxies can be ruled out as an explanation for the observed high velocity dispersions.\\

The SB profile of NGC\,1277 shows a flattened, regular structure with no significant changes in the PA with increasing distance from the centre ($\Delta$PA $\le$ 2\degree). A single \sersic\ fit to its \textit{V}-band image reveals a moderate \sersic\ index of $n = 2.24$, a small effective radius of $R_e$ = 3.9\arcsec\ and a projected axis ratio $(b/a)$ of 0.53. The total \textit{V}-band magnitude is $m_{V, Vega}$ = 13.39. This rather simple fitting scenario is, of course, an under-representation of NGC\,1277's stellar complexity, leading to strong residuals in the centre - where the luminosity profile shows an excess of light when compared to the \sersic\ - and at larger radii.

A further decomposition with two components improves the fit significantly. Here, a flat $(b/a = 0.52)$, inner ($R_{e} = 2.85\arcsec$) bulge-like component ($n = 2.25$) is embedded in a rather flat $(b/a = 0.50$), outer ($R_{e} = 10.35 \arcsec$) disk like component ($n = 0.37$). 
The fit has a bulge-to-total ratio of $B/T = 0.75$.

An acceptable fit can be obtained with (at least) 4 \sersic\/s, as listed in vdB12. An interpretation of the various components, however, is difficult, except for the outermost component which resembles a round stellar halo. None of the components has a high \sersic\ index, making it difficult to find any photometric evidence for the presence of a "bulge". Devoid of a distinct spheroid component in this multi-\sersic\ fit of NGC\,1277, we again adopt the "bulge" of the two component decomposition as an upper limit to the bulge luminosity ($B/T=0.75$) whereas the luminosity of the second innermost \sersic\ in the four component decomposition will serve as a lower limit ($B/T=0.24$).

For the stellar mass model we make use of the \textit{V}-band Multi-Gaussian-Expansion of vdB12 with a fixed PA of 92.7\degree. E13 provided an alternative MGE, based on the \textit{R}-band image of NGC\,1277. The difference between the two parameterisations, though, is of little account. Both MGEs reproduce the 2D surface brightness profile equally well, and we refer the reader to E13 for an illustration of the isophotes. There is basically not enough leverage between the wide \textit{R}-band and medium \textit{V}-band filter to obtain any colour information that would also minimise the effect of obscuration by the central dust disc. The MGE of E13 in \textit{R}-band yields a lower total luminosity while increasing the central surface brightness only slightly. These values however do not change the inferred stellar dynamical masses substantially (Sec. \hyperref[sec:ngc1277_schwarzschild]{\ref{sec:ngc1277_schwarzschild}}).

\section{Stellar Kinematics}
\label{sec:stellar_kinematics}

This section covers the \textit{HET} long-slit and \textit{PPAK} IFU observations. After sketching the reduction of the individual data sets, we extract and present the kinematics which in turn are used as input for our orbit-based dynamical models.

\subsection{PPAK}
\label{sec:ppak}

Large-field, medium-resolution (V1200) observations of both galaxies have been carried out at the 3.5 m telescope at Calar Alto, with the \textit{Potsdam Multi Aperture Spectrograph} (\textit{PMAS}) \citep{2005PASP..117..620R} in the \textit{PPAK} mode \citep{2004AN....325..151V,2006PASP..118..129K}. The observing details of this run are outlined in Sec. \hyperref[sec:mrk1216_kinematics]{\ref{sec:mrk1216_kinematics}}. The \textit{PPAK} IFU consists of 382 fibers, which are bundled to a hexagonal shape. Each fiber has a diameter of 2.7\arcsec, resulting in a FOV of roughly 1.3 arcmin$^{2}$. Using a 3 dither-pointing strategy, the 331 science fibers have a 100\% covering factor across the entire FOV. An additional 36 fibers are used to sample the sky, while the remaining 15 fibers are used for calibration purposes.
The V1200 grating has a resolving power of $R = 1650$, at 4000 \AA. The spectral resolution across the nominal
$3400 - 4840$ \AA\ spectral range and FoV is homogenised to 2.3 \AA\ FWHM based on measured line widths in the arc lamp exposure. This spectral resolution corresponds to an instrumental velocity dispersion of $\sigma = 85$ \kms. The low
sensitivity at the blue end and vignetting at the red end reduce the useful spectral range to $3650 - 4620$ \AA.\\

The reduction of the \textit{PPAK} data follows the reduction procedure of the Calar Alto Legacy Integral Field Spectroscopy Area \textsc{(CALIFA)} survey. An extensive overview of the reduction pipeline is given in \cite{2012A&A...538A...8S} and \citet{2013A&A...549A..87H}. The data reduction steps by the pipeline include bias subtraction, straylight subtraction, cosmic ray rejection with PyCosmic \citep{2012A&A...545A.137H}, optimal fiber extraction, fiber flat-fielding, flexure correction, wavelength calibration and flux calibration. The sky subtraction is done by averaging the spectra of 36 dedicated sky fibers which are located 72\arcsec\ away from the \textit{PPAK} FoV centre. Given our compact objects' sizes, the sky fibers should be free from any contamination of the galaxy itself. To exclude any potential contamination by field stars or low-surface brightness objects, the sky spectrum is constructed by taking the mean of only the 30 faintest sky fibers. The resulting sky spectrum is then subtracted from its associated science exposure. Finally, the 3 dither-pointings are resampled to the final data cube with a 1\arcsec\ sampling using a distance-weighted 
interpolation algorithm as described in \citet{2012A&A...538A...8S}.\\

To measure reliable stellar kinematics, we first spatially bin the data with the adaptive Voronoi tessellation technique, as implemented by \citet{2003MNRAS.342..345C}. At the cost of spatial resolution, we co-add spectra into (Voronoi-) zones to reach a minimum S/N of 25 in each bin, after applying a minimum S/N cut of 4 for each spaxel. The binning process, however, is not straightforward. Spectra of grouped pixels are assumed to be uncorrelated, which is not true for our data. Due to the three-point dither-pattern of the \textit{PPAK} observations, correlated errors during binning appear. In the most general case, each spaxel contains information from a number of different fibers (as each fiber contributes to more than just one pixel). Hence, the noise in adjacent spaxels \emph{is} correlated and spatial covariances have to be taken into account during the S/N estimates of co-added spectra.
A correction for the correlation of the S/N in the data has been applied by quantifying the ratio of the real error - directly estimated from residuals of full spectrum continuum fitting - to the analytically propagated error of binned spectra. The ratio can be characterised by a logarithmic function \citep{2013A&A...549A..87H}: 
\begin{equation}
\epsilon_{real} \:/\: \epsilon_{bin} = 1 + \alpha \; \text{log} \; n
\end{equation}
with $\alpha = 1.38$ and $n$ as the number of spaxels in each bin. The ratio increases rapidly for small bins, indicating a high correlation between adjacent spaxels, and flattens out for spatially bigger bins, where the correlation between spaxels becomes less.

Once the effect of noise correlation has been taken into account, we adopt the Indo-US stellar library with 328 spectral templates \citep{2004ApJS..152..251V}. A non-negative linear combination of these templates is then convolved with a Gaussian line-of-sight  velocity distribution (LOSVD) and fitted to each spectrum in the range of $3750 - 4550$ \AA\ (covering prominent stellar absorption line features such as the Balmer and Calcium H- and K lines), while using additive Legendre polynomials of 15th order. In this way, we derive the mean line-of-sight velocity $v$, velocity dispersion $\sigma$ and higher order Gauss-Hermite velocity moments $h_3$ and $h_4$ (which represent asymmetric and symmetric departures from a Gaussian LOSVD)  per bin on the plane of the sky. Sky lines are masked beforehand and corresponding uncertainties of the kinematic moments are determined by means of 100 Monte Carlo simulations (see also Falc\'{o}n-Barroso et al., in prep. for a full description of the \textsc{CALIFA} stellar kinematics pipeline).

\subsection{HET}
\label{sec:het}

In addition to the medium-resolution wide-field \textit{PPAK} data, \textit{HET} long-slit kinematics along the major axis are in hand. Observations were carried out, using the Marcario Low Resolution Spectrograph (\textit{LRS}) \citep{1998SPIE.3355..433H}. The \textit{LRS} is a classical long-slit spectrograph with a slit length of 4\arcmin. We made use of the G2 grating and a slit width of 1", covering a wavelength range of 4200-7400 \AA. This configuration has a resolution of $R = 1300$ which corresponds to a spectral resolution of 4.8 \AA\ FWHM (i.e. an instrumental velocity dispersion of $\sigma$ = 108 \kms), at a pixel scale of 0.475\arcsec\ . Single exposures of 900 s have been taken for each galaxy, in good weather and seeing conditions of 1 arcsec, resulting in a total of 3 individual (apparent) major axis profiles each for MRK\,1216 and NGC\,1277.

The reduction of the \textit{HET} data is accomplished by a dedicated and fully automated pipeline \citep{2015ApJS..218...10V}, following a standard reduction practice of bad pixel and cosmic ray masking, overscan and bias subtraction, flat fielding correction and wavelength correction.
From the reduced data, we then extract kinematic information by applying an updated version of the pPXF code \citep{2004PASP..116..138C} with a set of 120 spectral templates from the MILES stellar library \citep{2006MNRAS.371..703S, 2011A&A...532A..95F}.

The PSF and positioning of our observations are crucial for an accurate determination of the modelling parameters, such as black hole mass and stellar mass-to-light ratio. We recover a reliable PSF in each data set of both galaxies by iteratively fitting a PSF convolved, reconstructed (slit) image to the MGE of the high-resolution \textit{HST} data. The PSF in turn is expanded by multiple, round Gaussians and the (slit) images are well reproduced in all cases by a PSF with one or two components (Tab. \hyperref[tab:psf]{\ref{tab:psf}})

\begin{table}
	\caption{Reconstructed point-spread functions of MRK\,1216's and NGC\,1277's \textit{PPAK} and \textit{HET} data. The rows display the individual data sets with the slit numbers appended (1), the point-spread function of MRK\,1216 for this particular data set, expanded by a set of multiple Gaussians with weight and effective radius (2), and the same for NGC\,1277 (3).}
	\begin{center}
	\begin{tabular}{ c  c  c }
		\hline
		Data &  MRK\,1216 & NGC\,1277 \\
			& [weight] [arcsec] & [weight] [arcsec] \\
		\hline
		PPAK  &  0.77      1.34    &   0.75      1.21 \\
			&  0.23      3.72    &   0.25      2.44\\
		\hline
		HET 1  &  0.55      1.19   &        1.00      0.83 \\
		  &  0.45      3.39   & \\
		HET 2  &  0.78      0.69   &        1.00      0.83 \\
		  &  0.22      1.89   & \\
		HET 3  &  0.64      1.10   &        1.00      0.83 \\
		  &  0.36      3.77   & \\
		\hline
	\end{tabular}
	\vspace{2ex}
	\label{tab:psf}
	\end{center}
\end{table}

\subsection{MRK\,1216}
\label{sec:mrk1216_kinematics}

\begin{figure*}
	\includegraphics[width=1.\textwidth]{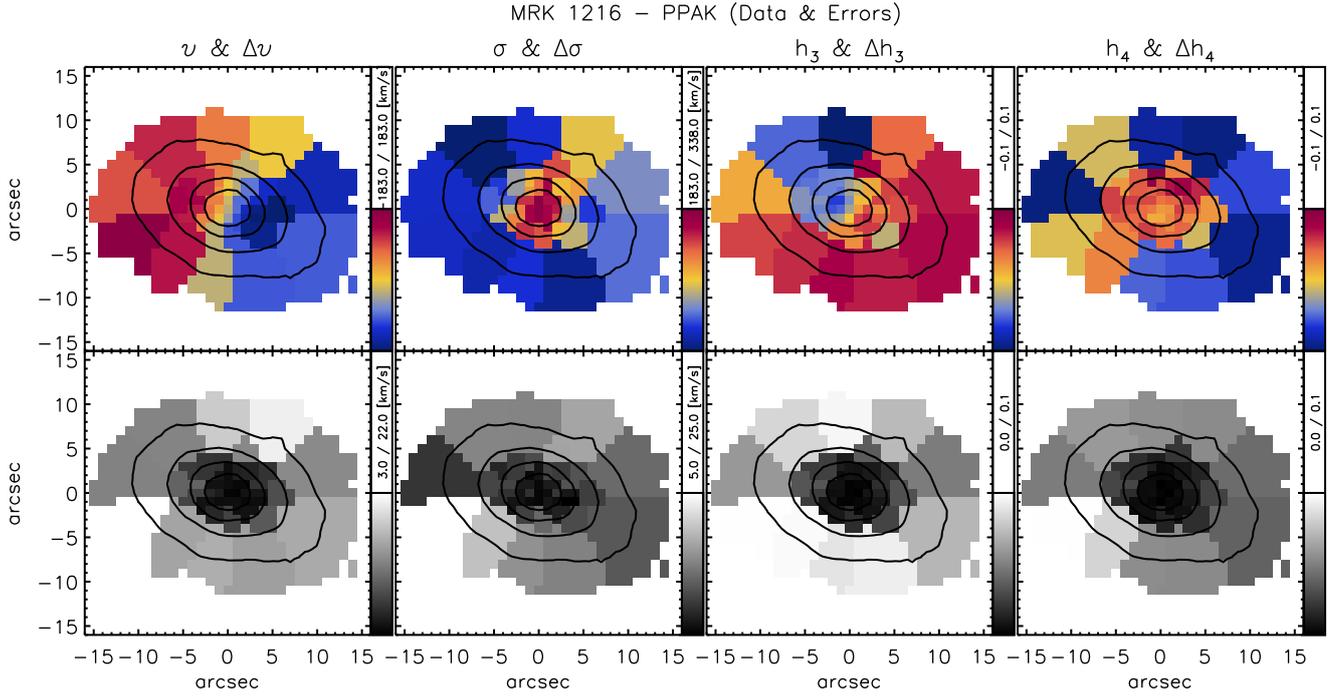}
	\caption{\textit{Top:} \textit{PPAK} IFU stellar kinematic maps of MRK\,1216, showing the mean velocity ($v$), velocity dispersion ($\sigma$), $h_{3}$ and $h_{4}$. The maps show fast rotation around the short axis of 182 \kms\ and a central velocity dispersion of 338 \kms. Given its effective radius of $\sim$ 5\arcsec, the \textit{PPAK} data cover the kinematics out to $\sim$ 3 \Reff. Overplotted are contours of constant surface brightness from the same observing run. \textit{Bottom:} Corresponding uncertainty maps. Maps are oriented N.-E., i.e. north is up and east is left.}
	\label{fig:mrk1216_ppak_maps}
\end{figure*}

\begin{figure*}
	\includegraphics[width=1.\textwidth]{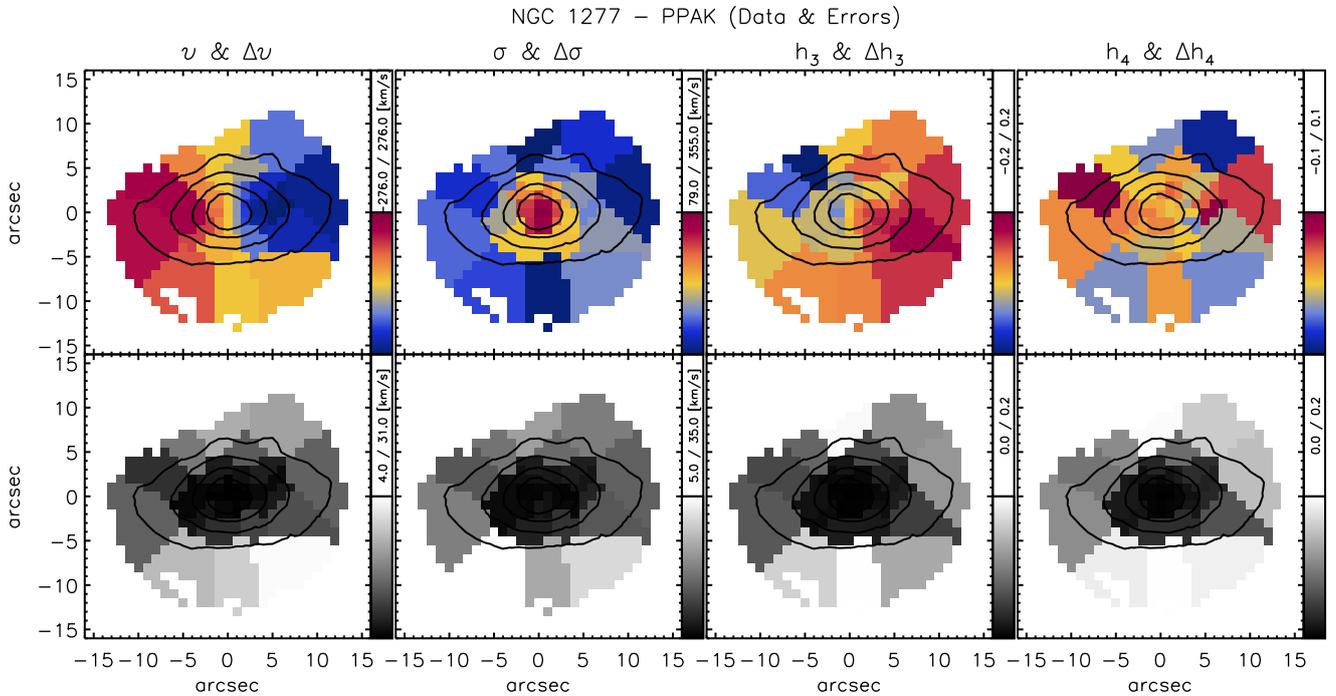}
	\caption{\textit{Top:} \textit{PPAK} IFU stellar kinematic maps of NGC\,1277 with contours of constant surface brightness, revealing fast rotation of 276 \kms\ and a peak in velocity dispersion of 355 \kms. NGC\,1277 has an effective radius of $\sim$ 3.5\arcsec, and the data thus cover the kinematics out to $\sim$ 3-4 \Reff. \textit{Bottom:} Corresponding uncertainty maps. All maps are oriented N.-E..}
	\label{fig:ngc1277_ppak_maps}
\end{figure*}

On December 5, 2011, we obtained \textit{PPAK} data using the medium-resolution V1200 grating. The seeing was $\sim$ 1 arcsec. Two science exposures, 900 seconds each, were taken per pointing, resulting in a total exposure time of 1.5 hours on-source.

Fig. \hyperref[fig:mrk1216_ppak_maps]{\ref{fig:mrk1216_ppak_maps}} displays the two-dimensional line-of-sight kinematics. Reliable data are available out to a major-axis radius of 15\arcsec. The kinematic maps show fast and regular rotation with a maximum velocity of 182 \kms. The velocity dispersion has a very pronounced peak of about 338 \kms\ in the centre, indicating a very high mass concentration and hinting at the presence of a SMBH. Superimposed are contours of constant surface brightness from the same data cube.

One out of our three \textit{HET} long-slit kinematics along the apparent major axis will be illustrated in Fig. \hyperref[fig:m1216_maps_models]{\ref{fig:m1216_maps_models}}. The velocity and velocity dispersion profiles are in agreement with the \textit{PPAK} data, revealing a rotation with a maximum velocity of 219 \kms\ and a peak in velocity dispersion of 345 \kms. Furthermore, we measure strong $h_{3}$ moments that appear to be anti-correlated with $v$, commonly observed in axisymmetric galaxies.

\subsection{NGC\,1277}
\label{sec:ngc1277_kinematics}

\textit{PPAK} data of NGC\,1277 have been obtained in the same run as data of MRK\,1216. The observing strategy, setup as well as data processing and reduction are also identical, resulting in the kinematic maps in Fig. \hyperref[fig:ngc1277_ppak_maps]{\ref{fig:ngc1277_ppak_maps}}. Similar to MRK\,1216, the kinematic data of NGC\,1277 are limited to a radius of $\sim$ 15\arcsec. The maps reveal very fast rotation around the short axis, peaking at 276 \kms, and an extraordinarily flat rotation curve out to several effective radii. The peak in velocity dispersion is about 355 \kms\ and hence considerably lower than the dispersion in the (three major axis) \textit{HET} slits (415 \kms) (Fig. \hyperref[fig:n1277_maps_models]{\ref{fig:n1277_maps_models}}). Moreover, the central $h_4$ measurements are also lower in the \textit{PPAK} data cube. The difference, though, is largely attributable to the difference in spatial resolution between both data sets (see Section \hyperref[sec:uncertainties]{\ref{sec:uncertainties}} for the reliability of the individual measurements and the recovery of the black hole mass).

We observe the same anti-correlation between $h_{3}$ and $v$, which is expected in the case of axial symmetry and reasonable given NGC\,1277's apparent flatness and strong rotation around its short axis. We superimpose its contours of constant surface brightness from the same data cube, with slight irregularities due to extensive masking of nearby objects and the presence of faint fore- and background stars.

\section{Dynamical Analysis}
\label{sec:dynamical_analysis}

We introduce our orbit-based dynamical modelling technique which fits the observed line-of-sight kinematics and the photometry, simultaneously. We hereby constrain the intrinsic contributions of black hole, stars and dark matter to the overall mass budget and infer the orbital structure of both galaxies.

\subsection{Schwarzschild's Method}
\label{sec:schwarzschild}

Schwarzschild's orbit superposition method \citep{1979ApJ...232..236S} has proven to be a reliable technique to recover in great detail the dynamical and structural properties of galaxies. The basic idea behind this modelling approach is as simple as it is striking: The motion of astronomical objects, e.g. stars, is governed by the underlying gravitational potential, which in turn can be a sum of not only visible matter but also any non-visible components. This means that once a gravitational potential is assumed, a representative library of orbits can be calculated in that potential that thoroughly samples all integrals of motion (4 in the case of spherical systems and 3 in axisymmetric or triaxial configurations).
By assigning weights to the orbits we can then compute their combined properties and compare them to present-day observables, which represent a snapshot of a certain gravitational and dynamical configuration. The implementation of Schwarzschild's method then probes a set of gravitational potentials and tests whether there is a steady-state superposition of orbits in that potential that matches the full LOSVD and the (intrinsic and projected) light/mass distribution.

A wealth of Schwarzschild codes exist. Starting with the modelling of spherical galaxies (e.g. \citealt{2003Sci...301.1696R}), to axisymmetric galaxies (e.g. \citealt{1999ApJS..124..383C,2003ApJ...583...92G,2004ApJ...602...66V,2004MNRAS.353..391T,2008ApJ...682..841C}) right up to the modelling of triaxial systems (\citealt{2008MNRAS.385..647V}, vdB08 hereafter). In what follows, we make use of the triaxial implementation of Schwarzschild's method. This code represents a very flexible way to reproduce not only all available data but at the same time to recover the internal dynamical structure of galaxies \citep{2008MNRAS.385..614V}, to constrain their intrinsic shapes \citep{2009MNRAS.398.1117V}, their SMBH masses \citep{2010MNRAS.401.1770V}, their (constant) mass-to-light ratios \citep{2013MNRAS.434L..31L}, as well as their dark matter fractions and profiles \citep{2009MNRAS.398..561W}. For a detailed overview of the working principles, we refer the reader to vdB08. Here, we confine ourselves to a brief description of the main steps:

\begin{enumerate}
	\item The implementation begins with a surface brightness distribution that has been parameterised with a set of Gaussians (see Section \hyperref[sec:mrk1216_photometry]{\ref{sec:mrk1216_photometry}} and \hyperref[sec:ngc1277_photometry]{\ref{sec:ngc1277_photometry}}). Once a set of viewing angles are chosen, a de-projection of the surface brightness, corresponding to the surface mass density, can be carried out which yields the intrinsic stellar mass distribution and hence the stellar gravitational potential of the galaxy. For a triaxial deprojection three viewing angles are needed to pin down the shape and orientation of the triaxial ellipsoid. On the other hand, in the axisymmetric case, the short- to long-axis ratio - i.e. the flattening ($q$), which is then directly related to the inclination ($i$) - remains the only free parameter.
	\item Within this potential, a representative library of orbits is calculated. In this work, the library consists of more than 7500 orbits (dithering excluded), given by the 9 starting points (in each of the radial and angular directions) at each of the 31 logarithmically sampled equipotential shells between 0.003\arcsec\ and 150\arcsec.
	\item During orbit integration, intrinsic and projected quantities are stored and then PSF convolved for comparison with the data.
	\item For a given potential, $\chi^2$ statistics is used to find a non-negative, linear superposition of orbits that matches the set of kinematic and photometric observables. We recover the spatially binned 2D LOSVD in a least-squares sense by finding the optimal set of orbital contributions to the Gauss-Hermite moments \citep{1993MNRAS.265..213G,1993ApJ...407..525V,1997ApJ...488..702R}. The orbital contributions, in turn, represent the mass in each orbit. To ensure self-consistency, they must be able to simultaneously reproduce the intrinsic and projected stellar masses, which are stored on a grid and given by the integration of the MGE model contribution over the respective area. The masses are constrained to an accuracy of 2 per cent, which reflects the usual uncertainties in the surface brightness parameterisation by the MGE, and are allowed to vary within the boundaries while finding the best-fitting kinematics.
	\item A reiteration of the steps (i - iv) is carried out for differing gravitational potentials, including the presence of a SMBH and dark matter.
\end{enumerate}

For multiple reasons we do not employ regularisation during the construction of the Schwarzschild models. First, we hereby make sure that our models are unbiased with respect to regularisation. Second and more importantly, it is not possible to accurately determine a proper level of regularisation that is needed a priori. Third, it has been shown that regularisation changes neither the values of the best-fitting parameters nor the orbital weights significantly, as long as reasonable values are chosen and an over-smoothing of the distribution function (DF) is prevented \citep{2002MNRAS.335..517V,2008MNRAS.385..647V,2010MNRAS.401.1770V}. And finally, while regularisation can be helpful in individual cases to find the set of orbital weights that best fits the velocity moments and to prevent the weights from varying rapidly, it decreases the degrees of freedom at the same time and leads to an artificial narrowing of the $\chi^2$ contours and thus to smaller confidence intervals for the recovered parameters (but see \citealt{2004ApJ...602...66V}, \citealt{2005MNRAS.360.1355T} and \citealt{2013MNRAS.431.3570M} for a more detailed discussion of the effects of regularisation in their individual models).\\

In the case of MRK\,1216 and NGC\,1277, the photometry and kinematics are consistent with oblate axial symmetry (see Section \hyperref[sec:uncertainties]{\ref{sec:uncertainties}}). In constructing dynamical models we will therefore restrict ourselves to an axisymmetric stellar system \footnote{The modelling machinery itself is triaxial, but oblate axisymmetric models can be run in the triaxial limit (i.e. intermediate- to long-axis ratio (b/a) = 0.99). Although the orbits in axisymmetric galaxies are dominated by tube orbits, our slightly triaxial models still benefit from a non-negligible amount of additional triaxial orbit families (e.g. box orbits).}.

\subsection{Mass Profiles}
\label{sec:mass_profiles}

\subsubsection{MRK\,1216}
\label{sec:mrk1216_schwarzschild}

We consider three gravitational sources; the central black hole mass \Mbh, the stellar mass \Mstar\ (which is the deprojected, intrinsic luminosity density times the constant stellar mass-to-light ratio \MLstar), and a spherically symmetric dark matter component with an NFW profile \citep{1996ApJ...462..563N} with concentration $c_{DM}$ and total virial mass $M_{DM} = M_{200}$. In the case of MRK\,1216, the final models will thus probe a four parameter space in log(\Mbh/$M_{\scriptscriptstyle \odot}$) $\in$ [7, 11], $\Upsilon_{H}$ $\in$ [0.5, 3], $c_{DM}$ $\in$ [5, 15] and log($f_{DM}$) = log($M_{DM}/\Mstar$) $\in$ [-9, 5]. The search in parameter space is mainly motivated by observational and theoretical constraints on: the stellar mass-to-light ratio for SSP models with a Kroupa and Salpeter IMF \citep{1996ApJS..106..307V}; the black hole mass from predictions of the black hole scaling relations \citep{2009ApJ...698..198G} and the dark halo parameters from investigations of \cite{2001MNRAS.321..559B}, \cite{2010ApJ...710..903M} and \cite{2008MNRAS.391.1940M} (see also Section \hyperref[sec:dm_halo]{\ref{sec:dm_halo}}).\\

By design, we do not explore the inclination space. As has been shown in \cite{2005MNRAS.357.1113K} and \cite{2009MNRAS.398.1117V}, it is not possible to infer the inclination angle by means of two-dimensional line-of-sight stellar kinematics, alone, unless kinematic features exist (e.g. kinematically decoupled cores) that put additional constraints on the intrinsic shape of galaxies. Even in the case of three integral axisymmetric orbit-based models, different inclinations above the lower limit that is given by the photometry are able to fit the LOSVD almost equally well. However, we can further constrain the inclination by simple observational arguments. Although the flattest Gaussian in our MGE limits the minimum possible inclination for the projection in an oblate axisymmetric case (Section \hyperref[sec:mrk1216_photometry]{\ref{sec:mrk1216_photometry}}), the deprojection of Gaussians close to the lower inclination limit of 59\degree\ generates intrinsically flat galaxies with unphysical axis ratios of $q=b/a \le 0.2$. On the other hand, an edge-on configuration (even though possible) appears to be unlikely, too. MRK\,1216 is much rounder than expected for a flat, oblate system that is observed at 90\degree, with an axis ratio that quickly converges to unity beyond 1 \Reff. We therefore choose an inclination of 70\degree\ that is in between these two extreme cases. To assess the reliability and robustness of our results with respect to changes in the inclination, we also explore models with a close to edge-on configuration of 85\degree\ and find that our parameter constraints are affected by less than 5 per cent. The upper limit of the stellar mass-to-light ratio increases by $\sim$ 0.1 when increasing the inclination, which in turn leads to insignificant changes in the derived values for the black hole and dark halo values. In general, the results are very robust with respect to variations in the inclination. Changes in the parameter estimation, and in particular in the stellar mass-to-light ratio, are only significant if a large range of inclinations is probed or the minimal observed axis ratio $q$ is larger than 0.7, which translates to lower inclination limits of  $i \le$ 45\degree\ \citep[see also][]{2006MNRAS.366.1126C}.\\

\begin{figure*}
	\includegraphics[width=0.49\textwidth]{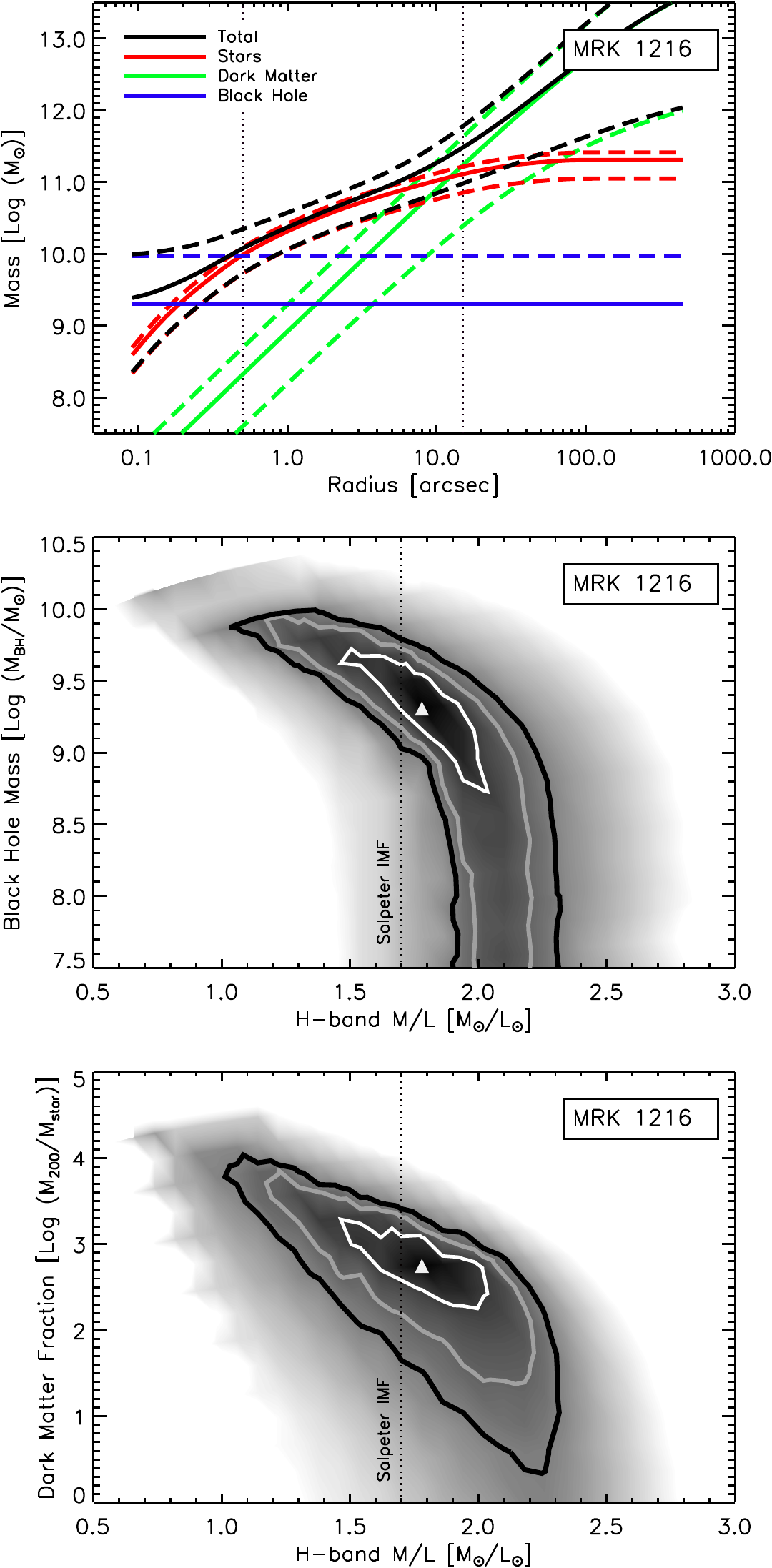}
	\includegraphics[width=0.49\textwidth]{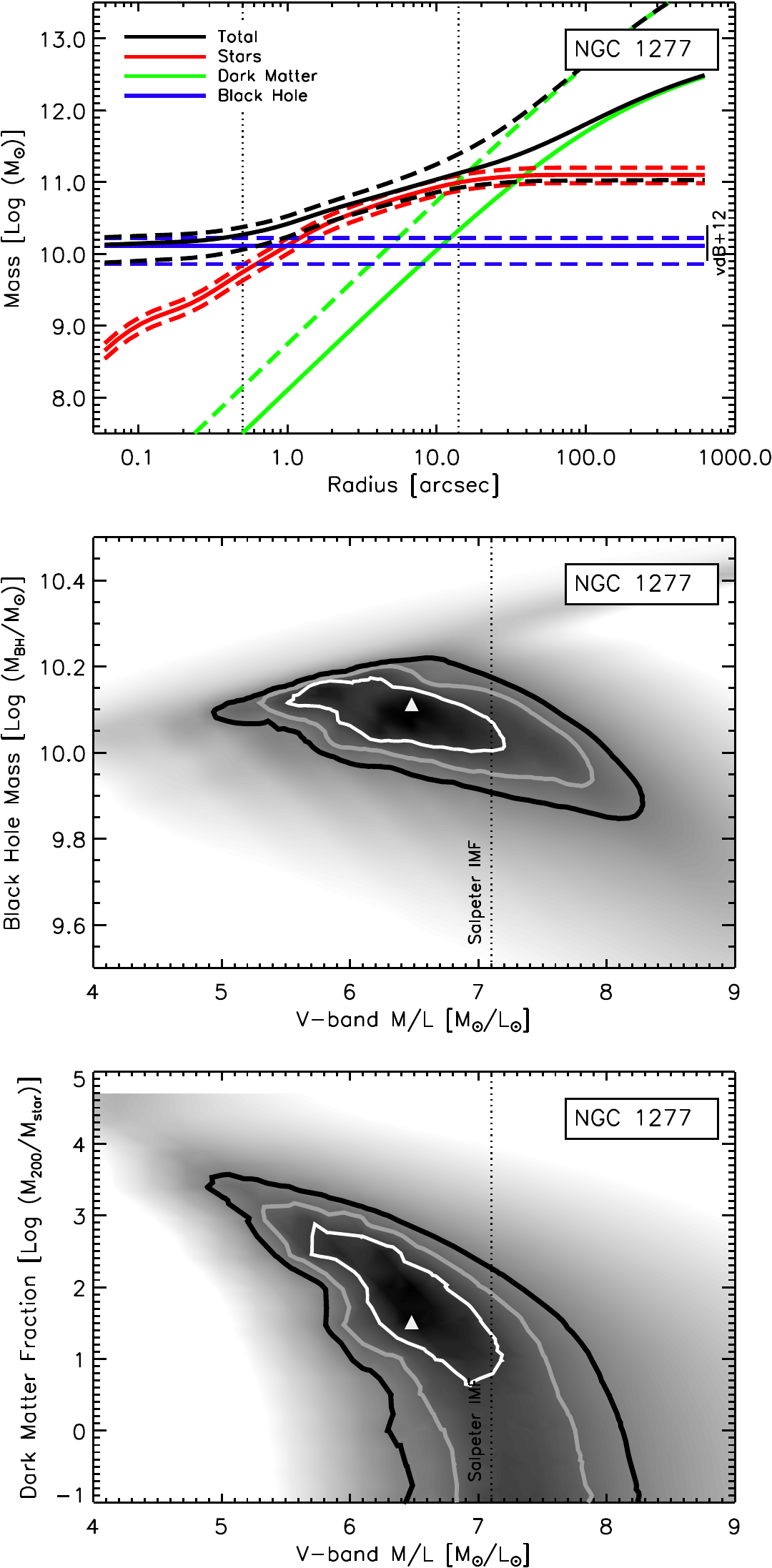}
	\caption{\textit{Top:} Intrinsic enclosed mass profiles of MRK\,1216 and NGC\,1277. Solid lines represent best-fitting values, dashed lines are $3\sigma$ confidence intervals. The dotted vertical lines indicate the minimum and maximum extent of the kinematic data for each galaxy. For NGC\,1277, we also depict an earlier estimate of the black hole mass from vdB12. \textit{Middle:} Confidence contours of black hole mass vs. stellar mass-to-light ratio. \textit{Bottom:} Confidence contours of dark matter fraction vs. stellar mass-to-light ratio. The lines denote the 68.3 (white), 95 (grey) and 99.7 (black) per cent quantiles of a $\chi^2$ distribution with two degrees of freedom. As a reference, we overplot stellar mass-to-light ratio predictions of SSP models with a (single) power-law Salpeter IMF \citep{1996ApJS..106..307V,2012MNRAS.424..157V} in the respective bands.}
	\label{fig:model_res}
\end{figure*}

\begin{figure*}
	\includegraphics[width=.7\textwidth,angle=90]{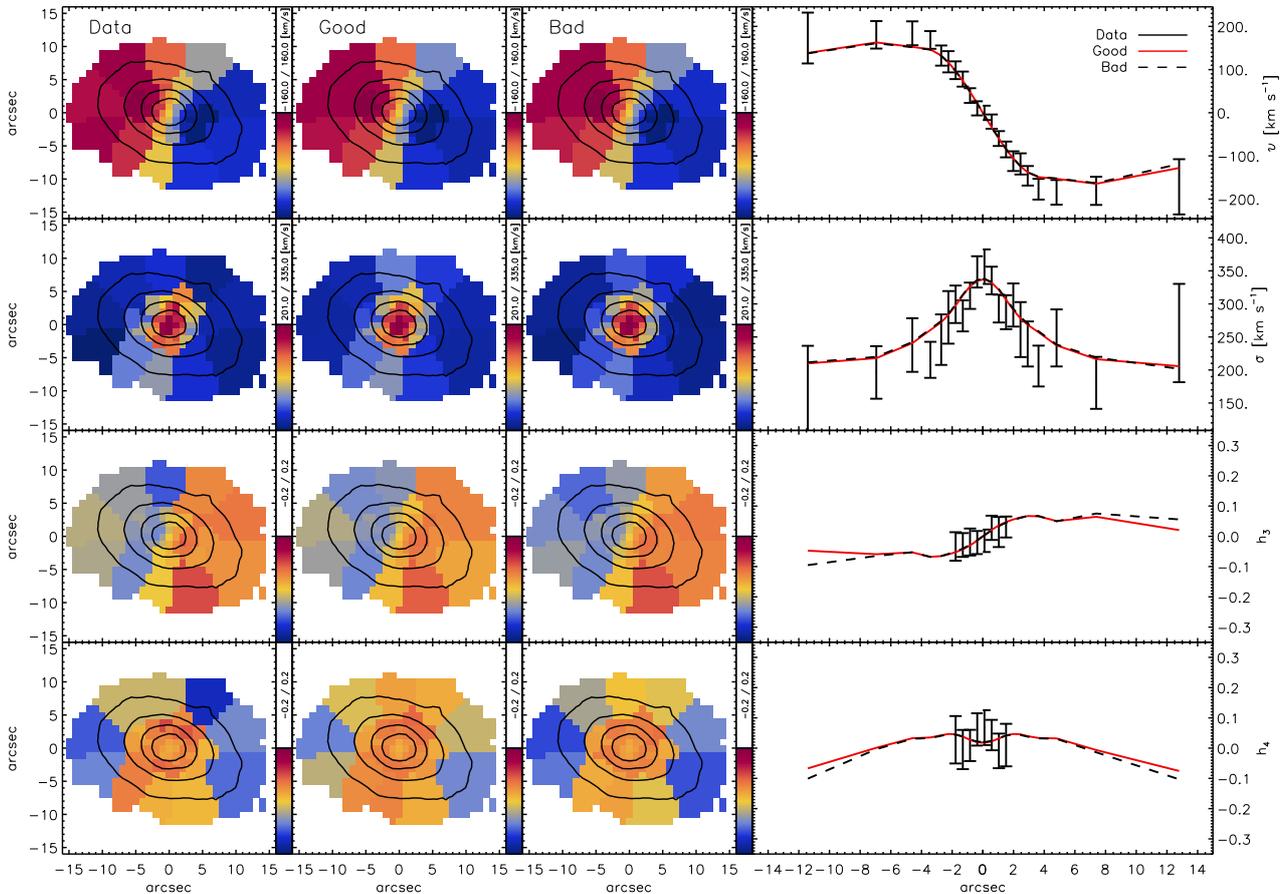}
	\caption{\textit{First column:} \textit{PPAK} IFU (bi-symmetrised) velocity, velocity dispersion, $h_3$ \& $h_4$ maps of MRK\,1216, overplotted with contours of constant surface brightness. \textit{Second column:} Best-fitting Schwarzschild model with a reduced $\chi^2$ of 0.31. \textit{Third column:} Bad model with log($M_{DM}$/$M_{\scriptscriptstyle \odot}$) = -9. \textit{Fourth column:} One out of the three simultaneously fitted \textit{HET} long-slit data with corresponding uncertainties, best-fitting Schwarzschild model (red) and bad model (dashed). The IFU maps are oriented N.-E., i.e. north is up and east is left.}
	\label{fig:m1216_maps_models}
\end{figure*}

We bi-symmetrise the observed kinematics beforehand. Although being fairly symmetric, the symmetrisation reduces noise and systematic effects which helps for the recovery of the higher order Gauss-Hermite moments in the models. The PA for the bi-symmetrisation is obtained by using the weighted first and second moments of the intensity distribution in the \textit{PPAK} data (PA$_{kin}$=70.7\degree), and turns out to be in excellent agreement with the PA that is inferred from the high-resolution imaging (PA$_{phot}$=70.2\degree).\\

We construct $\sim$ 200 000 models to constrain the best-fitting parameters as well as all parameters within a relative likelihood of three standard deviations. Figure \hyperref[fig:model_res]{\ref{fig:model_res}} (top left) shows the enclosed mass distribution of MRK\,1216, derived from our entire set of models. The solid lines represent the stellar mass (red), black hole mass (blue), dark matter content (green) and total mass (black) for the best fit.

The dashed lines indicate 3$\sigma$ confidence intervals for one degree of freedom. Based on these models, we obtain a total stellar mass of log(\Mstar/$M_{\scriptscriptstyle \odot}$) = 11.3$_{-0.2}^{+0.1}$, a black hole mass of log(\Mbh/$M_{\scriptscriptstyle \odot}$) = 9.4$_{-9.4.}^{+0.6}$ and a dark halo mass of log($M_{DM}/$$M_{\scriptscriptstyle \odot}$) = 14.2$_{-2.2}^{+1.1}$.
Neither the black hole nor the DM halo parameters are very well constrained. The best fitting dark halo dominates at radii larger than 15\arcsec\ (i.e. $\ge$ 7\,kpc), which is at the edge of the extent of our kinematic data. Interestingly, models without any dark matter are not able to recover the observations and can be ruled out.

Figure \hyperref[fig:model_res]{\ref{fig:model_res}} (middle left) is a slice through the \Mbh\ - $\Upsilon_{H}$ plane, i.e. we plot every combination of black hole mass and stellar mass-to-light ratio, marginalised over the dark halo parameters $c_{DM}$ and $f_{DM}$. As is already visible in the enclosed mass profile plot, the best-fitting black hole mass is log(\Mbh/$M_{\scriptscriptstyle \odot}$) = 9.4. While we obtain an upper limit of log(\Mbh/$M_{\scriptscriptstyle \odot}$) = 10.0, the black hole is unconstrained at the lower boundary of the grid, at log(\Mbh/$M_{\scriptscriptstyle \odot}$) = 7. We therefore carry out additional tests at the lower end of the parameter space (\Mbh/$M_{\scriptscriptstyle \odot}$ = 0) which show that the presence of a black hole is not required, as models with no black hole are able to match the data as well. The stellar mass-to-light ratio in \textit{H}-band spans a range of 1.0 - 2.3. The best-fitting model favours $\Upsilon_{H}$ = 1.8. For comparison, stellar population synthesis (SPS) models with a single power-law Salpeter stellar initial mass function (IMF) \citep{1996ApJS..106..307V} predict a stellar mass-to-light ratio of 1.7 (assuming solar metallicity and an age of $\sim$ 13 Gyr).

\begin{figure*}
	\includegraphics[width=.7\textwidth,angle=90]{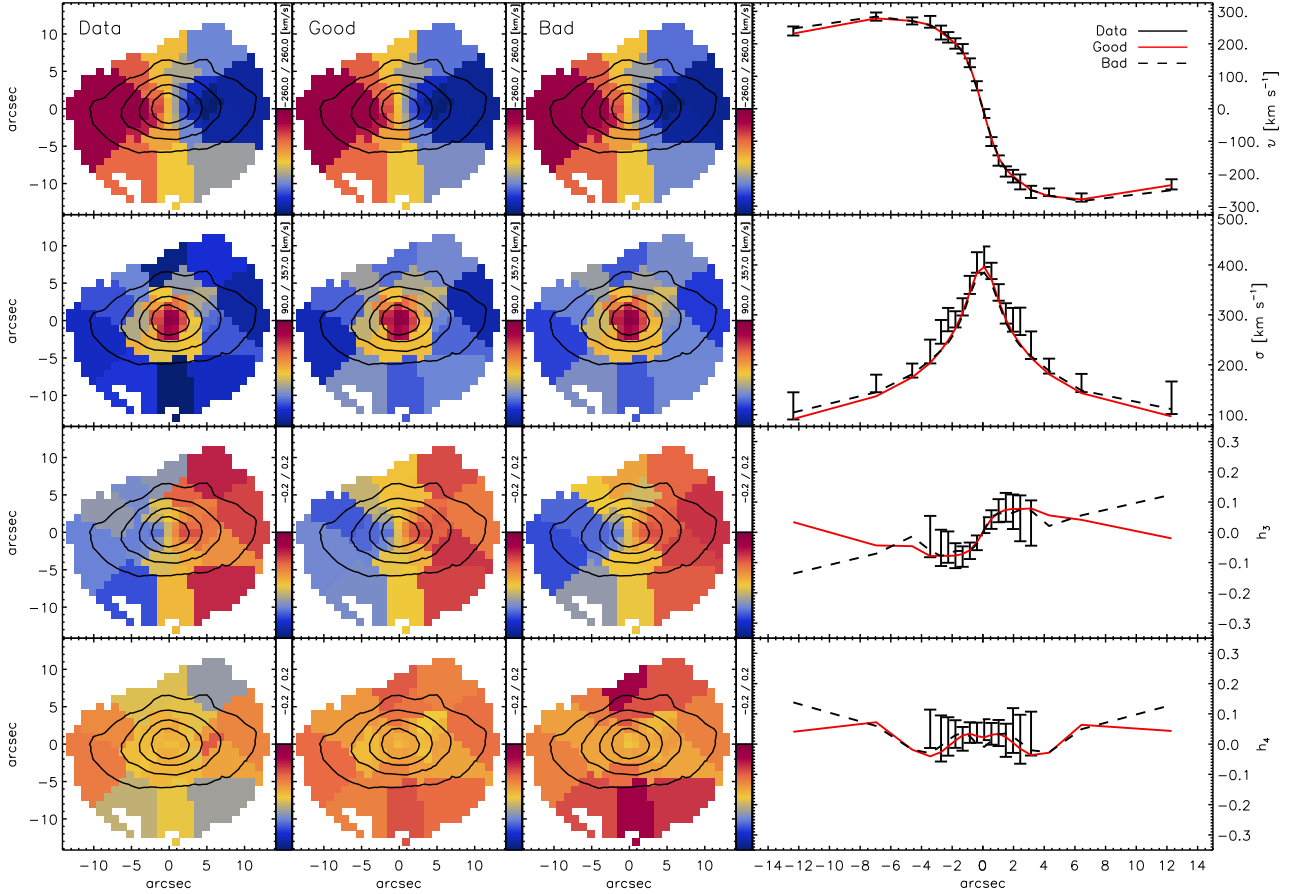}
	\caption{\textit{First column:} \textit{PPAK} IFU (bi-symmetrised) velocity, velocity dispersion, $h_3$ \& $h_4$ maps of NGC\,1277, oriented N-E with contours of constant surface brightness overplotted. \textit{Second column:} Best-fitting Schwarzschild model with a reduced $\chi^2$ of 0.55. \textit{Third column:} Bad model with log($\Mbh/M_{\scriptscriptstyle \odot}$) = 9.5 and $\Upsilon_{V} = 8.3$. \textit{Fourth column:} One out of the three simultaneously fitted \textit{HET} data with corresponding uncertainties, best-fitting Schwarzschild model (red) and bad model (dashed).}
	\label{fig:n1277_maps_models}
\end{figure*}

The bottom left panel in Figure \hyperref[fig:model_res]{\ref{fig:model_res}} is analogous to the middle panel and displays the goodness-of-fit contours for the dark matter fraction $f_{DM}$ and \textit{H}-band stellar mass-to-light ratio $\Upsilon_{H}$, marginalised over all remaining parameters. We observe closed contours that clearly call for a non-negligible amount of dark matter. The halo concentration in the models is unconstrained and can adopt any value within the range that is probed. The best fitting dark halo has a concentration of $c_{DM} = 10$, a mass of log($M_{DM}$/$M_{\scriptscriptstyle \odot}$) = 14.2 and a scale radius of $r_{s} = 110$\,kpc.

The corresponding best-fitting Schwarzschild model kinematics of MRK\,1216 are shown in Figure \hyperref[fig:m1216_maps_models]{\ref{fig:m1216_maps_models}}. The plots display fits to the first four kinematic moments of the \textit{PPAK} data, and one of the three simultaneously fitted individual \textit{HET} long-slits. Our models can accurately recover the kinematics of the \textit{PPAK} and \textit{HET} data, in particular the peak in the velocity dispersion profile and the flat and rapidly rotating velocity curve beyond 5\arcsec. For illustration purposes we add a "bad model" to the plots. The bad model was chosen by following the ridge of minimum $\chi^{2}$ beyond the 3$\sigma$ confidence level and displays the predicted kinematics for the best-fitting model without a dark halo. The relative likelihoods of the two models are separated by $\Delta \chi^{2} = \chi^{2}_{b} - \chi^{2}_{b\:w/o\:dm} = 15$ and the differences in the figures are barely distinguishable. Despite these similarities, we will show that dark matter is a necessary ingredient to successfully recover the observational constraints and is well in line with our current understanding of the stellar build-up and properties of elliptical galaxies (see Section \hyperref[sec:dm_halo]{\ref{sec:dm_halo}}).

\subsubsection{NGC\,1277}
\label{sec:ngc1277_schwarzschild}

\vspace{0.5em}
For the dynamical analysis of NGC\,1277, our models explore the parameter ranges in log(\Mbh/$M_{\scriptscriptstyle \odot}$) $\in$ [7,11], $\Upsilon_{V}$ $\in$ [2,10], $c_{DM}$ $\in$ [5, 15], log($f_{DM}$) = log($M_{DM}/\Mstar$) $\in$ [-9, 5] and $i \in$ [75]. In contrast to vdB12, the freedom of the models is further restrained by fitting the \textit{PPAK} and \textit{HET} data at the same time. As in the case of MRK\,1216, we present the mass distribution, the black hole mass vs. stellar mass-to-light ratio and dark matter fraction vs. stellar mass-to-light ratio plots, which outline the limits for the individual parameters.

Figure \hyperref[fig:model_res]{\ref{fig:model_res}} (top right) shows the enclosed mass profile with a total stellar mass of log($\Mstar/M_{\scriptscriptstyle \odot}) = 11.1_{-0.1}^{+0.1}$, a black hole mass of log(\Mbh/$M_{\scriptscriptstyle \odot}$) = 10.1$_{-0.2}^{+0.1}$ and a dark halo mass of ($M_{DM}$/$M_{\scriptscriptstyle \odot}$) = $12.6_{-12.6}^{+1.9}$, at a significance of 3$\sigma$. The kinematic data of NGC\,1277 show the same problems as the data of MRK\,1216, leading to poor constraints on the dark halo parameters. The vertical dotted line displays the extent of our kinematic information and illustrates the inability to constrain the dark matter halo, which becomes dominant only at larger radius for the best-fitting model. Here, the presence of dark matter is not necessary to fit the observed velocity moments.

We present contours of $\chi^{2}$ as a function of \Mbh\ and $\Upsilon_{V}$ in Figure \hyperref[fig:model_res]{\ref{fig:model_res}} (middle right). Despite the low resolution data, the black hole mass is well constrained (see also Sec. \hyperref[sec:masses]{\ref{sec:masses}} and \hyperref[sec:uncertainties]{\ref{sec:uncertainties}}). We obtain an upper limit of log(\Mbh/$M_{\scriptscriptstyle \odot}$) = 10.2 and a lower limit of log(\Mbh/$M_{\scriptscriptstyle \odot}$) = 9.9. The best-fitting \textit{V}-band stellar mass-to-light ratio is $\Upsilon_{V}$ = $6.5_{-1.5}^{+1.5}$ and consistent with the mass-to-light ratio that is predicted from spectral synthesis fits \citep{2010MNRAS.404.1639V,2012MNRAS.424..172R,2012MNRAS.424..157V} of NGC\,1277's deep, optical long-slit observations (T14), assuming a single power-law Salpeter IMF (see Section \hyperref[sec:uncertainties]{\ref{sec:uncertainties}}). Models with a black hole mass of log(\Mbh/$M_{\scriptscriptstyle \odot}$) $\sim$ 9 - as suggested by the scaling relation between black hole mass and bulge luminosity $\Mbh - L_{Bulge}$ \citep{2009ApJ...698..198G,2013ApJ...764..184M,2013ARA&A..51..511K} - are clearly disfavoured and not able to recover the data (see also Section \hyperref[sec:masses]{\ref{sec:masses}} and \hyperref[sec:scaling_relations]{\ref{sec:scaling_relations}}).

The confidence intervals for the dark halo are shown in the bottom right panel of Figure \hyperref[fig:model_res]{\ref{fig:model_res}}. In contrast to MRK\,1216, the 99.7 per cent contours cannot rule out models without dark matter. The best-fitting NFW profile has a concentration of $c_{DM} = 10$, a dark halo mass of log($M_{DM}$/$M_{\scriptscriptstyle \odot}$) = 12.6 and a scale radius of $r_{s} = 33$\,kpc.

In Fig. \hyperref[fig:n1277_maps_models]{\ref{fig:n1277_maps_models}}, we show the bi-symmetrised \textit{PPAK} kinematics, one of the three individual but simultaneously fitted \textit{HET} long-slit kinematics, the best-fitting model and a "bad model" of NGC\,1277. Note that one of the three long-slit kinematics is identical with the data presented in vdB12. In this case, the \textit{PPAK} velocities and dispersions are fitted exceptionally well. Problems arise in fitting the \textit{HET} velocity dispersion as the peak in the \textit{HET} data differs by $\sim$ 60 \kms\ from the peak in the \textit{PPAK} cube (Section \hyperref[sec:ngc1277_kinematics]{\ref{sec:ngc1277_kinematics}}). The best fit predicts a slightly lower central dispersion in NGC\,1277, which matches the \textit{PPAK} data but is in contrast to the \textit{HET} observations (see Section \hyperref[sec:uncertainties]{\ref{sec:uncertainties}}). We also emphasise that even though the $h_4$ values of the "good model" are slightly off along the minor axis, they are still within the measurement errors. For the illustration of a bad model we explicitly chose a model with a higher mass-to-light ratio and a black hole mass that is about a factor of 2 smaller than the lower limit. The $\Delta\chi^2$ of this fit is $\sim$ 50 and well beyond the 3$\sigma$ boundary. The difference between observed and modelled kinematics are most pronounced in the second and fourth column where the bad model clearly fails to fit $\sigma$ and $h_4$ in the centre. The deviation in $h_4$ along the minor axis is also much stronger. In contrast to the good model, the bad model fails to reproduce the kinematics within the measurement errors, by underestimating the number of stars with line-of-sight velocities close to the average velocity.

Overall, our models are in good agreement with the results of vdB12, with tighter constraints especially on the lower end of stellar mass-to-light ratios and hence a slightly decreased upper limit of the black hole mass. This effect is mainly driven by lower estimates of the dark halo content that is constrained by the wide-field IFU data and higher stellar mass-to-light ratios which then propagate towards the centre.

\section{Discussion}
\label{sec:discussion}

We summarise the findings and take a closer look into the results of our orbit-based dynamical models, their orbital structures and how they compare to the photometric analysis. In addition, we place the black hole masses back into the scaling relations, discuss the significance of our dark halo detections and finish with a concluding remark concerning the origin and evolutionary history of both galaxies
\subsection{Orbital Structure}
\label{sec:orbital_structure}

Apart from inferring the mass distribution, orbit-based dynamical models also allow a detailed probe of the orbital structure of galaxies. We can not only inspect the amount of mass that is assigned to each particular orbit, or orbit family in general, but also quantify the system's degree of anisotropy, which holds important clues about the processes that shaped its evolution \citep{1992ApJ...399..462B}. The anisotropy profiles of early-type galaxies have been investigated extensively. While data and techniques differ, ranging from long-slit observations and spherical models \citep{2000A&AS..144...53K,2001AJ....121.1936G} to more general axisymmetric models \citep{2003ApJ...583...92G} that make use of the full 2D spectral information \citep{2007MNRAS.379..418C}, there is a common agreement, namely that luminous, round and slowly rotating early-type galaxies are almost isotropic whereas oblate, fast-rotating galaxies span a large range of anisotropy profiles. The orbital structures in the dynamical models, though, have not been linked to the many and varied components that are observed via photometric decompositions of high resolution imaging, which also provide an independent record of a galaxy's evolutionary history. In the first part of this subsection we aim to provide this link by mapping the components in phase space to the multi-\sersic\ components in Section \hyperref[sec:mrk1216_photometry]{\ref{sec:mrk1216_photometry}} and \hyperref[sec:ngc1277_photometry]{\ref{sec:ngc1277_photometry}}. In the second part, we then present a direct comparison between the orbital distribution of the two compact objects in this work and a more general and representative sample of early-type galaxies.\\

\begin{figure*}
	\centering
	\includegraphics[width=0.42\textwidth]{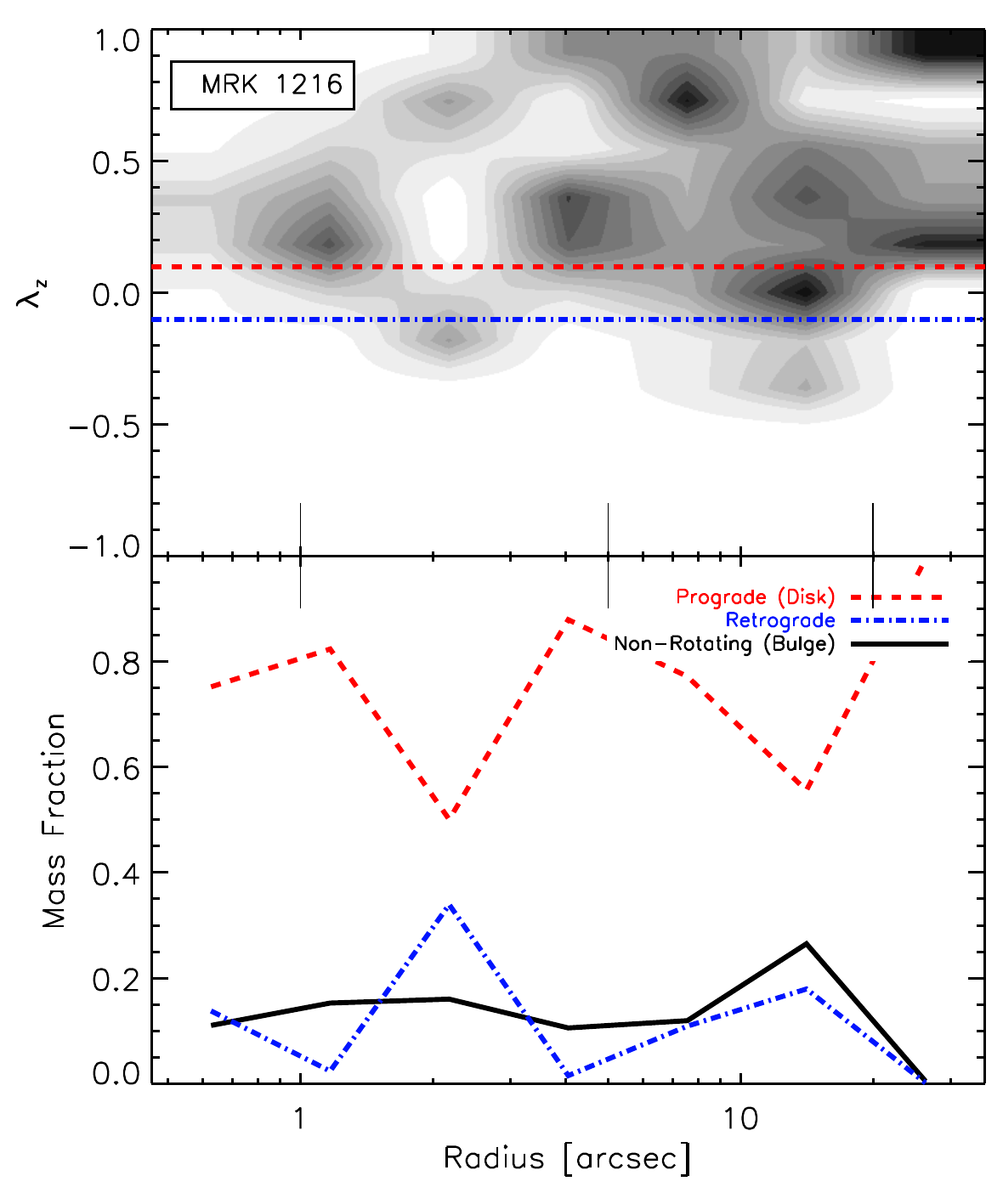}
	\includegraphics[width=0.42\textwidth]{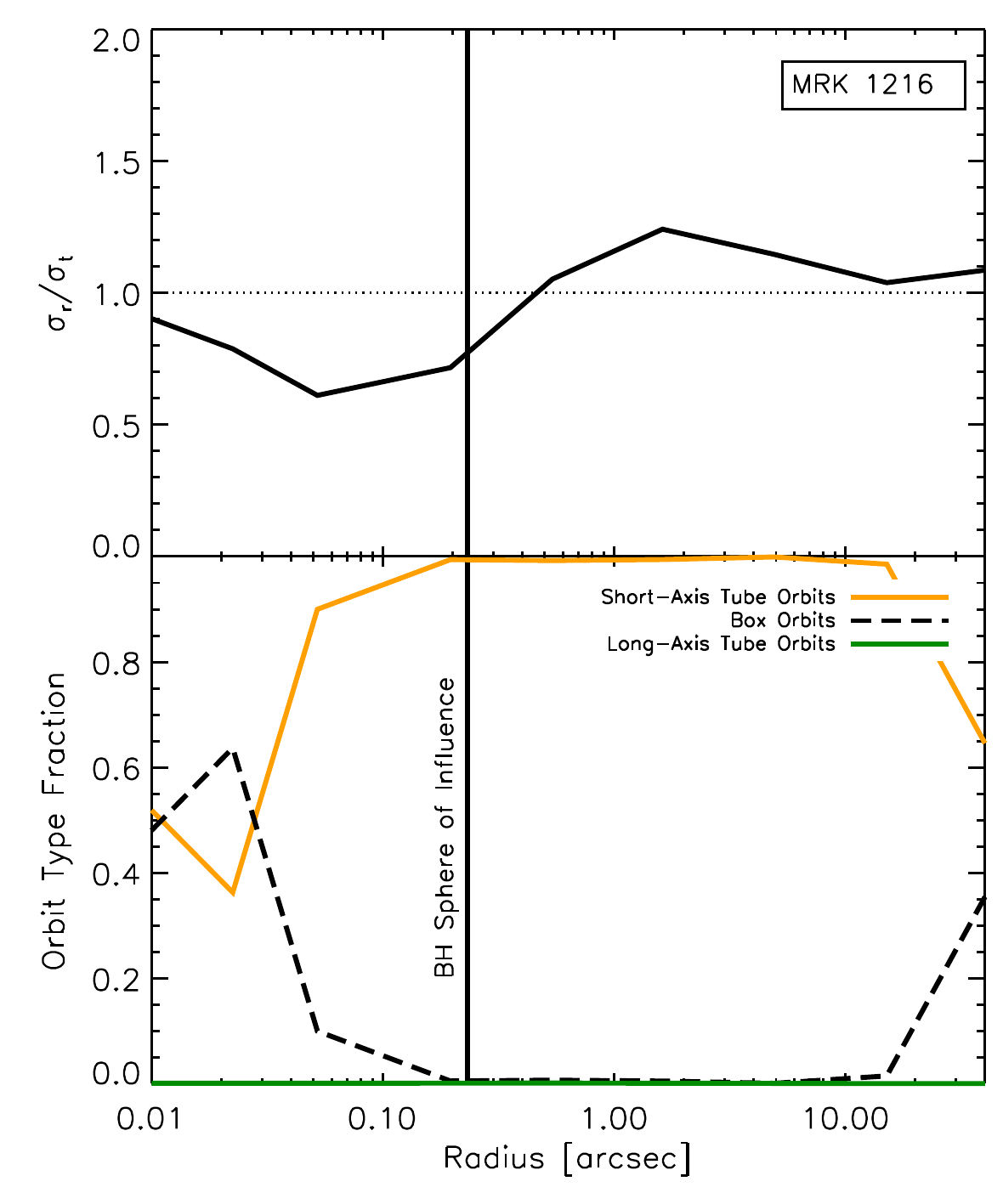}
	\caption{Orbital structure of MRK\,1216 for the best-fitting model. \textit{Top left:} Mass distribution along all orbits as a function of angular momentum along the short $z$-axis and radius. Mass located above (below) the red (blue) line is rotating strongly prograde (retrograde). \textit{Bottom left:} Local mass fraction as a function of average radius, divided into a non-rotating bulge-like $(|\lambda_z|<$ 0.1), prograde rotating disk-like ($\lambda_z>$ 0.1) and retrograde rotating ($\lambda_z<$ -0.1) component. This classification is based on the net angular momentum of the orbits along the z-axis. The long tick marks denote the effective radius of the photometric components in the multi-\sersic\ fit. \textit{Top right:} Profile of radial vs. tangential velocity dispersion. \textit{Bottom right:} Mass fraction per orbit type as a function of radius. The vertical solid line denotes the predicted black hole sphere of influence. The orbital distribution below 0.4\arcsec\ is not resolved by the data and merely extrapolated by the models.}
	\label{fig:m1216_df}
\end{figure*}

\begin{figure*}
	\centering
	\includegraphics[width=0.42\textwidth]{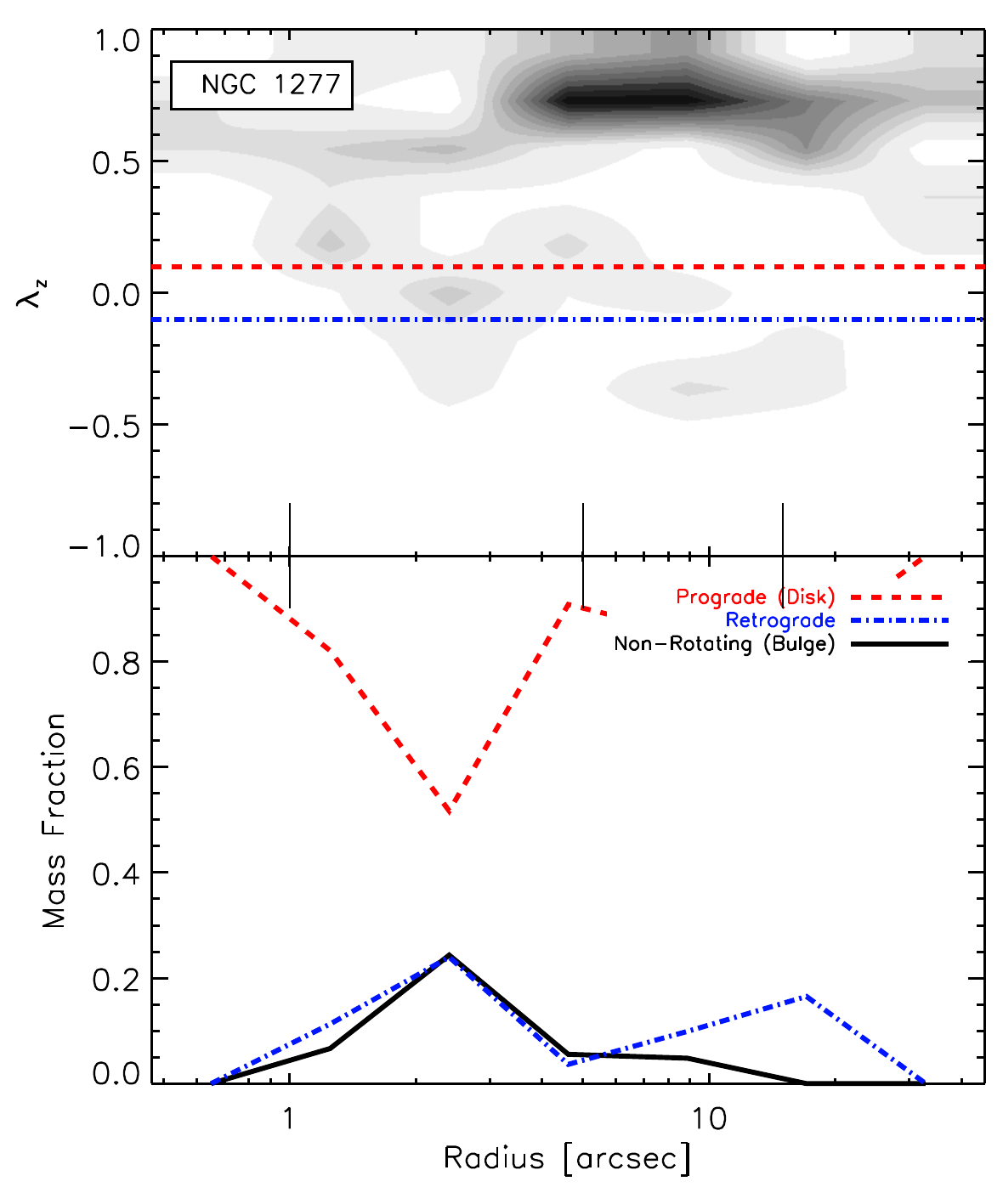}
	\includegraphics[width=0.42\textwidth]{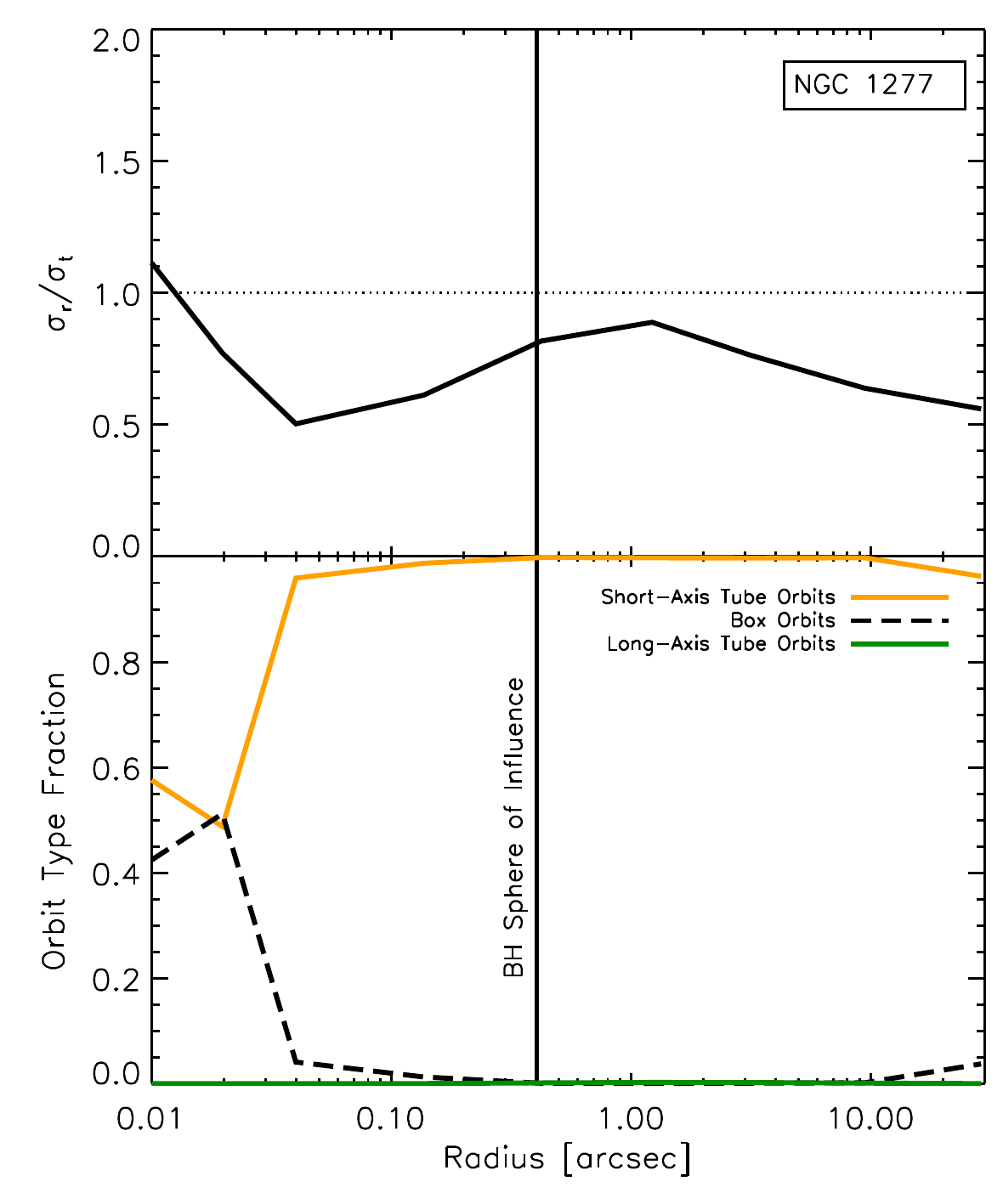}
	\caption{Orbital structure of NGC\,1277 for the best-fitting model. \textit{Top left:} Mass distribution along the orbits as a function of average angular momentum and radius. \textit{Bottom left:} Local mass fraction as a function of radius for the individual components (see caption in Figure \hyperref[fig:m1216_df]{\ref{fig:m1216_df}}). \textit{Top right:} Velocity anisotropy profile. \textit{Bottom right:} Mass distribution along individual orbit types. Here again, radii below 0.4\arcsec\ are not resolved by our observations and have therefore to be treated with care.}
	\label{fig:n1277_df}
\end{figure*}

In Figure \hyperref[fig:m1216_df]{\ref{fig:m1216_df}} and \hyperref[fig:n1277_df]{\ref{fig:n1277_df}}, we show the orbital mass weights as a function of average radius ($\bar r$) and spin ($\bar\lambda_{z} = \bar{J_z} / (\bar{r} \times \bar{\sigma}$) - where $\bar J_z$ is the average specific angular momentum of the orbits along the short $z$-axis and $\bar \sigma$ their average dispersion - and further examine the orbital structures by inspecting the ratio of radial to tangential velocity dispersion ($\sigma_{r}/\sigma_{t} = \sqrt{2\sigma_{r}^{2}/(\sigma_{\phi}^{2} + \sigma_{\theta}^{2}})$) and the occupation fractions of the individual orbit families \citep{1985MNRAS.216..273D,1987ApJ...321..113S,1991ARA&A..29..239D}. The averages are time averages per single orbit, which the Schwarzschild models keep track of.

Deciphering the mass distribution among the orbits as a function of angular momentum will provide the necessary link to the photometric components. Hitherto, only two other galaxies have been investigated in a similar manner. \citet{2012ApJ...753...79W} presented the S0 galaxy NGC 3998 which showed a very clear non-rotating bulge and a non-maximal rotating disk. \citet{2013MNRAS.431.3364L} presented the E5 galaxy FCC 277 with a nuclear star cluster, which showed both a pro- and retrograde disk and a non-rotating component. A direct comparison to the photometric structures, though, was not within the scope of those investigations and 
has therefore been omitted so far.\\

In what follows, we present the orbital configuration only for the best-fitting models, but the general trend is preserved for most models that are within the statistical 3$\sigma$ uncertainties.\\

\subsubsection{MRK\,1216}
\label{sec:mrk1216_orb}

MRK\,1216 rotates rapidly ($\sim$ 220 \kms) around the short axis. It is thus a fast-rotating, oblate early-type galaxy.
The two-component photometric decomposition contains a small, flattened and massive bulge and an outer exponential envelope, with a bulge-to-total ratio of $B/T = 0.69$. However, the dynamical decomposition is not as straightforward as matching directly to these two \sersic\ components; there is a large, extended, rotating structure ($\lambda_z = 0.0 - 0.5$) beyond 10\arcsec; two, more centrally located, moderately rotating structures ($\lambda_z = 0.1 - 0.5$) between 1\arcsec and 5\arcsec; an inner (8\arcsec) rapidly ($\lambda_z \sim 0.7$) and outer ($\ge$ 20\arcsec) maximally rotating component ($\lambda_z \sim 1$), as well as some mass in two counter-rotating and one outer ($\ge$ 10\arcsec) non-rotating structure (Fig. \hyperref[fig:m1216_df]{\ref{fig:m1216_df}}, left panel). The orbital structure is devoid of a massive, non-rotating component which harbours a major fraction of the stellar mass and hence we conclude that MRK\,1216 does not contain a classical, non-rotating bulge. If anything, we identify the moderately rotating component at $\sim$ 1\arcsec\ as the bulge, with a dynamical $B/T$ of 13 per cent, which however is in sharp contrast to the photometric $B/T$ of the two-component decomposition.

Remarkably, the multiple component photometric fit of MRK~1216 with four \sersic\/s (Section \hyperref[sec:mrk1216_photometry]{\ref{sec:mrk1216_photometry}}) is in much better agreement with the dynamical decomposition. The outer three photometric components can be mapped onto the 3 mildly rotating distinct components in the orbital configuration at $\sim$ 1\arcsec, 4\arcsec and 15\arcsec. All these photometric components have low ($n\sim1$) \sersic\ indices that are normally associated with an exponential disk, except for the outermost component that resembles an envelope with close to zero ellipticity. Adopting \sersic\ component number two as the bulge gives a photometric $B/T$ of 12 per cent (Sec. \hyperref[sec:mrk1216_photometry]{\ref{sec:mrk1216_photometry}}), which is in very good agreement with the dynamical $B/T$ of 13 per cent. Moreover, the component around 15\arcsec\ carries a mass of 30 per cent and is almost as massive as the large, outer component in the photometry with a contribution of 35 per cent to the total mass. The most notable difference is between the third and most massive \sersic\ in the photometry and the dynamical substructure at 4\arcsec, which ought to contribute 45 per cent of the total stellar mass but only constitutes 20 per cent in the orbital configuration. We also note that the innermost photometric component is too small (0.4\arcsec) to be resolved by the dynamics, while the rapidly rotating structure ($\lambda_z \sim 0.7$) at 8\arcsec\ has no photometric counterpart at all. Finally, the maximally rotating structure ($\lambda_z \sim 1$) at 20\arcsec\ is beyond the reach of our kinematic data and merely a result of an extrapolation, but is expected to correspond to a disky component whereas the photometry at those radii are dominated by the round outer halo.\\

MRK\,1216's radial anisotropy profile is simple and almost isotropic with oscillations of only 30 per cent around $\sigma_{r}/\sigma_{t} = 1$ (Fig. \hyperref[fig:m1216_df]{\ref{fig:m1216_df}}, right panel). The largest deviation from isotropy appears within the central 1\arcsec, which is attributable to the strong gravitational perturbation of the axisymmetric potential due to the presence of a black hole and hence the requirement of a non-negligible amount of stars in box orbits. Most of the mass resides in short-axis tube orbits whereas long-axis tube orbits, which are vital orbit types for triaxial and prolate systems \citep{2008MNRAS.385..614V,2009MNRAS.398.1117V,2010MNRAS.401.1770V}, are totally absent.\\

\subsubsection{NGC\,1277}
\label{sec:ngc1277_orb}

NGC\,1277 is also a flat, fast-rotating ( $\sim$ 300 \kms) early-type galaxy and, as expected, our best-fitting dynamical model reveals a simple structure where most of its stars ($\ge 80$ per cent) reside in strongly rotating orbits between 1 and 20\arcsec\ (Fig. \hyperref[fig:n1277_df]{\ref{fig:n1277_df}}, left panel). We distinguish at least three individual components in phase space; a highly rotating one ($\lambda_{z} = 0.7 - 1$) between 5\arcsec\ and 10\arcsec, containing 65 per cent of all stars; one moderately rotating and very extended component ($\lambda_{z} = 0.1 - 0.5$) between 0.5\arcsec\ and 3\arcsec, which contains 23 per cent of the stellar mass, and a centrally located non-rotating one at 2\arcsec, that harbours 3 per cent of all stars. The remaining 9 per cent are distributed among the tiny substructures at various positions. The lack of a massive, non-rotating ($\lambda_z = 0$) component in the dynamical decomposition suggests that this galaxy does not contain a pressure supported bulge. Moreover, due to the absence of a distinct, central, non-rotating component, this result also does not match with what we see and anticipate based on our results in the photometric two-component decomposition (Section \hyperref[sec:ngc1277_photometry]{\ref{sec:ngc1277_photometry}}), which has a bulge and an exponential disk with a $B/T = 0.75$. The massive bulge in the two-component fit is located around $\sim$ 3\arcsec, where most of the mass in the orbital decomposition resides in rotating structures. While the massive bulge in the photometry could indeed be mildly rotating, similar to our identification of the dynamical bulge in the orbital structure of MRK\,1216, the mass fractions at these radii are simply at odds and disfavour the simple two-component decomposition.

The orbital structure is also not consistent with the 1D photometric analysis of NGC~1277 from \cite{2013ARA&A..51..511K}, as they connected their inner flattened bulge with the outer round halo ($> 20\arcsec$) as a single component, which is more luminous than the disk in their analysis ($B/T = 0.55$). In the dynamical decomposition, however, these two components (i.e. bulge and outer halo) do not appear to be connected and most of the mass resides in the extended flat and rapidly spinning component ($\lambda_{z} \ge 0.7$).

In contrast, the overall dynamical structure has an intriguing resemblance to the multi-component \sersic\ fit, which hints at an inner, exponential disk ($n \sim 1$) with a small contribution to the overall stellar mass (24 per cent); and most of the mass (53 per cent) settled in a flat, outer disk-like component that resembles the extended highly rotating structure in our orbital decomposition. Even though the low \sersic\ index of the second innermost component in the photometric fit is usually associated with a disky component, we cannot rule out the existence of a mildly rotating, flattened spheroidal component. Given the match in the orbital and photometric decomposition, we therefore adopt this component as a lower limit to the bulge, with a $B/T = 0.24$ (see Sec. \hyperref[sec:ngc1277_photometry]{\ref{sec:ngc1277_photometry}} and \hyperref[sec:scaling_relations]{\ref{sec:scaling_relations}}).  The innermost photometric component ($\le$ 0.5\arcsec) is not resolved by our observations and, unfortunately, our stellar kinematics do not reach out to large enough radii to determine the dynamical structure of the outer halo that is expected at 15\arcsec\ and beyond. However, these substructures contain only a minor fraction (23 per cent) of the total stellar mass of this object.\\

Considering that the dynamical models of NGC\,1277 are also axisymmetric, with deviations from axial symmetry close to the black hole, the resemblance between both galaxies in the mass weights of the different orbit types is no surprise.
Unlike MRK\,1216, however, NGC\,1277 is mildly radially anisotropic in the immediate vicinity of the black hole and becomes strongly tangentially anisotropic beyond (Fig. \hyperref[fig:n1277_df]{\ref{fig:n1277_df}}, right panel).\\

Dynamical orbit based decompositions are a good tool for unravelling components in phase space. In this work, we could trace back the spinning orbital components of MRK\,1216 and NGC\,1277 to the flattened, low-\sersic\ components in the photometry, which are commonly associated with rotating structures. Moreover, our models show that the two compact galaxies are rotationally supported while both the orbital and photometric structures indicate the lack of a central pressure supported, massive spheroidal component. A small, mildly rotating bulge could be present, given the match in mass and location of the second component in the photometric multi-\sersic\ fits and the dynamical substructures. However, taking into account their flattening, low \sersic\ index, small mass fraction and rotational support, these structures more likely correspond to rotating discs (maybe even thick disks), and can only be considered as a modest lower limit of a putative bulge (see e.g. Sec. \hyperref[sec:scaling_relations]{\ref{sec:scaling_relations}}).

It is also worth noting that although the models rely on the more general MGE, which is completely independent from the \sersic\ fits and devoid of any physical interpretation, the orbital substructures bear no resemblance to it. The location and mass weights of the orbital structures do not match the individual Gaussians, which is most pronounced in Fig. \hyperref[fig:n1277_df]{\ref{fig:n1277_df}} where the overall structure is clearly comprised of less than ten components.

Nevertheless, more decompositions are needed to get a better understanding. For instance, drawing boundaries between the substructures in phase space is not trivial, while the photometric models are often plagued by strong degeneracies between the individual components in the fit. In our case, the issue of connecting photometric and dynamical structures is most prominent in MRK\,1216, where the mismatch in mass between the most massive photometric component in the multi-\sersic\ fit and the mildly rotating orbital component at 4\arcsec\ is worrisome, while some minor components in the dynamical structure do not seem to have a photometric counterpart at all. Given that our Schwarzschild method is capable of recovering the distribution function and hence the rich, internal dynamical structure of ETGs \citep{2008MNRAS.385..614V}, this might hint at difficulties in a) recovering the stellar build-up by multi-component \sersic\ fits to the photometry and/or in b) simply linking flat, (high-) low-\sersic\ components with (non-)rotating dynamical structures. Currently, there is no a priori definition of what range of $\lambda_{z}$ values can be associated with these structures. The connection between the individual dynamical and photometric components here is therefore driven by the agreement between both, and our intuitive understanding of associating low $\lvert\lambda_{z}\rvert$ components with (non-)rotating bulges and high $\lvert\lambda_{z}\rvert$ components with highly rotating disks.

Surely, modelling limitations, such as the assumption of axisymmetry (but see also Sec. \hyperref[sec:uncertainties]{\ref{sec:uncertainties}}) will also have a non-negligible effect on the recovery of the internal dynamics and their subsequent interpretation. More tests are therefore necessary, optimally of mock galaxy kinematics of purely rotational or pressure supported dynamical systems, to assess the robustness of our approach. This, in turn, will yield valuable information regarding the reliability and physical interpretation of photometric decompositions.

\subsubsection{Classification And Comparison}
\label{sec:comp_class}

In \cite{2007MNRAS.379..401E,2011MNRAS.414..888E}, early-type galaxies were separated into two classes of systems based on their specific stellar angular momentum. Fast rotators reveal a high specific angular momentum, comprise the majority of early-type galaxies, are close to axisymmetric in most cases and span a large range of anisotropy profiles \citep{2007MNRAS.379..418C}, in contrast to slow rotators which appear to be nearly isotropic. MRK\,1216 and NGC\,1277 are both fast rotators, as is expected by their rapid rotation around the apparent short axis, with a specific angular momentum $\lambda_R$ of 0.34 (0.41) and 0.25 (0.53) within one (three) effective radii.

To facilitate a comparison between the dynamical structure of the two compact galaxies in this work and a more general and representative sample of galaxies, such as presented within the \sauron\ framework, we follow the procedure and notation in \cite{2007MNRAS.379..418C} and show the relation between the global anisotropy parameter $\delta = 1 - \Pi_{zz}/\Pi_{xx}$ \citep{1987gady.book.....B} and the anisotropy parameter $\beta_z = 1 - \Pi_{zz}/\Pi_{RR}$, which describes the shape of the velocity dispersion tensor in the meridional plane. The values in this work have been measured within 3 \Reff, i.e. $\sim$ 6 kpc and 3.5 kpc for MRK\,1216 and NGC\,1277 respectively. The measurements are based on a larger relative scale, in contrast to the \sauron\ sample which usually covers the kinematics only out to $\sim$ 1 \Reff, but corresponds much better to the \sauron\ measurements of typically larger ETGs in an absolute sense.

\begin{figure}
	\centering
	\includegraphics[width=0.48\textwidth]{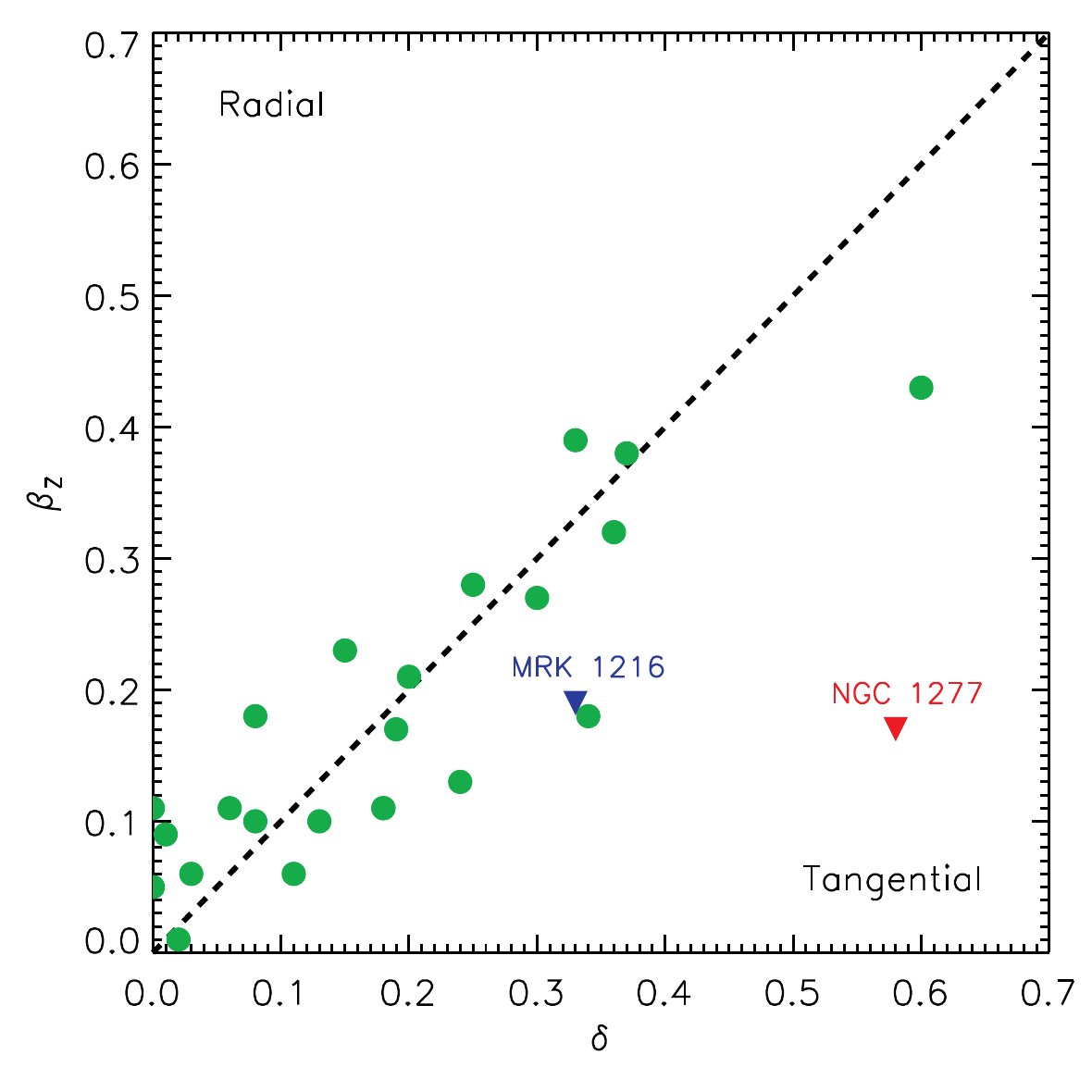}
	\caption{Anisotropy in the meridional plane ($\beta_z$) vs. global anisotropy ($\delta$) of MRK\,1216 (blue) and NGC\,1277 (red), measured by our orbit-based dynamical models of the wide-field \textit{PPAK} IFU and long-slit \textit{HET} data within 3 \Reff. MRK\,1216 follows the bulk of axisymmetric, fast-rotating ETGs in the \sauron\ sample (green), with a flattened velocity dispersion tensor in z-direction. NGC\,1277 exhibits a distinct kinematic structure. Besides a flattening in the meridional plane, NGC\,1277 is highly tangentially anisotropic in the plane orthogonal to the symmetry axis.}
	\label{fig:anisotropy}
\end{figure}

We confirm the picture of diverse anisotropy profiles of fast-rotating systems in Fig. \hyperref[fig:anisotropy]{\ref{fig:anisotropy}}. Here, MRK\,1216 is located in a region that is populated by the bulk of fast-rotating galaxies in the \sauron\ sample. It's only slightly tangential anisotropic in the  $\phi-r$ plane, which leads to the conclusion that most of its anisotropy can be traced back to a flattening of the velocity dispersion tensor in the meridional plane. While MRK\,1216 follows the trend presented in \cite{2007MNRAS.379..418C}, that fast-rotating early-type galaxies are mainly flattened oblate systems, NGC\,1277 is an outlier in every aspect and appears to belong (kinematically) to a totally different class of objects. It is flattened in z-direction, but also shows a substantial amount of tangential anisotropy in the plane orthogonal to the symmetry axis, which is necessary to account for the high and extended amplitude in rotational velocity. In the \sauron\ sample only one galaxy, NGC\,4550 ($\beta_z = 0.43$ and $\delta = 0.60$), is highly dominated by tangential dispersion. In contrast to NGC\,1277, though, NGC\,4550 consists of two massive counter-rotating disks.

The difference between the two compact galaxies in our sample is not only a difference of orbital structure but also of sheer size (see Table \hyperref[tab:photometric_props]{\ref{tab:photometric_props}}), with MRK\,1216 being almost twice as large as NGC\,1277. Taking into account the similarity between MRK\,1216 and the \sauron\ galaxies, this may indicate that MRK\,1216 has already entered a path of becoming a "regular", fast-rotating elliptical whereas NGC\,1277 is still in its infancy.\\

\subsection{Black Hole}
\label{sec:black_hole}

\subsubsection{Masses}
\label{sec:masses}

A credible determination of \Mbh\ requires a very thorough analysis. In the most optimal case, this is done by dynamical modelling of high-spatial resolution data that can resolve the black hole sphere of influence ($R_{SOI} = G\Mbh/\sigma^{2}$), i.e. the region where the gravitational pull of the black hole dominates. However, even state-of-the-art adaptive optics can resolve $R_{SOI}$ only for a limited number of galaxies, unless the black hole is either very nearby or very massive. The sphere of influence of NGC\,1277 is about 1.6\arcsec - as measured from the best-fitting black hole mass of log(\Mbh/$M_{\scriptscriptstyle \odot}$) $\simeq$ 10.1 and the effective velocity dispersion in the \textit{PPAK} data (Table \hyperref[tab:photometric_props]{\ref{tab:photometric_props}}) - and hence at the edge of being resolved by the \textit{HET} kinematics. Yet, this measurement of the sphere of influence is based on the assumption that the stellar density is well approximated by an isothermal sphere, and changes drastically if we adopt a more conservative estimate based on the region where the enclosed stellar mass equals the black hole mass, which yields $R_{SOI} = 0.9$\arcsec\ (Fig. \hyperref[fig:model_res]{\ref{fig:model_res}}). Moreover, even if the sphere of influence is resolved, our measurements still rely on the seeing limited dispersion peak and $h_4$ values within the central 1\arcsec\ and caution should be exercised regarding the black hole mass reliability in NGC\,1277 (but see also Section \hyperref[sec:uncertainties]{\ref{sec:uncertainties}} for a more in-depth discussion of the black hole mass).

Nevertheless, the gravitational, and consequently the dynamical, influence of the black hole is clearly imprinted in the observed velocity moments. The rapid rise and distinct peak in the velocity dispersion profile as well as the positive values in $h_4$ indicate a strong mass excess within the central arcseconds. The high $h_4$ values imply an LOSVD with heavy tails and a considerable amount of rapidly rotating stars in the very centre, and these features can - as far as the models are concerned - be solved best with a central black hole mass of log(\Mbh/$M_{\scriptscriptstyle \odot}$) $\simeq$ 10.1. Models with an "ordinary" SMBH of log(\Mbh/$M_{\scriptscriptstyle \odot}$) $\sim$ 9, as suggested by $\Mbh - L_{Bulge}$, are not able to recover the photometric and kinematic properties, as they fail to either fit the dispersion profile and/or the fourth Gauss-Hermite moment. In particular, the robustness of the $h_4$ measurement eliminates the possibility of a more moderate black hole measurement in favour of a higher mass-to-light ratio, as illustrated in Fig. \hyperref[fig:n1277_maps_models]{\ref{fig:n1277_maps_models}}.

The same, however, cannot be said for MRK\,1216. Although the best-fitting model favours an over-massive SMBH, the total absence of a black hole cannot be ruled out. Models with and without a black hole provide an almost equally good fit to the kinematics and thus are not able to discern between the various black hole mass scenarios, which is why (for the time being) our measurements can only be regarded as an upper limit. Upcoming high spatial resolution spectroscopic observations with \textit{NIFS} (PI: Walsh) that resolve the sphere of influence will be able to tell the difference and show whether or not MRK\,1216 follows the trend of NGC\,1277.\\

\subsubsection{Scaling Relations}
\label{sec:scaling_relations}

\begin{figure}
	\centering
	\includegraphics[width=0.48\textwidth]{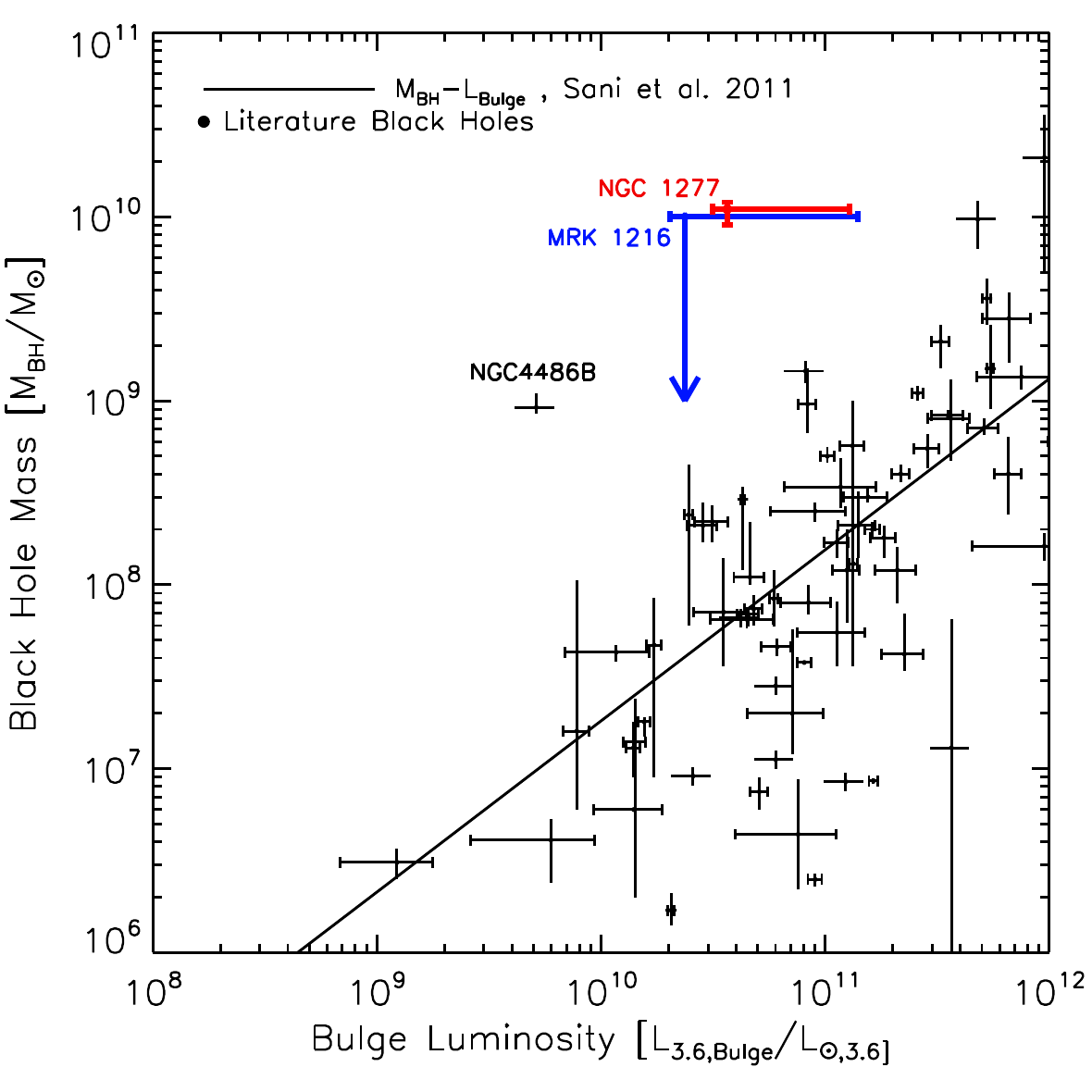}
	\caption{$\Mbh-L_{3.6, Bulge}$ plot from \citet{2011MNRAS.413.1479S}, with \textit{Spitzer}/IRAC 3.6 $\mu$m bulge-disk decompositions and dynamical black hole mass measurements (including 1$\sigma$ errors) for 57 galaxies. The red and blue error bars represent 3$\sigma$ statistical uncertainties of NGC\,1277's and MRK\,1216's black hole mass. The lower limit for the black hole mass in MRK\,1216 is consistent with no black hole. For the \textit{K}-band bulge luminosities we adopt their total \textsc{2MASS} \textit{K}-band luminosities and the bulge-to-total ratios from our photometric multi-component decompositions of the \textit{HST} \textit{V}- and \textit{H}-band images.}
	\label{fig:sani_plot_black_new}
\end{figure}

We place both black hole masses back onto the $\Mbh - L_{Bulge}$ relation (Fig. \hyperref[fig:sani_plot_black_new]{\ref{fig:sani_plot_black_new}}). To this end, we utilise the compilation of \cite{2011MNRAS.413.1479S} with bulge-disk decompositions of 57 galaxies, based on \textit{Spitzer}/IRAC 3.6 $\mu$m mid-infrared imaging. The use of mid-infrared data has not only the advantage of less dust extinction susceptibility but is also a better tracer of the underlying stellar mass. To this sample, we add 10 black hole masses with \textsc{2MASS} \textit{K}-band luminosities and bulge-to-total ratios if available; five from disk galaxies, as presented in \cite{2011ApJ...727...20K} and \cite{2010ApJ...721...26G}; two in brightest cluster galaxies (BCG), published in \cite{2011Natur.480..215M}; one from a low-luminosity elliptical \citep{1997ApJ...482L.139K}; one from a high velocity dispersion lenticular galaxy \citep{2011MNRAS.410.1223R} and one from a recent merger galaxy \citep{2009ApJS..182..216K}, investigated in \cite{2011ApJ...741...38G}. The solid line in Figure \hyperref[fig:sani_plot_black_new]{\ref{fig:sani_plot_black_new}} represents the black hole mass-bulge luminosity relation ($\Mbh - L_{Bulge}$) derived in \cite{2011MNRAS.413.1479S}, based on their bulge-disk decompositions of literature black hole host galaxies and a linear regression fit to the data. The blue and red error bars mark our findings for MRK\,1216's and NGC\,1277's black hole mass with a statistical uncertainty of 3$\sigma$. Their \textit{K}-band bulge luminosities are based on the photometric decompositions in Section \hyperref[sec:mrk1216_photometry]{\ref{sec:mrk1216_photometry}} and \hyperref[sec:ngc1277_photometry]{\ref{sec:ngc1277_photometry}} of the \textit{HST} \textit{H}- and \textit{V}-band images (using the second innermost component of the multi-\sersic\ fit as a lower limit to the bulge luminosity while the bulge in the two-component \sersic\ fit serves as an upper limit) and their total \textsc{2MASS} \textit{K}-band luminosities. The figure illustrates the exceptional position of NGC\,1277. The best-fitting black hole mass remains a significant outlier from this relation and the 3$\sigma$ lower bound, log(\Mbh/$M_{\scriptscriptstyle \odot}$) = 9.9, still overshoots the upper 99.7 per cent confidence envelope of the relation by at least one order of magnitude. Similarly, and given the difficulties in identifying a bulge in both the dynamical and photometric decompositions (Sec. \hyperref[sec:ngc1277_photometry]{\ref{sec:ngc1277_photometry}} and \hyperref[sec:ngc1277_orb]{\ref{sec:ngc1277_orb}}), NGC\,1277 strongly deviates from the  black hole mass-total luminosity relation in \textit{K}-band \citep{2014ApJ...780...70L}, where the lower 3$\sigma$ limit of the black hole mass is marginally consistent with upper 3$\sigma$ bound of this relation.

Interestingly, the black hole measurement in NGC\,1277 is consistent with estimates of the scatter in the $\Mbh - L_{Bulge}$ and $\Mbh - \sigma$ relation in the optical \citep{2009ApJ...698..198G,2013ApJ...764..184M,2013ARA&A..51..511K}. The best-fitting black hole mass is an outlier by a factor of $\sim$ 8 (4) with respect to the mean predicted black hole mass in the $\Mbh - L_{Bulge}$ ($\Mbh - \sigma$) relation, but still within a 3$\sigma$ (2$\sigma$) confidence if the intrinsic/cosmic scatter of 0.44 (0.38) dex is taken into account.\\

The consistency between the black hole mass of NGC\,1277 and the $\Mbh - L_{Bulge}$ relation in the optical is a result of a larger intrinsic scatter when compared to the relation in the mid-infrared, and expected if the black hole mass-bulge luminosity relation is just a tracer of a more fundamental link between black hole mass and bulge mass \citep{2003ApJ...589L..21M,2004ApJ...604L..89H}. As NGC\,1277 becomes an outlier in the tighter relation in the mid-infrared this could be interpreted as a hint for a different formation channel that lacks the physical interplay and causal link between the black hole and the spheroidal component of its host \citep{1998A&A...331L...1S}. An alternative but speculative example for such a channel is presented by \cite{2013ApJ...772L...5S}. Based on gravitational radiation recoil during the final stages of two massive black hole mergers and the accompanied velocity kicks \citep{2004ApJ...607L...9M}, they argue that a massive black hole in a nearby galaxy could have been ejected and recaptured by NGC\,1277. The probability of mergers which could produce kicks that exceed the escape velocity of the host is non-negligible \citep{2010PhRvD..81h4023L}, but the ejected black hole would be accompanied by a hypercompact stellar system (HCSS) with a stellar mass of $M_{HCCS} \le 10^{-2} \times$ \Mbh\ \citep{2009ApJ...699.1690M}. Consequently, we ought to observe a considerable number of these free floating, compact stellar systems already in the Virgo cluster. The lack of any such observational evidence questions the likeliness of this scenario.\\

A different idea has been put forward by E13, to reconcile the black hole in NGC\,1277 with predictions of the scaling relations. Here, individual dynamical models from $N$-body realisations have been chosen to match the \textit{HET} and \textit{HST} data in the very centre and over a wide radial range. A hand-picked model without dark matter and a black hole mass of log(\Mbh/$M_{\scriptscriptstyle \odot}$) = 9.7 shows an acceptable fit to the kinematics. However, no parameter search was done to find a best-fit model and confidence intervals. In particular, the models fail to fit all kinematic moments simultaneously and especially the fourth Gauss-Hermite moment, which seems to be the key discriminator between the various black hole mass scenarios.

The presence of a bar was also discussed briefly as an alternative explanation of the very distinctive kinematic moments. For instance, a model with no black hole but an edge-on bar was able to overcome the problem of fitting $h_4$ while a model with an end-on bar was a good fit to the remaining data. The truth could lie somewhere in between these two opposing bar configurations, with a more moderate black hole mass in addition. However, it is worth noting that we were not able to find any evidence for the presence of a bar in any of the data sets. Although limited by the spatial resolution of our kinematic observations, we see a clear trend for an anti-correlation between $h_3$ and $v$. The presence of a bar should break this trend over its projected length, as has been shown by N-body simulations of bar-unstable disks by \cite{2005ApJ...626..159B} and observations of edge-on spiral galaxies \citep{2004AJ....127.3192C}. We also thoroughly inspected the high-resolution \textit{HST} data and performed photometric decompositions with \textsc{Galfit} that included a bar. The decompositions, however, resulted in visually and statistically worse fits. Even though we do not rule out the possibility of a small (i.e. $\le$ 1\arcsec) end-on-bar, which would not be resolved by the currently available data, we raise concerns that a) this would be a special and unlikely case and b) that the same argument could be easily applied to a number of other dynamical black hole measurements. Finally, high-resolution spectroscopic observations with NIFS (PI: Richstone) have already been carried out for NGC\,1277, which will shed light on this argument.\\

\subsection{Dark Matter Halo}
\label{sec:dm_halo}

\subsubsection{Dark Halo Detection}
\label{sec:dark_halo_detection}

Based on our orbit-based models of the wide-field IFU data and the \textit{HET} long-slit kinematics, we have no clear evidence for the presence of a dark halo in NGC\,1277. In MRK\,1216, on the other hand, the data can only be recovered with the addition of dark matter.  We note, though, that the detection is only of a weak statistical nature. The best-fitting model without a dark halo deviates by $\Delta\chi^{2} =15$ from the overall best-fitting model, which is slightly beyond the 3$\sigma$ confidence limit, as has been shown in Fig. \hyperref[fig:model_res]{\ref{fig:model_res}}. Given the four kinematic moments of the \textit{PPAK} data and the three \textit{HET} slits that are fitted simultaneously, the mean deviation per kinematic moment and bin is $\sim$ 0.04 between both models and the predicted velocity moments are barely distinguishable in the IFU maps as well as in the major axis profiles of the long slits (Fig. \hyperref[fig:m1216_maps_models]{\ref{fig:m1216_maps_models}}). The difference in the relative likelihoods of both models is mostly attributable to the IFU kinematics, which account for 2/3 of the $\Delta\chi^{2}$. This is in contrast to e.g. the statistically stronger black hole detection in NGC\,1277, where the difference between our best-fitting model and a model with a black hole mass of log(\Mbh/$M_{\scriptscriptstyle \odot}$) = 9.7 is driven by a few central bins with a $\Delta\chi^2$ of 25 and is visible in the mismatch of the central velocity moments (Fig. \hyperref[fig:n1277_maps_models]{\ref{fig:n1277_maps_models}}).

\begin{figure}
	\centering
	\includegraphics[width=0.48\textwidth]{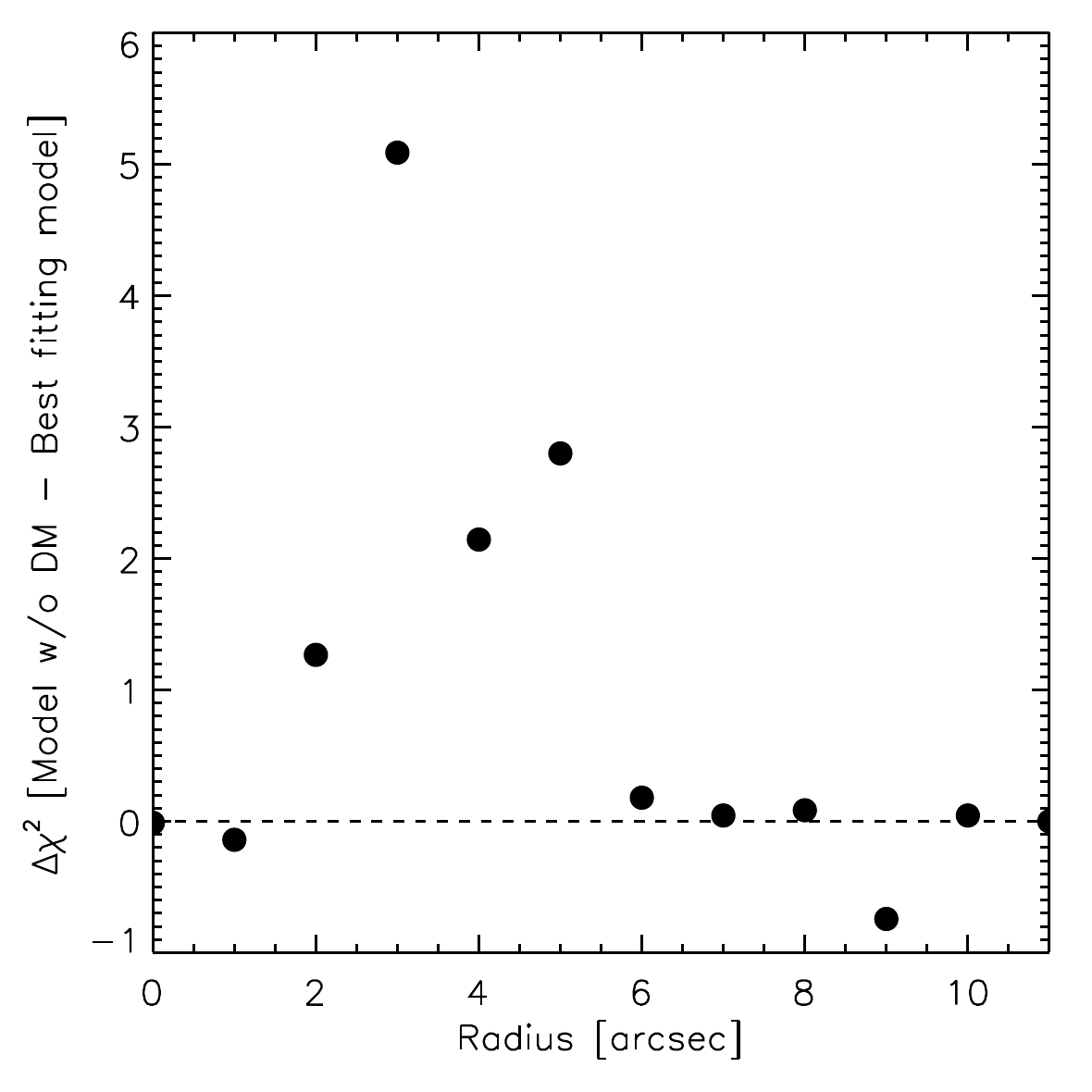}
	\caption{Total $\chi^{2}$ difference of MRK\,1216's \textit{PPAK} data, between the best-fitting model without a dark halo and the overall best-fitting model, as a function of distance from the centre.}
	\label{fig:chi2diff}
\end{figure}

In the case of MRK\,1216, one would expect the outer bins to be the driver of the $\chi^2$ difference, where the lack of the dark halo should lead to the most prominent deviation between a model with and without a dark halo. We show that this is not the case. In Fig. \hyperref[fig:chi2diff]{\ref{fig:chi2diff}} we present the $\Delta\chi^{2}$ of the \textit{PPAK} data between the best-fitting model without a halo and the overall best-fitting model as a function of radius. The plot reveals the central region ($\le 5\arcsec$) as the cause of the $\Delta \chi^{2}$ difference. In addition to the statistical claim, this is a clear indication for additional dark mass that can be explained as follows: The absence of a dark halo naturally leads to an increase in the stellar mass-to-light ratio which mitigates the effects of missing mass in the outer parts. This is illustrated in Fig. \hyperref[fig:chi2diff]{\ref{fig:chi2diff}}, where the simple mass-follows-light model presents an equally good fit to the outer kinematics as the overall best-fitting model with a halo. The rise in \textit{constant} mass-to-light ratio however leads to a mismatch between data and model (or best-fitting model and best-fitting model w/o a halo) in the central regions.

A natural way to resolve this issue and to make the mass-follows-light models fit the outer and inner data points would be a radially increasing mass-to-light ratio that adopts the best-fitting value for the central regions and steadily increases towards the outer parts to account for the outer bins. Indications for such a trend should be imprinted in the colour profiles of galaxies \citep{2001ApJ...550..212B,2003MNRAS.344.1000B}, attributable to variations in age and/or metallicity of the galaxy's stellar population as is expected if galaxies grow inside-out \citep{2013ApJ...764L...1P,2013ApJ...766...15P}. We have therefore inspected the colour profiles of NGC\,1277 and MRK\,1216 based on \textit{SDSS} $g-i$ and \textit{HST} F814W-F160W imaging, but the lack of a significant trend with increasing distance from the centre in both does not promote the use of a stellar mass-to-light-ratio gradient in our models. This is also in accordance with the spectroscopic results of T14 for NGC\,1277, which suggest a uniformly old stellar population with almost constant metallicity and $\alpha$/Fe values. Even though spatial gradients in the colours (e.g. \citealt{1989AJ.....98..538F,2003AJ....126..596T}) and stellar population properties (e.g. \citealt{2013ApJ...776...64G}) of individual ETGs have been observed, which would justify the assumption of a radially varying \MLstar, the analysis of a large sample of late- and early-type galaxies suggests that gradients for \MLstar\ are in general negative \citep{2011MNRAS.418.1557T}, which in turn would further increase the dark mass and hence the discrepancy between our models with and without a dark halo \citep[see e.g.][]{2013ApJ...768L..21M}.

In principle, variations in the IMF could conceal a colour gradient in both compact objects while effectively increasing the stellar mass-to-light ratio. A recent study of radial trends in the IMF of individual, massive, high-dispersion galaxies however argues the converse and indicates that the observed trend of a bottom-heavy IMF is only a local property - confined to the central region of a galaxy - followed by a decrement of the IMF slope with increasing distance from the centre, and hence a radially decreasing stellar mass-to-light ratio (see \citealt{2015MNRAS.447.1033M,2015arXiv150501485M})

Even if MRK\,1216 and NGC\,1277 did not assemble in the same way as the most massive ellipticals did and just evolved passively (see Section \hyperref[sec:high_redshift]{\ref{sec:high_redshift}}), there is currently no comprehensive theory of star formation that could explain the tendency of a more bottom heavy IMF in the less dense outskirts of galaxies. 

\subsubsection{Dark Halos In Elliptical Galaxies}
\label{sec:dark_halo_detection}

The results for the dark matter halos in our analyses are puzzling, in particular in the light of other orbit-based dynamical models with a similar extent in the stellar kinematic information of the full LOSVD (e.g. \citealt{1997ApJ...488..702R,2007MNRAS.382..657T}). While those investigations provided unambiguous evidence for the presence of dark matter in elliptical galaxies, we can only partially confirm this trend. For instance, \cite{2009MNRAS.398..561W} examined the two early-type galaxies NGC 3379 and NGC 821. Based on \sauron\ data out to four effective radii, they obtained a dark matter contribution of at least 8 and 18 per cent to the total mass budget within one \Reff. They also predicted a dark matter fraction of 30-50 per cent within four \Reff\ and concluded that dark matter is necessary to explain the observed kinematics.

In particular NGC\,3379, with a small effective radius of $\sim$ 2\,kpc, is easily comparable to our compact objects, where we provide a similar relative and absolute coverage of the LOSVD. However, we can detect a dark halo only in MRK\,1216. For NGC\,1277, the reverse is true as the models are able to recover the observations without the need of any dark matter and predict a maximal dark matter fraction of only 13 per cent within one effective radius. Interestingly, the analysis of a larger sample of these compact, high central velocity dispersion galaxies (Y{\i}ld{\i}r{\i}m et al., in prep.) indicates a dominance of the stellar mass distribution within one effective radius, owed to the decrease of the effective radius which encompasses less of the dark volume. While the contribution of dark mass to the total mass budget within one effective radius is around 10 per cent for these objects, with stellar masses above log($\Mstar/M_{\scriptscriptstyle \odot}) = 11.1$, and thus systematically lower than inferred for the population of massive (log($\Mstar/M_{\scriptscriptstyle \odot}) \ge 11.1$), local ETGs in e.g. the \textsc{SAURON} and \textsc{ATLAS$^{3D}$} sample \citep{2006MNRAS.366.1126C,2013MNRAS.432.1709C}, it appears to be consistent with the dark matter content of compact, high central velocity dispersion galaxies at redshift $z=2$ \citep{2013ApJ...771...85V} (see also Sec. \hyperref[sec:high_redshift]{\ref{sec:high_redshift}}).

\subsubsection{NFW Profiles}
\label{sec:nfw_profiles}

The aforementioned numbers are based on the assumption that the dark halo profile in both galaxies is well described by a spherically symmetric NFW profile. As a further check of this hypothesis, we compare our results with a semi-analytic approach of \citet{2010ApJ...710..903M} that links the stellar mass of a galaxy to the mass of its dark matter halo. By comparing the galaxy mass function with the halo mass function they obtained a well-defined stellar-to-halo mass (SHM) relation, which enables the determination of a halo mass for a given stellar mass and vice versa.

\begin{figure}
	\centering
	\includegraphics[width=0.48\textwidth]{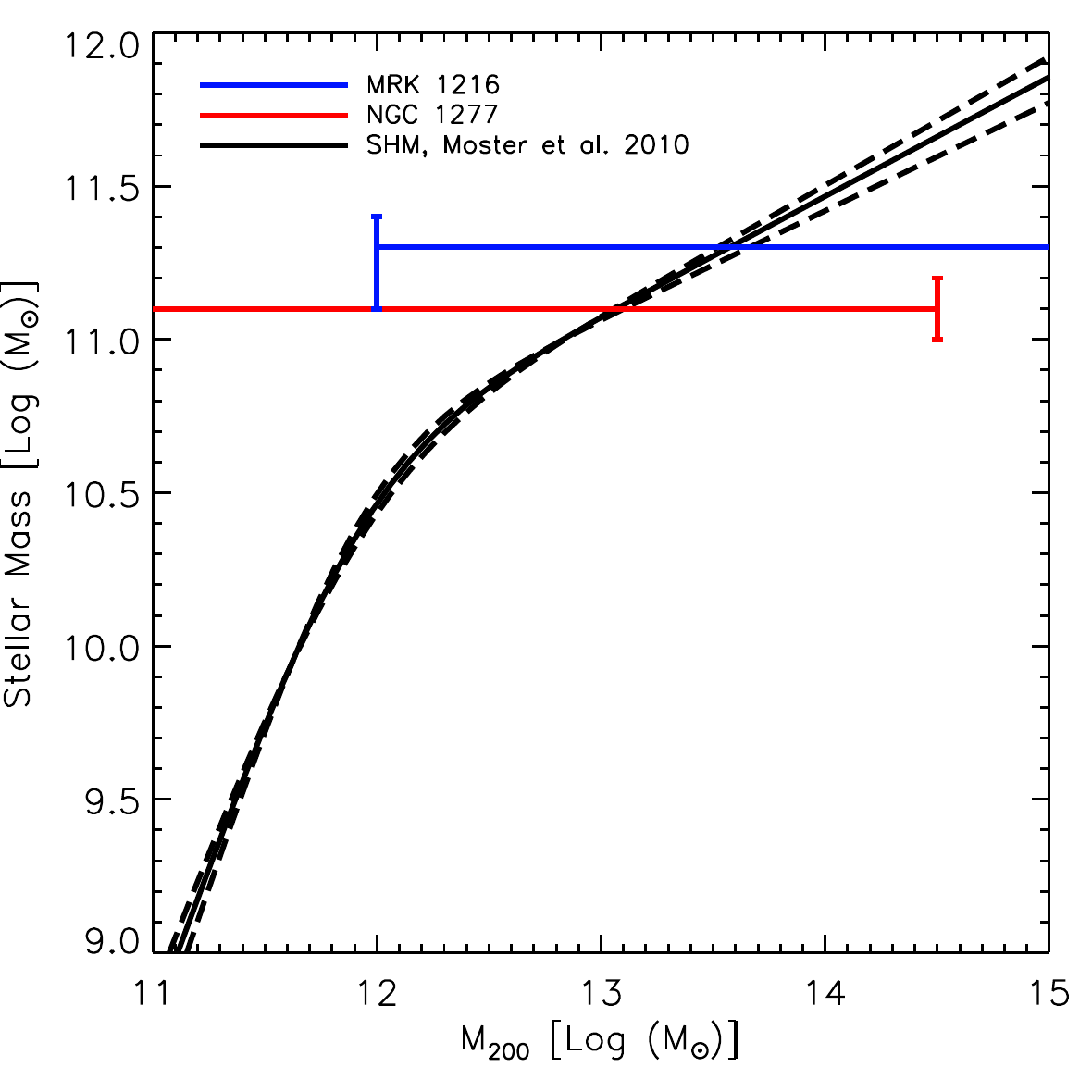}
	\caption{Stellar-to-halo mass (SHM) relation from \citet{2010ApJ...710..903M} (black line). Overplotted are stellar-to-halo mass predictions of MRK\,1216 (blue) and NGC\,1277 (red), derived from our entire set of Schwarzschild models. The plot displays upper and/or lower limits in the relevant range of $M_{200} = 10^{11} - 10^{15}$ \Msun. In the case of NGC\,1277, the lower limit is consistent with no dark matter.}
	\label{fig:moster_comb}
\end{figure}

In Figure \hyperref[fig:moster_comb]{\ref{fig:moster_comb}} we overplot all results of our orbit-based dynamical models that are enclosed by the 99.7 per cent confidence limit and thus immediately test the consistency of our models with the standard cold dark matter paradigm ($\Lambda$CDM), which is the underlying cosmological model that defines the halo mass function. There is a small range of overlap between the predictions of our models and the SHM relation that would imply consistency with $\Lambda$CDM, but we also see a wide coverage of allowed halo masses due to the inability of our models to constrain the parameter space in $c_{DM}$ and $f_{DM}$ effectively. A different quantification of the SHM relation in terms of late- and early-type galaxies \citep{2010MNRAS.407....2D} does not change anything in this respect, as the halo masses of both MRK\,1216 and NGC\,1277 still overshoot the upper and lower bound of these relations by about one order of magnitude.\\

The difficulty in detecting a dark halo in both galaxies, in particular in NGC\,1277, and in constraining the dark halo parameters cannot simply be attributed to the use of larger 3$\sigma$ confidence intervals in our study. More probably, the obstacle can be traced back to their compactness and high stellar masses within the small spatial extent that is probed by the available kinematic data. Given our best-fitting results, stellar masses are of the order of log($\Mstar/M_{\scriptscriptstyle \odot}) = 11.1$ within 7 and 5 kpc for MRK\,1216 and NGC\,1277 respectively. The contribution of a NFW halo to the total mass profile within the same range can be estimated to be of the order of log($\Mstar/M_{\scriptscriptstyle \odot}) = 10.5$ and 10.2 for MRK\,1216 and NGC\,1277, assuming that the mass-concentration relation \citep[e.g.][]{2001MNRAS.321..559B,2008MNRAS.390L..64D,2008MNRAS.391.1940M} and stellar-to-halo mass relation \citep[e.g.][]{2010ApJ...710..903M,2010MNRAS.404.1111G,2010ApJ...717..379B} hold. Accordingly, the dark halo would constitute $\sim$ 25 per cent of the total mass budget in MRK\,1216 and only $\sim$ 13 per cent in NGC \,1277, which would explain the statistically weak detection in the former and our struggle to verify the presence of dark matter in the latter, as the additional mass is easily compensated by a marginal increase in the stellar M/L.

\subsection{The Origin Of Compact, High Velocity Dispersion Galaxies}
\label{sec:high_redshift}

MRK\,1216 and NGC\,1277 are unusual and rare galaxies in the nearby universe. Their detection was a result of the selection criteria of the \textsc{HETMGS} \citep{2015ApJS..218...10V} which, based on the sphere of influence argument, naturally looked for dense, high-dispersion objects that could possibly host very massive SMBHs. Still, the number of objects that are similar to both, even in the \textsc{HETMGS}, are limited and questions regarding their origin and evolution arise. Typically, the stellar populations are the first resort for exploring the (stellar) evolutionary history of a galaxy, but this would only be feasible for the long-slit spectroscopic data due to the short wavelength coverage of our IFU observations. We therefore focus on the already available data and find hints for a rather unremarkable and quiet past in their photometric and structural properties.\\

MRK\,1216 is an isolated galaxy in the field, which has only two other galaxies within a search radius of 1 Mpc at its distance. It has no tidal signatures or asymmetries and any recent galaxy-galaxy interaction can therefore be ruled out. Given its compact shape and rotationally supported dynamical structure, an active merging history seems to be unlikely, too. Violent relaxation due to collisionless 1:1 or 1:2 mergers, for instance, commonly yields boxy, slow rotating ellipticals \citep[e.g][]{2003LNP...626..327B,2003ApJ...597..893N,2006MNRAS.372..839N}, which is at odds with the rapid rotation and dynamical characteristics of MRK\,1216 and NGC\,1277. Note, though, that this formation scenario also fails in reproducing the detailed dynamical properties of massive ETGs in general \citep{2008ApJ...685..897B,2014MNRAS.444.3357N}. Likewise, violent relaxation due to "dry" unequal mass mergers have been shown to be able to recover the photometric and kinematic properties of disky, fast-rotating ellipticals \citep{2003ApJ...597..893N,2014MNRAS.444.3357N}. However, unequal mass mergers tend to increase the galaxy size drastically \citep{2009ApJ...699L.178N,2010ApJ...725.2312O,2013MNRAS.429.2924H}, and are thus hard to reconcile with the sizes observed in both compact objects. On the other hand, dissipative equal mass mergers can reproduce fast-rotating ETGs, while also recovering the tilt in the FP \citep[e.g.][]{2006ApJ...641...21R,2006ApJ...650..791C}. However, both MRK\,1216 and NGC\,1277 are still outliers in the FP (Y{\i}ld{\i}r{\i}m et al., in prep.) and the non-negligible gas fractions involved in the merging process are expected to boost the star formation activity, which is in contrast to the uniformly old age and star formation history of NGC\,1277 (T14 and \citealt[]{2015arXiv150501485M}), unless the merging event has taken place more than 10 Gyr ago. As a result, the aforementioned simulated merger scenarios - which actually have been tailored to test and recover the formation and evolution mechanisms of today's population of ETGs - fail to fully explain the two compact galaxies in this work.\\

This brings up the idea whether these two objects are representatives of a galaxy population that has (at some point) taken a significantly different path than the present-day massive galaxy population, which has grown in mass and size since $z=2$ \citep{2010ApJ...709.1018V} presumably through successive (minor and major) merging events. In fact, stellar age estimates of the present-day massive galaxy population \citep{2015MNRAS.448.3484M} are consistent with the inferred stellar ages of NGC\,1277, and the range of allowed stellar mass-to-light ratios in our dynamical models cannot rule out the trend of a more bottom-heavy IMF with increasing stellar velocity dispersion, which is also commonly observed for the most massive ellipticals. Accordingly, both galaxies would present unaltered and passively evolved analogues of the massive, quiescent galaxy population at much earlier times, which are thought to constitute the cores of today's massive ellipticals.\\

Indeed, the two galaxies are quantitatively similar to the quiescent galaxies at $z = 2$ . Those are also found to be small \citep{2005ApJ...626..680D,2006ApJ...650...18T,2007ApJ...656...66Z,2008ApJ...677L...5V,2008ApJ...688...48V,2014ApJ...788...28V}, possess extremely high dispersions \citep{2009Natur.460..717V} and generally have a disk-like structure \citep{2011ApJ...730...38V}. T14 were able to go beyond a simple structural, photometric and kinematic comparison by carrying out a stellar population analysis of NGC\,1277. Based on long-slit spectra out to $\sim$ 3 \Reff\ they found that NGC\,1277 consists of a uniformly old stellar population ($\ge$ 12 Gyr), formed during a very short-lived era at $z \ge 3$ with an intense star formation rate. This again is in good agreement with spectroscopic investigations of \cite{2006ApJ...649L..71K,2009ApJ...700..221K} and \cite{2012ApJ...754....3T} for individual quiescent galaxies at $z \sim 2$. Those have also very old stellar populations, with the bulk of their mass already assembled at $z \ge 3$, and are absent of any significant star formation. Recently, evidence has even mounted for a further evolutionary link with the sub-millimeter galaxies (SMGs) at $z \ge 3$ \citep{2014ApJ...782...68T}. The SMGs not only provide the necessary ages and compact sizes, but also the intense star formation rates - which could have been triggered by gas-rich (major) mergers at high redshifts \citep{2007ApJ...658..710N,2010ApJ...722.1666W} - to explain the old, compact stellar populations of the quiescent galaxies at $z = 2$ (but see also \citealt{2014ApJ...780....1W,2014MNRAS.438.1870D} and \cite{2014ApJ...791...52B} for an alternative formation channel).\\

The resemblance between NGC\,1277 and MRK\,1216 and the quiescent galaxies at higher redshifts is remarkable. Nevertheless, we need to go beyond single anecdotal examples if we want to underpin the claim that the compact galaxies, found in the \textsc{HETMGS}, are passively evolved descendants of the quiescent population at $z \le 2$. It is encouraging though that we have found 18 compact, high-dispersion, early-type galaxies in total, which will enable us to investigate in detail their photometric, structural, kinematic and stellar evolutionary properties.

\section{Uncertainties}
\label{sec:uncertainties}

Our orbit-based dynamical analysis and its implications are afflicted by a number of moderate concerns, which we would like to highlight here.
\begin{itemize}
\item
During the construction of our dynamical models we have assumed axisymmetric stellar systems. The models are robust with respect to changes in the inclination (Section \hyperref[sec:mrk1216_schwarzschild]{\ref{sec:mrk1216_schwarzschild}}) but the orbital structures can change rapidly when the assumption of axial symmetry is relaxed. Even mild triaxiality would alter the observed phase space structures in Fig. \hyperref[fig:m1216_df]{\ref{fig:m1216_df}} and \hyperref[fig:n1277_df]{\ref{fig:n1277_df}} noticeably, leading also to variations in the derived values of e.g. the black hole mass \citep{2010MNRAS.401.1770V}. In this respect, even the slightest twist in the PA can be interpreted as a deviation from axisymmetry. An MGE with a fixed PA for all Gaussians (Sec. \hyperref[sec:mrk1216_photometry]{\ref{sec:mrk1216_photometry}}) is a necessary but insufficient condition for the assumption of axial symmetry, as triaxial deprojections cannot be ruled out. However, the body of evidence that has been presented throughout this paper, namely the fast and regular rotation around the short axis, the anti-correlation between $v$ and $h_3$, the negligible mis-alignment between the kinematic and photometric PA and results from shape inversions of a large sample of fast-rotating early-type galaxies \citep{2014MNRAS.444.3340W}, show that axial symmetry is a justified assumption of the intrinsic shape of both compact objects.\\
\item
Tightly linked to the black hole mass is the stellar mass-to-light ratio which in turn is degenerate with the dark matter halo \citep{2009ApJ...700.1690G}. The determination of \MLstar\ is therefore crucial in constraining the black hole, if the black hole sphere of influence is not resolved \citep{2013AJ....146...45R}. In our set of dynamical models, the stellar mass-to-light ratio is assumed to be constant throughout the observed range of kinematics. This is also supported by the lack of colour gradients in both galaxies and an only mild change in the stellar population properties of NGC\,1277 within a radial extent of $\sim$ 3 effective radii (T14). We emphasise, though, that a change in the stellar M/L at smaller radii ($\le$ 1\arcsec) might be present, which would neither be resolved by the \textit{SDSS} photometry nor by our long-slit and wide-field IFU kinematics of NGC\,1277. In fact, \cite{2015arXiv150501485M} found a slight increase in the stellar M/L in NGC\,1277, by tracing gravity sensitive features in their NIR, long-slit spectroscopic data. Limited by the spatial resolution of their data set, however, the stellar M/L increases only marginally from 7.0 to 7.5 between their outermost ($\sim$ 6\arcsec) and innermost ($\sim$ 1\arcsec) data points, which is still consistent with the stellar M/L inferred in our dynamical models.\\

Despite emerging evidence for strong systematic variations in the IMF of early-type galaxies \citep[e.g][]{2010ApJ...721L.163A,2011MNRAS.416..322D,2012Natur.484..485C,2012ApJ...753L..32S}, predictions of SSP models with a single power-law Salpeter IMF \citep{1996ApJS..106..307V,2012MNRAS.424..157V} are consistent with our orbit-based dynamical models of both NGC\,1277 and MRK\,1216. This has formerly been excluded at the 3$\sigma$ level for NGC\,1277 in vdB12. Whereas those conclusions based on spectral synthesis fits of NGC\,1277's single SDSS aperture \citep{2005MNRAS.358..363C} with Bruzual\&Charlot models \citep{2003MNRAS.344.1000B}, our values for the stellar mass-to-light ratios are derived from fits to the spatially resolved long-slit spectra (T14) based on MIUSCAT \citep{2012MNRAS.424..157V,2012MNRAS.424..172R} SPS models. Both approaches show that NGC\,1277 is comprised of an uniformly old stellar population. Since more standard variations of the inferred stellar population parameters (i.e. metallicity and $\alpha$-abundance) do not seem to be able to explain the difference between the \MLstar\ values quoted in vdB12 and the values in this work, we assume that the difference might be attributable to the choice of a non-standard IMF in the former. We consider the values reported here as a conservative estimate. Hence the strong tendency towards a more bottom-heavy IMF in high dispersion galaxies \citep{2010ApJ...709.1195T,2012ApJ...760...70V,2012ApJ...760...71C,2013MNRAS.429L..15F,2013MNRAS.433.3017L,2014MNRAS.438.1483S} cannot be ruled out by our orbit-based dynamical analysis of MRK\,1216 and NGC\,1277, and would indeed favour the presence of a more moderate black hole mass. However, according to our Schwarzschild models, the upper range of possible mass-to-light ratios implies that the IMF in these two objects can only be more massive by $\sim$ 15 and 25 per cent at most in NGC\,1277 and MRK\,1216, respectively, with respect to a Salpeter IMF. While this is largely consistent with the observed scatter in the relation between IMF slope and velocity dispersion in the aforementioned studies, exotic variations of the IMF -  as predicted e.g. by the best-fitting relation in \cite{2010ApJ...709.1195T} and \cite{2014MNRAS.438.1483S}, which implies a shift in \MLstar\ by $\sim$ 40 per cent for a stellar velocity dispersion of $\sim$ 300 \kms\ - can be excluded.\\

\item
A major concern in the modelling of NGC\,1277 remains the nuclear dust ring. Although contaminated regions have been generously masked while constructing the luminous mass model, this is by no means an appropriate physical account of dust extinction. Since our orbit-based dynamical models measure the enclosed mass within a given radius, an underprediction of the stellar mass in the nucleus will obviously bias the measurement towards higher black hole masses, although it would take a considerable amount of mass to be screened by dust ($\ge 50\%$ of the stellar mass within 1\arcsec) to bring the black hole in line with the scaling relations.\\
\item
The biggest concern in the recovery of the individual mass contributions and in particular for the black hole mass in NGC\,1277, though, remains the accuracy of the kinematic measurements. E13 has shown that the detection of an over-massive SMBH entirely hinges on the dispersion peak and the positive $h_4$ values in the centre. This is easily verified by our models, where most of the $\chi^2$ difference between the best-fitting model and a model with a more moderate black hole mass of log(\Mbh/$M_{\scriptscriptstyle \odot}$) = 9.5 (Fig. \hyperref[fig:n1277_maps_models]{\ref{fig:n1277_maps_models}}) is attributable to the fits to $\sigma$ and $h_4$ that contribute 3/4 of the $\Delta \chi^2$. In particular, the seeing limited measurements within 1\arcsec\ in the \textit{HET} long-slits are the driver of this difference and question the reliability of the black hole mass estimate. Moreover, dust obscuration could also affect the measurement of the LOSVD, even though modest dust mass assumptions show that this is only significant for the large scale kinematics \citep{2001ApJ...563L..19B}.

The \textit{PPAK} observations, although limited by their spatial resolution, provide an independent way to assess the accuracy of the \textit{HET} measurements and the models in vdB12. As has been shown in Section \hyperref[sec:ngc1277_kinematics]{\ref{sec:ngc1277_kinematics}}, the central dispersion in the \textit{PPAK} cube is considerably lower than the peak observed in the \textit{HET} data. As a result, models that fit the combined data set predict a velocity dispersion that matches the \textit{PPAK} data but slightly fails to do so for the peak in the \textit{HET} data (Section \hyperref[sec:ngc1277_schwarzschild]{\ref{sec:ngc1277_schwarzschild}}). In addition, the central $h_4$ moments in the \textit{PPAK} data, while still positive, are slightly below the \textit{HET} measurements (Fig. \hyperref[fig:ppak_het_h4]{\ref{fig:ppak_het_h4}}), which is of concern considering that those values have been key in discriminating between the various black hole mass scenarios (Sec. \hyperref[sec:masses]{\ref{sec:masses}}). In principle, we could try to reconcile both data sets and let them "meet in the middle".  However, quantifying the offset between the long-slit and IFU kinematics is a non-trivial task, which is why we follow a different route and check the inter-consistency between both by fitting the \textit{PPAK} data individually.

\begin{figure}
	\centering
	\includegraphics[width=0.48\textwidth]{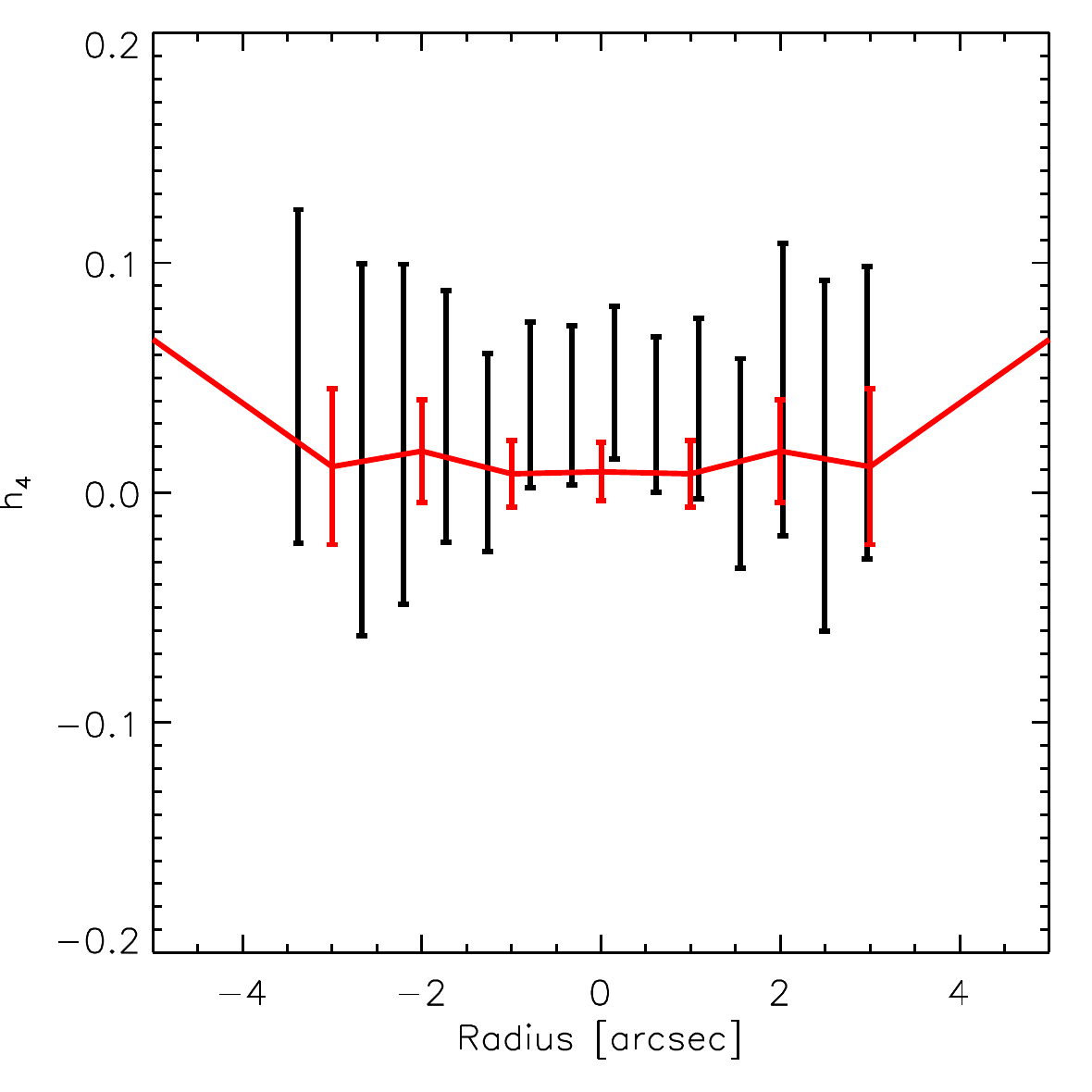}
	\caption{Comparison between one of the three \textit{HET} long-slit (black) and \textit{PPAK} IFU measurements (red) of the fourth Gauss-Hermite moment in NGC\,1277. The \textit{HET} measurements have been obtained along the apparent major axis. The \textit{PPAK} measurements correspond to the values of the Voronoi-binned data, for which the bin centroids are located within a 1\arcsec\ wide strip along the major axis.}
	\label{fig:ppak_het_h4}
\end{figure}

We display the results of these test models in Figure \hyperref[fig:n1277_consistency]{\ref{fig:n1277_consistency}}, and show $\Delta \chi^2 = \chi^2 - \chi^{2}_{min}$ as a function of black hole mass and dark halo mass (marginalising over all remaining parameters). The model predictions for NGC\,1277 based on the \textit{PPAK} data are shown on top with the predictions based on both data sets - already illustrated in Fig. \hyperref[fig:model_res]{\ref{fig:model_res}} - plotted below. While the inferred values for the dark halo are identical, the \textit{PPAK} only models of NGC\,1277 yield a black hole mass of log(\Mbh/$M_{\scriptscriptstyle \odot}$) = $10.0^{+0.2}_{-0.4}$. This is largely in agreement with the values derived in Section \hyperref[sec:ngc1277_schwarzschild]{\ref{sec:ngc1277_schwarzschild}} and vdB12, although with a decreased lower limit for the black hole mass by a factor of two, and now consistent with the black mass in E13, which was previously ruled out. Here again, the main contribution to the $\Delta \chi^2$ between our best-fitting model and models that are ruled out by the statistical 3$\sigma$ uncertainties comes from the velocity dispersion and the fourth Gauss-Hermite moment. In contrast to our fiducial models, however, which fitted both data sets simultaneously, the main driver in the fits to $\sigma$ and $h_4$ cannot be traced back to the seeing limited innermost data points but is more uniformly distributed, as highlighted in the second and fourth row of Fig. \hyperref[fig:n1277_maps_models]{\ref{fig:n1277_maps_models}}.

\begin{figure} 
	\begin{minipage}[b]{0.5\linewidth}
		\centering
		\includegraphics[width=.95\linewidth]{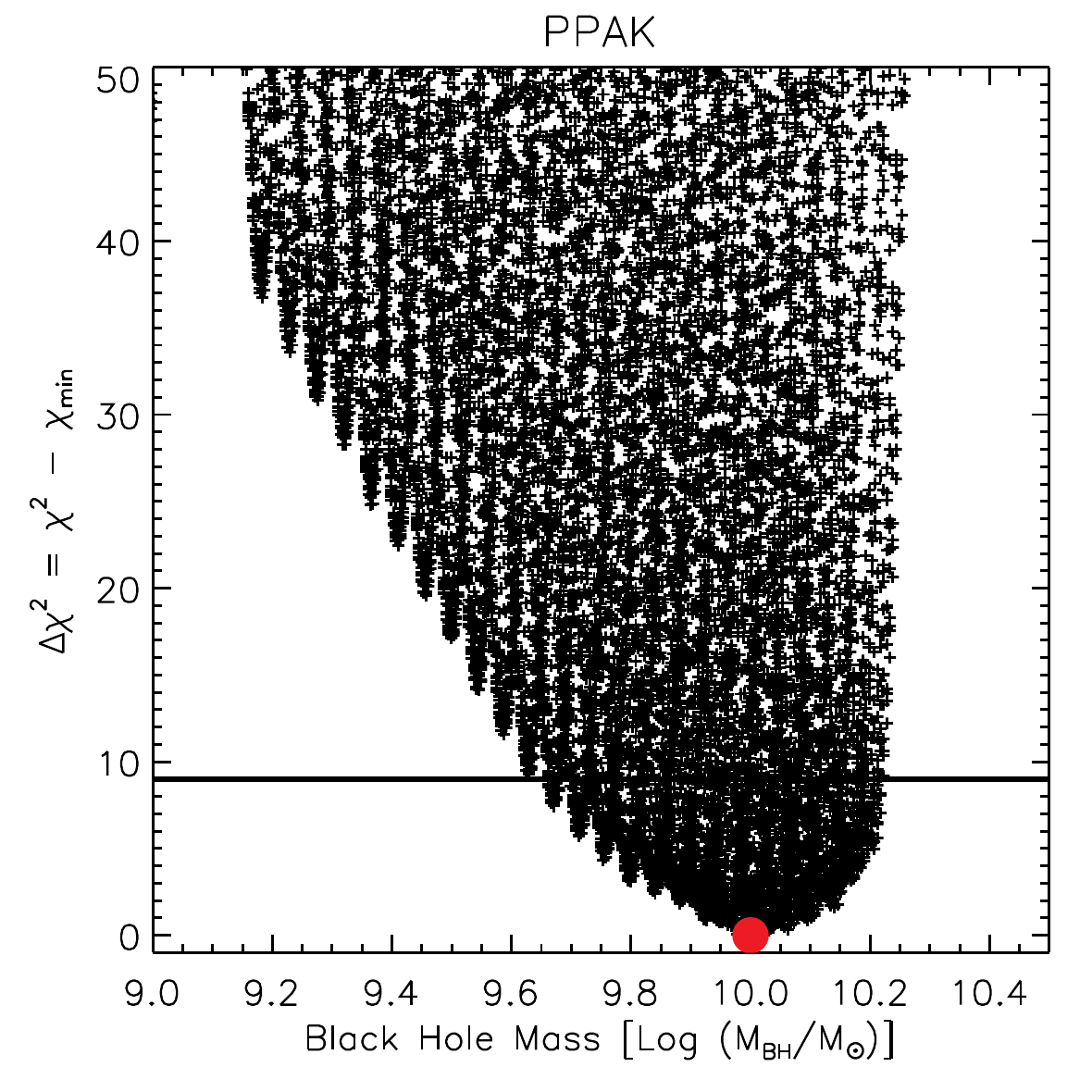} 
		\hspace{4ex}
	\end{minipage}
	\begin{minipage}[b]{0.5\linewidth}
		\centering
		\includegraphics[width=.95\linewidth]{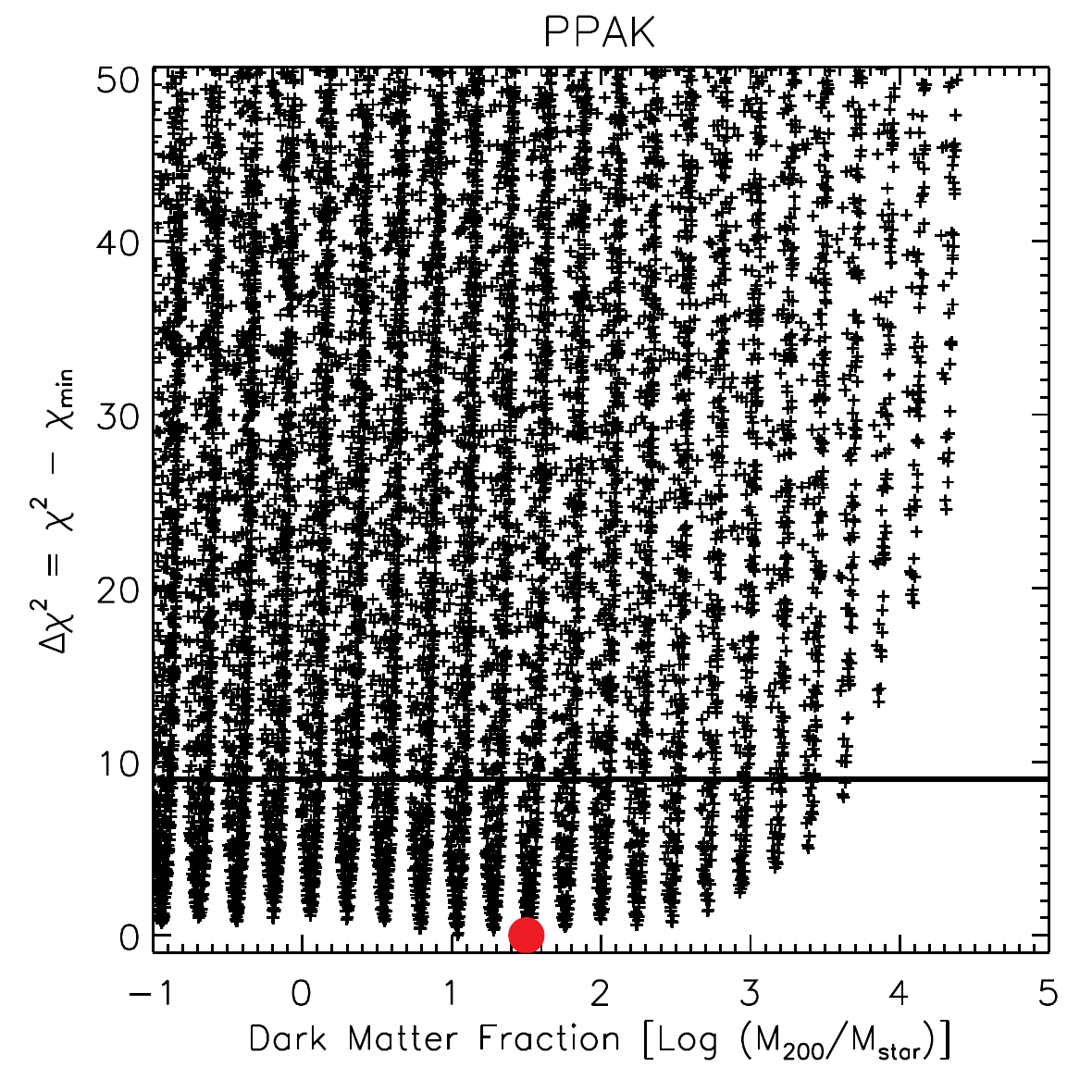} 
		\hspace{4ex}
	\end{minipage} 
	\begin{minipage}[b]{0.5\linewidth}
		\centering
		\includegraphics[width=.95\linewidth]{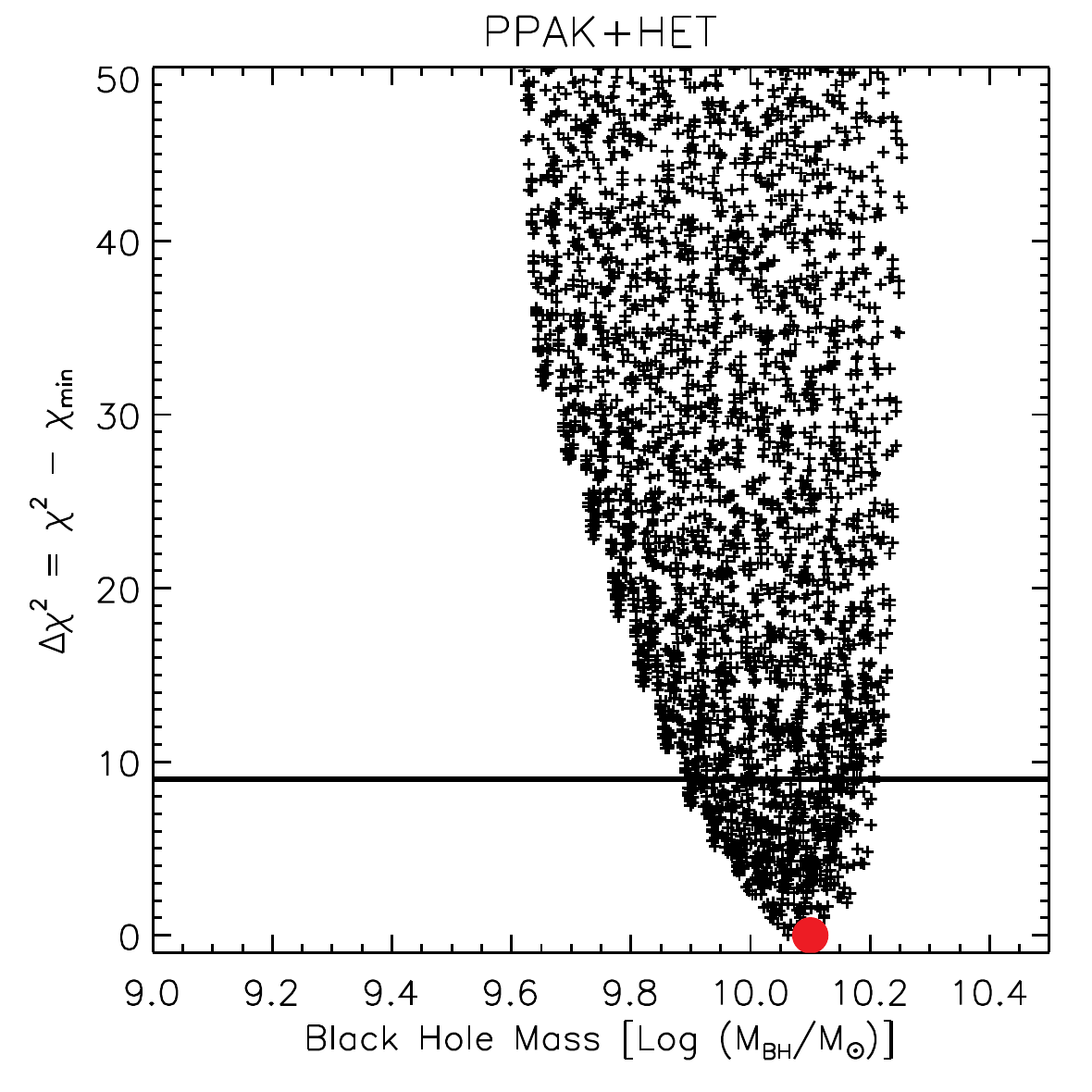} 
	\end{minipage}
	\begin{minipage}[b]{0.5\linewidth}
		\centering
		\includegraphics[width=.95\linewidth]{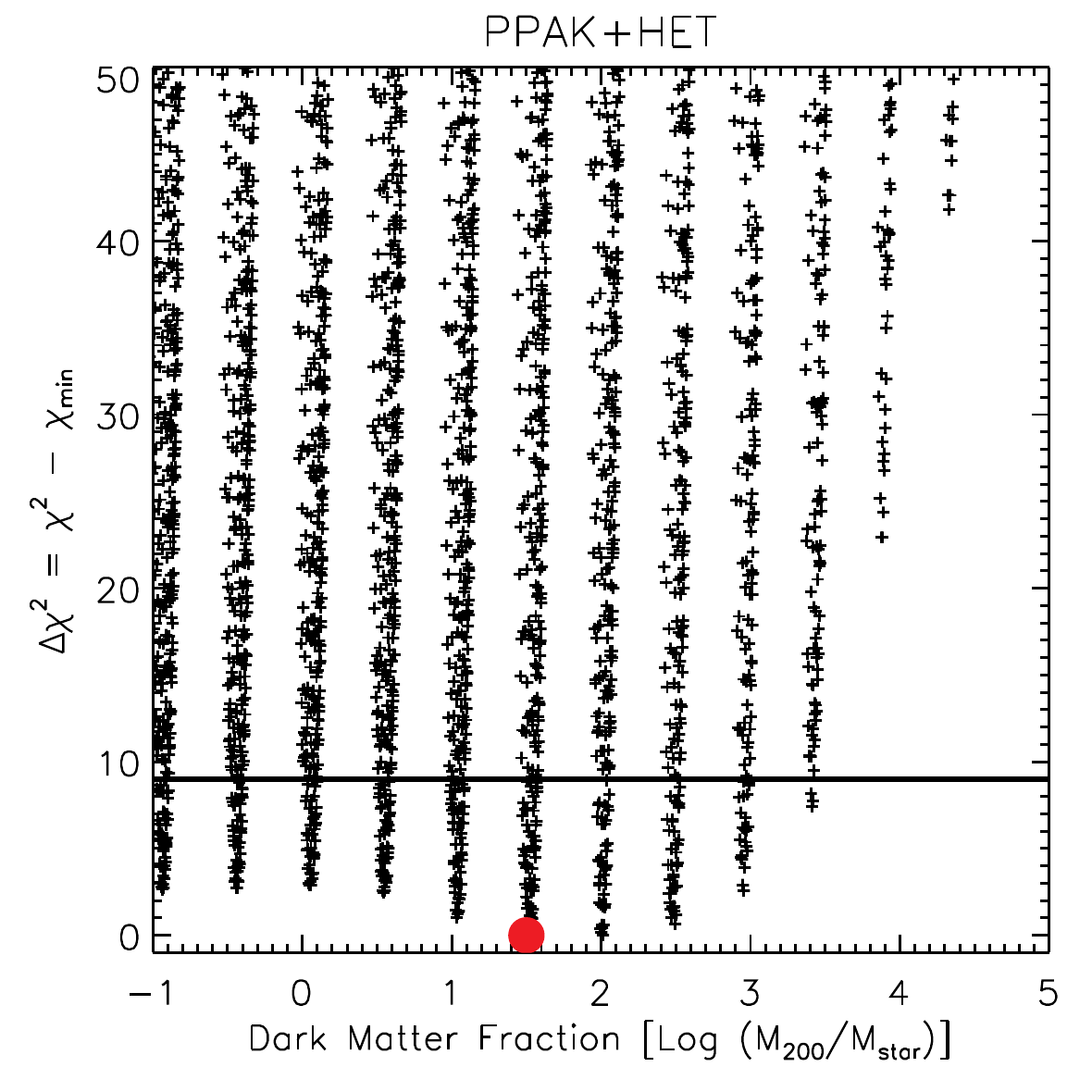} 
	\end{minipage}
\caption{Comparison of the inferred values for black hole mass and dark matter halo based on NGC\,1277's orbit-based models of the \textit{PPAK} data only (top) and the combined \textit{PPAK}+\textit{HET} data set (bottom). The red dot marks the best-fitting value. The horizontal line denotes a $\Delta \chi^2$ difference of 9, which corresponds to statistical 3$\sigma$ uncertainties for one degree of freedom.}
\label{fig:n1277_consistency}
\end{figure}

We thus ascribe the decreased lower limit to the larger spatial resolution of the \textit{PPAK} kinematics which is not able to resolve the sphere of influence of the massive black hole. For our conclusions we choose to give no preference to either one of the measurements and stick to the fiducial models in Section \hyperref[sec:ngc1277_schwarzschild]{\ref{sec:ngc1277_schwarzschild}}. Due to the lingering issues between both data sets, though, the very careful reader can adopt a lower limit of log(\Mbh/$M_{\scriptscriptstyle \odot}$) = 9.6 (but see also the next two points).\\

In the case of MRK\,1216, the difference between the \textit{PPAK} and \textit{HET} kinematics is marginal (Sec. \hyperref[sec:mrk1216_kinematics]{\ref{sec:mrk1216_kinematics}}). Although the best-fitting model in Fig. \hyperref[fig:m1216_maps_models]{\ref{fig:m1216_maps_models}} seems to be slightly off the measured \textit{HET} dispersion, it is still well within the measurement errors. Hence, fits to the \textit{PPAK} data alone do not show any difference in the derived values for black hole mass, stellar mass-to-light ratio and dark halo mass. We therefore do not present a comparison similar to Fig. \hyperref[fig:n1277_consistency]{\ref{fig:n1277_consistency}} but rather refer the reader to the detailed analysis and modelling results in Section \hyperref[sec:mrk1216_schwarzschild]{\ref{sec:mrk1216_schwarzschild}} and \hyperref[sec:dark_halo_detection]{\ref{sec:dark_halo_detection}}.\\

\item
In \cite{2013MNRAS.431.3570M}, Monte Carlo simulations of mock galaxy kinematics are utilised to estimate appropriate confidence intervals. Based on their made-to-measure particle code \textsc{nmagic}, they advocate the use of larger $\Delta\chi^2$ values to be able to recover their model galaxy parameters. While these findings are certainly interesting, there are significant differences in the modelling approach as well as in the kinematic data sets. The investigations of \cite{2013MNRAS.431.3570M} are based on a single case study and more extensive tests are necessary to verify the reliability of their adopted confidence levels. The uncertainties in our parameter estimation are based on the commonly used $\Delta\chi^2$ values. After marginalising over the orbital weights as well as over e.g. \MLstar, $c$ and $f$, we obtain formal 3$\sigma$ errors of the black hole mass with a $\Delta\chi^2$ of 9. Alternatively, we can make use of the expected standard deviation of $\chi^2$ for our parameter estimation, as promoted by e.g. \cite{2009MNRAS.398.1117V} in the case of IFU kinematics. The standard deviation in $\chi^2$ is $\sqrt(2\times(N-M))$, where $N$ is the number of kinematic constraints (i.e. the four kinematic moments $v,\, \sigma,\, h_3$ and $h_4$ in each bin) and $M$ the number of free parameters in our models, namely the dark halo parameters $c$ and $f$ as well as the black hole mass \Mbh\ and stellar mass-to-light ratio \MLstar. As a result, we obtain a lower limit for the black hole mass in NGC\,1277 of log(\Mbh/$M_{\scriptscriptstyle \odot}$) = 9.6, which again is well in line with the estimate of the black hole mass based on models of the \textit{PPAK} only data and the black hole mass that was put forward by E13.
Applying the same argument to MRK\,1216, however, would imply consistency with models which do not contain a dark halo.\\

\item
Finally, we point out that all uncertainties presented here are solely statistical errors. A large uncertainty factor in any measurement of the black hole mass remains the estimate of systematic errors. These are hard to quantify and can arise not only through the use of modelling assumptions such as a constant mass-to-light ratio, axial symmetry and the adoption of a spherical NFW halo, but also through technical limitations as for instance a stellar template mismatch and the influence of the wavelength range that is used to infer the LOSVD. It is beyond the scope of this paper to derive an assessment of each of these factors but we acknowledge that their total contribution most likely overshoots our statistical errors. Still, to provide a conservative estimate of the black hole mass - by taking into account the effects of systematic uncertainties - we simply follow the practice of \cite{2013ARA&A..51..511K} and adopt a $1\sigma$ error of $\Delta$ log \Mbh = 0.117 (the mean of all literature 1$\sigma$ black hole mass uncertainties). For NGC\,1277, this yields a black hole mass of log($\Mbh/M_{\scriptscriptstyle \odot}) = 10.1_{-0.4}^{+0.4}$ within a $3\sigma$ confidence interval, or log($\Mbh/M_{\scriptscriptstyle \odot}) = 10.0_{-0.4}^{+0.4}$ if the fits to the \textit{PPAK} data only are taken into account.

\end{itemize}

\section{Summary}
\label{sec:summary}

We have performed a detailed analysis of a suite of kinematic and photometric information of the two compact, nearby, high velocity dispersion galaxies MRK\,1216 and NGC\,1277. Our analysis combined three different but complementary data sets; high spatial resolution imaging with the \textit{HST}, low-resolution, long-slit spectroscopic observations with the \textit{HET} and medium-resolution spectroscopic observations with the \textit{PPAK} IFU. 

We first analysed the reduced and combined \textit{HST} images with multiple \sersic\ components to infer the structure and morphology of each galaxy. Both galaxies show a very compact, early-type structure without any noticeable substructures. By means of a multi-component decomposition, we obtained estimates for a bulge luminosity. A decent fit was obtained with at least four components in both cases. We further parameterised the observed light distribution with a set of multiple Gaussians, which in turn was used to build axisymmetric dynamical models.

Kinematic information was extracted by fitting the binned spectra with a set of stellar libraries. The observations revealed a distinct central peak in the velocity dispersion - hinting at a very high mass concentration in the nucleus of both galaxies - and fast and regular rotation around the short axis that is consistent with axial symmetry.

Our dynamical models rely on a triaxial implementation of Schwarzschild's orbit superposition method. Probing a wide range of parameters, we infer upper and lower limits for the individual gravitational contributions of black hole mass, stellar mass and dark matter halo. For NGC\,1277 we obtained good constraints on the black hole mass of log(\Mbh/$M_{\scriptscriptstyle \odot}$) = 10.1$_{-0.2}^{+0.10}$ for the best fitting model, consistent with former measurements of vdB12. Even for high stellar mass-to-light ratios, the lower limit on the black hole is considerably higher than predictions of the $\Mbh - L_{Bulge}$ relation in the mid-infrared. In the case of MRK\,1216, we only obtain an upper limit of log(\Mbh/$M_{\scriptscriptstyle \odot}$) = 10.0. High-resolution spectroscopic observations are thus needed to resolve the sphere of influence and to place firm constraints on its black hole mass.

Despite kinematic information out to $\ge$ 5\,kpc, we were not able to constrain the dark halo parameters significantly. The models predict a dark matter contribution of up to 52 per cent in MRK\,1216 and 13 per cent in NGC\,1277 within one effective radius. Models without a dark halo are formally excluded at the 3$\sigma$ level in MRK\,1216, only. The difference between the best-fitting model without a halo and the overall best-fitting model is mainly driven by data points within 5\arcsec. We show that this difference is due to an increase in the constant mass-to-light ratio in the dark-halo-free models to account for the outer kinematics, which then propagates towards the centre and leads to the observed mismatch. A radially increasing mass-to-light ratio could indeed recover the data without the need of a dark halo. But, if anything, recent investigations of massive early-type galaxies suggest a radially decreasing stellar mass-to-light ratio.

The stellar mass-to-light ratios span a range of 5.0 - 8.0 in \textit{V}-band in NGC\,1277 and 1.0 - 2.3 in \textit{H}-band in MRK\,1216. The best-fitting models are in good agreement with predictions of SSP models with a single power-law Salpeter IMF. Higher mass-to-light ratios - as have been observed in high dispersion galaxies - cannot be excluded. Nevertheless we place upper limits on possible deviations from the derived values.

The orbital structure is rotationally supported in both galaxies, which is consistent with the multi-component \sersic\ decompositions of the deep \textit{HST} images. This is highly indicative that MRK\,1216 and NGC\,1277 do not possess any pressure supported classical bulges that have formed through violent relaxation in late (i.e. $z \le 2$), equal mass mergers. Recent, successive minor and dissipative major merging events are unlikely too, as these tend to increase the galaxy size drastically and should yield more recent star formation activities, which are in contrast to their uniformly old stellar age estimates. Taking into account their compact, featureless and regular structures as well as their high dispersions and rapid rotation, these compact objects might well be unaltered descendants of the quiescent galaxy population at $z = 2$, which in turn are thought to be remnants of highly dissipative submillimeter mergers at even higher redshifts.

\section*{Acknowledgements}
The authors thank the anonymous referee for a very thorough review and helpful list of comments which improved the paper appreciably. A.\,Y. thanks Ronald L\"asker and Arjen van der Wel for lively and fruitful discussions and acknowledges support from the International Max Planck Research School for Astronomy \& Cosmic Physics at the University of Heidelberg and the Heidelberg Graduate School of Fundamental Physics. J.\,L.\,W. has been supported by an NSF Astronomy and Astrophysics Postdoctoral Fellowship under Award No. 1102845. The data presented here is based on observations collected at the Centro Astronómico Hispano Alemán (CAHA) at Calar Alto, operated jointly by the Max Planck Institute for Astronomy and the Instituto de Astrofísica de Andalucía (CSIC).  The Hobby-Eberly Telescope (\textit{HET}) is a joint project of the University of Texas at Austin, the Pennsylvania State University, Ludwig-Maximilians-Universit\"{a}t M\"{u}nchen, and Georg-August-Universität G\"{o}ttingen. The \textit{HET} is named in honor of its principal benefactors, William P. Hobby and Robert E. Eberly. The Marcario Low Resolution Spectrograph (\textit{LRS}) is named after Mike Marcario of High Lonesome Optics who fabricated several optics for the instrument but died before its completion. The \textit{LRS} is a joint project of the Hobby-Eberly Telescope partnership and the Instituto de Astronomía de la Universidad Nacional Autónoma de México. This research also based on observations made with the NASA/ESA \textit{Hubble Space Telescope}, obtained at the Space Telescope Science Institute, which is operated by the Association of Universities for Research in Astronomy, Inc., under NASA contract NAS 5-26555. The observations are associated with program \#13050. This research has made use of the NASA/IPAC Extragalactic Database (NED) which is operated by the Jet Propulsion Laboratory, California Institute of Technology, under contract with the National Aeronautics and Space Administration.

%


\bibliographystyle{yahapj}
\bibliography{mn2e}

\begin{thebibliography}{158}
\providecommand\natexlab[1]{#1}
\providecommand\JournalTitle[1]{#1}

\bibitem[{{Auger} {et~al.}(2010){Auger}, {Treu}, {Gavazzi}, {Bolton},
  {Koopmans}, \& {Marshall}}]{2010ApJ...721L.163A}
{Auger}, M.~W., {Treu}, T., {Gavazzi}, R., {et~al.} 2010,
  \href{http://dx.doi.org/10.1088/2041-8205/721/2/L163}{\JournalTitle{\apjl},
  721, L163}

\bibitem[{{Baes} \& {Dejonghe}(2001)}]{2001ApJ...563L..19B}
{Baes}, M., \& {Dejonghe}, H. 2001,
  \href{http://dx.doi.org/10.1086/338502}{\JournalTitle{\apjl}, 563, L19}

\bibitem[{{Barro} {et~al.}(2014){Barro}, {Faber}, {P{\'e}rez-Gonz{\'a}lez},
  {Pacifici}, {Trump}, {Koo}, {Wuyts}, {Guo}, {Bell}, {Dekel}, {Porter},
  {Primack}, {Ferguson}, {Ashby}, {Caputi}, {Ceverino}, {Croton}, {Fazio},
  {Giavalisco}, {Hsu}, {Kocevski}, {Koekemoer}, {Kurczynski}, {Kollipara},
  {Lee}, {McIntosh}, {McGrath}, {Moody}, {Somerville}, {Papovich}, {Salvato},
  {Santini}, {Tal}, {van der Wel}, {Williams}, {Willner}, \&
  {Zolotov}}]{2014ApJ...791...52B}
{Barro}, G., {Faber}, S.~M., {P{\'e}rez-Gonz{\'a}lez}, P.~G., {et~al.} 2014,
  \href{http://dx.doi.org/10.1088/0004-637X/791/1/52}{\JournalTitle{\apj}, 791,
  52}

\bibitem[{{Behroozi} {et~al.}(2010){Behroozi}, {Conroy}, \&
  {Wechsler}}]{2010ApJ...717..379B}
{Behroozi}, P.~S., {Conroy}, C., \& {Wechsler}, R.~H. 2010,
  \href{http://dx.doi.org/10.1088/0004-637X/717/1/379}{\JournalTitle{\apj},
  717, 379}

\bibitem[{{Bell} \& {de Jong}(2001)}]{2001ApJ...550..212B}
{Bell}, E.~F., \& {de Jong}, R.~S. 2001,
  \href{http://dx.doi.org/10.1086/319728}{\JournalTitle{\apj}, 550, 212}

\bibitem[{{Bender} {et~al.}(1992){Bender}, {Burstein}, \&
  {Faber}}]{1992ApJ...399..462B}
{Bender}, R., {Burstein}, D., \& {Faber}, S.~M. 1992,
  \href{http://dx.doi.org/10.1086/171940}{\JournalTitle{\apj}, 399, 462}

\bibitem[{{Binney} \& {Merrifield}(1998)}]{1998gaas.book.....B}
{Binney}, J., \& {Merrifield}, M. 1998, {Galactic Astronomy}

\bibitem[{{Binney} \& {Tremaine}(1987)}]{1987gady.book.....B}
{Binney}, J., \& {Tremaine}, S. 1987, {Galactic Dynamics}

\bibitem[{{Bruzual} \& {Charlot}(2003)}]{2003MNRAS.344.1000B}
{Bruzual}, G., \& {Charlot}, S. 2003,
  \href{http://dx.doi.org/10.1046/j.1365-8711.2003.06897.x}{\JournalTitle{\mnras},
  344, 1000}

\bibitem[{{Bullock} {et~al.}(2001){Bullock}, {Kolatt}, {Sigad}, {Somerville},
  {Kravtsov}, {Klypin}, {Primack}, \& {Dekel}}]{2001MNRAS.321..559B}
{Bullock}, J.~S., {Kolatt}, T.~S., {Sigad}, Y., {et~al.} 2001,
  \href{http://dx.doi.org/10.1046/j.1365-8711.2001.04068.x}{\JournalTitle{\mnras},
  321, 559}

\bibitem[{{Bureau} \& {Athanassoula}(2005)}]{2005ApJ...626..159B}
{Bureau}, M., \& {Athanassoula}, E. 2005,
  \href{http://dx.doi.org/10.1086/430056}{\JournalTitle{\apj}, 626, 159}

\bibitem[{{Burkert} \& {Naab}(2003)}]{2003LNP...626..327B}
{Burkert}, A., \& {Naab}, T. 2003,
  \href{http://dx.doi.org/10.1007/978-3-540-45040-5_27}{in Lecture Notes in
  Physics, Berlin Springer Verlag, Vol. 626, Galaxies and Chaos, ed.
  G.~{Contopoulos} \& N.~{Voglis}}, 327

\bibitem[{{Burkert} {et~al.}(2008){Burkert}, {Naab}, {Johansson}, \&
  {Jesseit}}]{2008ApJ...685..897B}
{Burkert}, A., {Naab}, T., {Johansson}, P.~H., \& {Jesseit}, R. 2008,
  \href{http://dx.doi.org/10.1086/591632}{\JournalTitle{\apj}, 685, 897}

\bibitem[{{Cappellari}(2002)}]{2002MNRAS.333..400C}
{Cappellari}, M. 2002,
  \href{http://dx.doi.org/10.1046/j.1365-8711.2002.05412.x}{\JournalTitle{\mnras},
  333, 400}

\bibitem[{{Cappellari} \& {Copin}(2003)}]{2003MNRAS.342..345C}
{Cappellari}, M., \& {Copin}, Y. 2003,
  \href{http://dx.doi.org/10.1046/j.1365-8711.2003.06541.x}{\JournalTitle{\mnras},
  342, 345}

\bibitem[{{Cappellari} \& {Emsellem}(2004)}]{2004PASP..116..138C}
{Cappellari}, M., \& {Emsellem}, E. 2004,
  \href{http://dx.doi.org/10.1086/381875}{\JournalTitle{\pasp}, 116, 138}

\bibitem[{{Cappellari} {et~al.}(2006){Cappellari}, {Bacon}, {Bureau}, {Damen},
  {Davies}, {de Zeeuw}, {Emsellem}, {Falc{\'o}n-Barroso}, {Krajnovi{\'c}},
  {Kuntschner}, {McDermid}, {Peletier}, {Sarzi}, {van den Bosch}, \& {van de
  Ven}}]{2006MNRAS.366.1126C}
{Cappellari}, M., {Bacon}, R., {Bureau}, M., {et~al.} 2006,
  \href{http://dx.doi.org/10.1111/j.1365-2966.2005.09981.x}{\JournalTitle{\mnras},
  366, 1126}

\bibitem[{{Cappellari} {et~al.}(2007){Cappellari}, {Emsellem}, {Bacon},
  {Bureau}, {Davies}, {de Zeeuw}, {Falc{\'o}n-Barroso}, {Krajnovi{\'c}},
  {Kuntschner}, {McDermid}, {Peletier}, {Sarzi}, {van den Bosch}, \& {van de
  Ven}}]{2007MNRAS.379..418C}
{Cappellari}, M., {Emsellem}, E., {Bacon}, R., {et~al.} 2007,
  \href{http://dx.doi.org/10.1111/j.1365-2966.2007.11963.x}{\JournalTitle{\mnras},
  379, 418}

\bibitem[{{Cappellari} {et~al.}(2012){Cappellari}, {McDermid}, {Alatalo},
  {Blitz}, {Bois}, {Bournaud}, {Bureau}, {Crocker}, {Davies}, {Davis}, {de
  Zeeuw}, {Duc}, {Emsellem}, {Khochfar}, {Krajnovi{\'c}}, {Kuntschner},
  {Lablanche}, {Morganti}, {Naab}, {Oosterloo}, {Sarzi}, {Scott}, {Serra},
  {Weijmans}, \& {Young}}]{2012Natur.484..485C}
{Cappellari}, M., {McDermid}, R.~M., {Alatalo}, K., {et~al.} 2012,
  \href{http://dx.doi.org/10.1038/nature10972}{\JournalTitle{\nat}, 484, 485}

\bibitem[{{Cappellari} {et~al.}(2013){Cappellari}, {Scott}, {Alatalo}, {Blitz},
  {Bois}, {Bournaud}, {Bureau}, {Crocker}, {Davies}, {Davis}, {de Zeeuw},
  {Duc}, {Emsellem}, {Khochfar}, {Krajnovi{\'c}}, {Kuntschner}, {McDermid},
  {Morganti}, {Naab}, {Oosterloo}, {Sarzi}, {Serra}, {Weijmans}, \&
  {Young}}]{2013MNRAS.432.1709C}
{Cappellari}, M., {Scott}, N., {Alatalo}, K., {et~al.} 2013,
  \href{http://dx.doi.org/10.1093/mnras/stt562}{\JournalTitle{\mnras}, 432,
  1709}

\bibitem[{{Chanam{\'e}} {et~al.}(2008){Chanam{\'e}}, {Kleyna}, \& {van der
  Marel}}]{2008ApJ...682..841C}
{Chanam{\'e}}, J., {Kleyna}, J., \& {van der Marel}, R. 2008,
  \href{http://dx.doi.org/10.1086/589429}{\JournalTitle{\apj}, 682, 841}

\bibitem[{{Chung} \& {Bureau}(2004)}]{2004AJ....127.3192C}
{Chung}, A., \& {Bureau}, M. 2004,
  \href{http://dx.doi.org/10.1086/420988}{\JournalTitle{\aj}, 127, 3192}

\bibitem[{{Cid Fernandes} {et~al.}(2005){Cid Fernandes}, {Mateus}, {Sodr{\'e}},
  {Stasi{\'n}ska}, \& {Gomes}}]{2005MNRAS.358..363C}
{Cid Fernandes}, R., {Mateus}, A., {Sodr{\'e}}, L., {Stasi{\'n}ska}, G., \&
  {Gomes}, J.~M. 2005,
  \href{http://dx.doi.org/10.1111/j.1365-2966.2005.08752.x}{\JournalTitle{\mnras},
  358, 363}

\bibitem[{{Cole} {et~al.}(2001){Cole}, {Norberg}, {Baugh}, {Frenk},
  {Bland-Hawthorn}, {Bridges}, {Cannon}, {Colless}, {Collins}, {Couch},
  {Cross}, {Dalton}, {De Propris}, {Driver}, {Efstathiou}, {Ellis},
  {Glazebrook}, {Jackson}, {Lahav}, {Lewis}, {Lumsden}, {Maddox}, {Madgwick},
  {Peacock}, {Peterson}, {Sutherland}, \& {Taylor}}]{2001MNRAS.326..255C}
{Cole}, S., {Norberg}, P., {Baugh}, C.~M., {et~al.} 2001,
  \href{http://dx.doi.org/10.1046/j.1365-8711.2001.04591.x}{\JournalTitle{\mnras},
  326, 255}

\bibitem[{{Conroy} \& {van Dokkum}(2012)}]{2012ApJ...760...71C}
{Conroy}, C., \& {van Dokkum}, P.~G. 2012,
  \href{http://dx.doi.org/10.1088/0004-637X/760/1/71}{\JournalTitle{\apj}, 760,
  71}

\bibitem[{{Cox} {et~al.}(2006){Cox}, {Dutta}, {Di Matteo}, {Hernquist},
  {Hopkins}, {Robertson}, \& {Springel}}]{2006ApJ...650..791C}
{Cox}, T.~J., {Dutta}, S.~N., {Di Matteo}, T., {et~al.} 2006,
  \href{http://dx.doi.org/10.1086/507474}{\JournalTitle{\apj}, 650, 791}

\bibitem[{{Cretton} {et~al.}(1999){Cretton}, {de Zeeuw}, {van der Marel}, \&
  {Rix}}]{1999ApJS..124..383C}
{Cretton}, N., {de Zeeuw}, P.~T., {van der Marel}, R.~P., \& {Rix}, H.-W. 1999,
  \href{http://dx.doi.org/10.1086/313264}{\JournalTitle{\apjs}, 124, 383}

\bibitem[{{Daddi} {et~al.}(2005){Daddi}, {Renzini}, {Pirzkal}, {Cimatti},
  {Malhotra}, {Stiavelli}, {Xu}, {Pasquali}, {Rhoads}, {Brusa}, {di Serego
  Alighieri}, {Ferguson}, {Koekemoer}, {Moustakas}, {Panagia}, \&
  {Windhorst}}]{2005ApJ...626..680D}
{Daddi}, E., {Renzini}, A., {Pirzkal}, N., {et~al.} 2005,
  \href{http://dx.doi.org/10.1086/430104}{\JournalTitle{\apj}, 626, 680}

\bibitem[{{de Zeeuw}(1985)}]{1985MNRAS.216..273D}
{de Zeeuw}, T. 1985, \JournalTitle{\mnras}, 216, 273

\bibitem[{{de Zeeuw} \& {Franx}(1991)}]{1991ARA&A..29..239D}
{de Zeeuw}, T., \& {Franx}, M. 1991,
  \href{http://dx.doi.org/10.1146/annurev.aa.29.090191.001323}{\JournalTitle{\araa},
  29, 239}

\bibitem[{{Dekel} \& {Burkert}(2014)}]{2014MNRAS.438.1870D}
{Dekel}, A., \& {Burkert}, A. 2014,
  \href{http://dx.doi.org/10.1093/mnras/stt2331}{\JournalTitle{\mnras}, 438,
  1870}

\bibitem[{{Duffy} {et~al.}(2008){Duffy}, {Schaye}, {Kay}, \& {Dalla
  Vecchia}}]{2008MNRAS.390L..64D}
{Duffy}, A.~R., {Schaye}, J., {Kay}, S.~T., \& {Dalla Vecchia}, C. 2008,
  \href{http://dx.doi.org/10.1111/j.1745-3933.2008.00537.x}{\JournalTitle{\mnras},
  390, L64}

\bibitem[{{Dutton} {et~al.}(2010){Dutton}, {Conroy}, {van den Bosch}, {Prada},
  \& {More}}]{2010MNRAS.407....2D}
{Dutton}, A.~A., {Conroy}, C., {van den Bosch}, F.~C., {Prada}, F., \& {More},
  S. 2010,
  \href{http://dx.doi.org/10.1111/j.1365-2966.2010.16911.x}{\JournalTitle{\mnras},
  407, 2}

\bibitem[{{Dutton} {et~al.}(2011){Dutton}, {Conroy}, {van den Bosch}, {Simard},
  {Mendel}, {Courteau}, {Dekel}, {More}, \& {Prada}}]{2011MNRAS.416..322D}
{Dutton}, A.~A., {Conroy}, C., {van den Bosch}, F.~C., {et~al.} 2011,
  \href{http://dx.doi.org/10.1111/j.1365-2966.2011.19038.x}{\JournalTitle{\mnras},
  416, 322}

\bibitem[{{Emsellem}(2013)}]{2013MNRAS.433.1862E}
{Emsellem}, E. 2013,
  \href{http://dx.doi.org/10.1093/mnras/stt840}{\JournalTitle{\mnras}, 433,
  1862}

\bibitem[{{Emsellem} {et~al.}(1994){Emsellem}, {Monnet}, \&
  {Bacon}}]{1994A&A...285..723E}
{Emsellem}, E., {Monnet}, G., \& {Bacon}, R. 1994, \JournalTitle{\aap}, 285,
  723

\bibitem[{{Emsellem} {et~al.}(2007){Emsellem}, {Cappellari}, {Krajnovi{\'c}},
  {van de Ven}, {Bacon}, {Bureau}, {Davies}, {de Zeeuw}, {Falc{\'o}n-Barroso},
  {Kuntschner}, {McDermid}, {Peletier}, \& {Sarzi}}]{2007MNRAS.379..401E}
{Emsellem}, E., {Cappellari}, M., {Krajnovi{\'c}}, D., {et~al.} 2007,
  \href{http://dx.doi.org/10.1111/j.1365-2966.2007.11752.x}{\JournalTitle{\mnras},
  379, 401}

\bibitem[{{Emsellem} {et~al.}(2011){Emsellem}, {Cappellari}, {Krajnovi{\'c}},
  {Alatalo}, {Blitz}, {Bois}, {Bournaud}, {Bureau}, {Davies}, {Davis}, {de
  Zeeuw}, {Khochfar}, {Kuntschner}, {Lablanche}, {McDermid}, {Morganti},
  {Naab}, {Oosterloo}, {Sarzi}, {Scott}, {Serra}, {van de Ven}, {Weijmans}, \&
  {Young}}]{2011MNRAS.414..888E}
---. 2011,
  \href{http://dx.doi.org/10.1111/j.1365-2966.2011.18496.x}{\JournalTitle{\mnras},
  414, 888}

\bibitem[{{Falc{\'o}n-Barroso} {et~al.}(2011){Falc{\'o}n-Barroso},
  {S{\'a}nchez-Bl{\'a}zquez}, {Vazdekis}, {Ricciardelli}, {Cardiel}, {Cenarro},
  {Gorgas}, \& {Peletier}}]{2011A&A...532A..95F}
{Falc{\'o}n-Barroso}, J., {S{\'a}nchez-Bl{\'a}zquez}, P., {Vazdekis}, A.,
  {et~al.} 2011,
  \href{http://dx.doi.org/10.1051/0004-6361/201116842}{\JournalTitle{\aap},
  532, A95}

\bibitem[{{Ferreras} {et~al.}(2013){Ferreras}, {La Barbera}, {de la Rosa},
  {Vazdekis}, {de Carvalho}, {Falc{\'o}n-Barroso}, \&
  {Ricciardelli}}]{2013MNRAS.429L..15F}
{Ferreras}, I., {La Barbera}, F., {de la Rosa}, I.~G., {et~al.} 2013,
  \href{http://dx.doi.org/10.1093/mnrasl/sls014}{\JournalTitle{\mnras}, 429,
  L15}

\bibitem[{{Franx} {et~al.}(1989){Franx}, {Illingworth}, \&
  {Heckman}}]{1989AJ.....98..538F}
{Franx}, M., {Illingworth}, G., \& {Heckman}, T. 1989,
  \href{http://dx.doi.org/10.1086/115157}{\JournalTitle{\aj}, 98, 538}

\bibitem[{{Gebhardt} \& {Thomas}(2009)}]{2009ApJ...700.1690G}
{Gebhardt}, K., \& {Thomas}, J. 2009,
  \href{http://dx.doi.org/10.1088/0004-637X/700/2/1690}{\JournalTitle{\apj},
  700, 1690}

\bibitem[{{Gebhardt} {et~al.}(2003){Gebhardt}, {Richstone}, {Tremaine},
  {Lauer}, {Bender}, {Bower}, {Dressler}, {Faber}, {Filippenko}, {Green},
  {Grillmair}, {Ho}, {Kormendy}, {Magorrian}, \&
  {Pinkney}}]{2003ApJ...583...92G}
{Gebhardt}, K., {Richstone}, D., {Tremaine}, S., {et~al.} 2003,
  \href{http://dx.doi.org/10.1086/345081}{\JournalTitle{\apj}, 583, 92}

\bibitem[{{Gerhard} {et~al.}(2001){Gerhard}, {Kronawitter}, {Saglia}, \&
  {Bender}}]{2001AJ....121.1936G}
{Gerhard}, O., {Kronawitter}, A., {Saglia}, R.~P., \& {Bender}, R. 2001,
  \href{http://dx.doi.org/10.1086/319940}{\JournalTitle{\aj}, 121, 1936}

\bibitem[{{Gerhard}(1993)}]{1993MNRAS.265..213G}
{Gerhard}, O.~E. 1993, \JournalTitle{\mnras}, 265, 213

\bibitem[{{Gerhard} \& {Binney}(1996)}]{1996MNRAS.279..993G}
{Gerhard}, O.~E., \& {Binney}, J.~J. 1996, \JournalTitle{\mnras}, 279, 993

\bibitem[{{Gonzaga} {et~al.}(2012){Gonzaga}, {Hack}, {Fruchter}, \&
  {Mack}}]{2012drzp.book.....G}
{Gonzaga}, S., {Hack}, W., {Fruchter}, A., \& {Mack}, J. 2012, {The DrizzlePac
  Handbook}

\bibitem[{{Greene} {et~al.}(2013){Greene}, {Murphy}, {Graves}, {Gunn},
  {Raskutti}, {Comerford}, \& {Gebhardt}}]{2013ApJ...776...64G}
{Greene}, J.~E., {Murphy}, J.~D., {Graves}, G.~J., {et~al.} 2013,
  \href{http://dx.doi.org/10.1088/0004-637X/776/2/64}{\JournalTitle{\apj}, 776,
  64}

\bibitem[{{Greene} {et~al.}(2010){Greene}, {Peng}, {Kim}, {Kuo}, {Braatz},
  {Impellizzeri}, {Condon}, {Lo}, {Henkel}, \& {Reid}}]{2010ApJ...721...26G}
{Greene}, J.~E., {Peng}, C.~Y., {Kim}, M., {et~al.} 2010,
  \href{http://dx.doi.org/10.1088/0004-637X/721/1/26}{\JournalTitle{\apj}, 721,
  26}

\bibitem[{{G{\"u}ltekin} {et~al.}(2011){G{\"u}ltekin}, {Richstone}, {Gebhardt},
  {Faber}, {Lauer}, {Bender}, {Kormendy}, \& {Pinkney}}]{2011ApJ...741...38G}
{G{\"u}ltekin}, K., {Richstone}, D.~O., {Gebhardt}, K., {et~al.} 2011,
  \href{http://dx.doi.org/10.1088/0004-637X/741/1/38}{\JournalTitle{\apj}, 741,
  38}

\bibitem[{{G{\"u}ltekin} {et~al.}(2009){G{\"u}ltekin}, {Richstone}, {Gebhardt},
  {Lauer}, {Tremaine}, {Aller}, {Bender}, {Dressler}, {Faber}, {Filippenko},
  {Green}, {Ho}, {Kormendy}, {Magorrian}, {Pinkney}, \&
  {Siopis}}]{2009ApJ...698..198G}
---. 2009,
  \href{http://dx.doi.org/10.1088/0004-637X/698/1/198}{\JournalTitle{\apj},
  698, 198}

\bibitem[{{Guo} {et~al.}(2010){Guo}, {White}, {Li}, \&
  {Boylan-Kolchin}}]{2010MNRAS.404.1111G}
{Guo}, Q., {White}, S., {Li}, C., \& {Boylan-Kolchin}, M. 2010,
  \href{http://dx.doi.org/10.1111/j.1365-2966.2010.16341.x}{\JournalTitle{\mnras},
  404, 1111}

\bibitem[{{H{\"a}ring} \& {Rix}(2004)}]{2004ApJ...604L..89H}
{H{\"a}ring}, N., \& {Rix}, H.-W. 2004,
  \href{http://dx.doi.org/10.1086/383567}{\JournalTitle{\apjl}, 604, L89}

\bibitem[{{Hill} {et~al.}(1998){Hill}, {Nicklas}, {MacQueen}, {Mitsch},
  {Wellem}, {Altmann}, {Wesley}, \& {Ray}}]{1998SPIE.3355..433H}
{Hill}, G.~J., {Nicklas}, H.~E., {MacQueen}, P.~J., {et~al.} 1998, in Society
  of Photo-Optical Instrumentation Engineers (SPIE) Conference Series, Vol.
  3355, Society of Photo-Optical Instrumentation Engineers (SPIE) Conference
  Series, ed. S.~{D'Odorico}, 433

\bibitem[{{Hilz} {et~al.}(2013){Hilz}, {Naab}, \&
  {Ostriker}}]{2013MNRAS.429.2924H}
{Hilz}, M., {Naab}, T., \& {Ostriker}, J.~P. 2013,
  \href{http://dx.doi.org/10.1093/mnras/sts501}{\JournalTitle{\mnras}, 429,
  2924}

\bibitem[{{Hinshaw} {et~al.}(2009){Hinshaw}, {Weiland}, {Hill}, {Odegard},
  {Larson}, {Bennett}, {Dunkley}, {Gold}, {Greason}, {Jarosik}, {Komatsu},
  {Nolta}, {Page}, {Spergel}, {Wollack}, {Halpern}, {Kogut}, {Limon}, {Meyer},
  {Tucker}, \& {Wright}}]{2009ApJS..180..225H}
{Hinshaw}, G., {Weiland}, J.~L., {Hill}, R.~S., {et~al.} 2009,
  \href{http://dx.doi.org/10.1088/0067-0049/180/2/225}{\JournalTitle{\apjs},
  180, 225}

\bibitem[{{Husemann} {et~al.}(2012){Husemann}, {Kamann}, {Sandin},
  {S{\'a}nchez}, {Garc{\'{\i}}a-Benito}, \& {Mast}}]{2012A&A...545A.137H}
{Husemann}, B., {Kamann}, S., {Sandin}, C., {et~al.} 2012,
  \href{http://dx.doi.org/10.1051/0004-6361/201220102}{\JournalTitle{\aap},
  545, A137}

\bibitem[{{Husemann} {et~al.}(2013){Husemann}, {Jahnke}, {S{\'a}nchez},
  {Barrado}, {Bekerait\`{e}}, {Bomans}, {Castillo-Morales},
  {Catal{\'a}n-Torrecilla}, {Cid Fernandes}, {Falc{\'o}n-Barroso},
  {Garc{\'{\i}}a-Benito}, {Gonz{\'a}lez Delgado}, {Iglesias-P{\'a}ramo},
  {Johnson}, {Kupko}, {L{\'o}pez-Fernandez}, {Lyubenova}, {Marino}, {Mast},
  {Miskolczi}, {Monreal-Ibero}, {Gil de Paz}, {P{\'e}rez}, {P{\'e}rez},
  {Rosales-Ortega}, {Ruiz-Lara}, {Schilling}, {van de Ven}, {Walcher}, {Alves},
  {de Amorim}, {Backsmann}, {Barrera-Ballesteros}, {Bland-Hawthorn}, {Cortijo},
  {Dettmar}, {Demleitner}, {D{\'{\i}}az}, {Enke}, {Florido}, {Flores},
  {Galbany}, {Gallazzi}, {Garc{\'{\i}}a-Lorenzo}, {Gomes}, {Gruel}, {Haines},
  {Holmes}, {Jungwiert}, {Kalinova}, {Kehrig}, {Kennicutt}, {Klar}, {Lehnert},
  {L{\'o}pez-S{\'a}nchez}, {de Lorenzo-C{\'a}ceres}, {M{\'a}rmol-Queralt{\'o}},
  {M{\'a}rquez}, {Mendez-Abreu}, {Moll{\'a}}, {del Olmo}, {Meidt}, {Papaderos},
  {Puschnig}, {Quirrenbach}, {Roth}, {S{\'a}nchez-Bl{\'a}zquez}, {Spekkens},
  {Singh}, {Stanishev}, {Trager}, {Vilchez}, {Wild}, {Wisotzki}, {Zibetti}, \&
  {Ziegler}}]{2013A&A...549A..87H}
{Husemann}, B., {Jahnke}, K., {S{\'a}nchez}, S.~F., {et~al.} 2013,
  \href{http://dx.doi.org/10.1051/0004-6361/201220582}{\JournalTitle{\aap},
  549, A87}

\bibitem[{{Jones} {et~al.}(2009){Jones}, {Read}, {Saunders}, {Colless},
  {Jarrett}, {Parker}, {Fairall}, {Mauch}, {Sadler}, {Watson}, {Burton},
  {Campbell}, {Cass}, {Croom}, {Dawe}, {Fiegert}, {Frankcombe}, {Hartley},
  {Huchra}, {James}, {Kirby}, {Lahav}, {Lucey}, {Mamon}, {Moore}, {Peterson},
  {Prior}, {Proust}, {Russell}, {Safouris}, {Wakamatsu}, {Westra}, \&
  {Williams}}]{2009MNRAS.399..683J}
{Jones}, D.~H., {Read}, M.~A., {Saunders}, W., {et~al.} 2009,
  \href{http://dx.doi.org/10.1111/j.1365-2966.2009.15338.x}{\JournalTitle{\mnras},
  399, 683}

\bibitem[{{Kelz} {et~al.}(2006){Kelz}, {Verheijen}, {Roth}, {Bauer}, {Becker},
  {Paschke}, {Popow}, {S{\'a}nchez}, \& {Laux}}]{2006PASP..118..129K}
{Kelz}, A., {Verheijen}, M.~A.~W., {Roth}, M.~M., {et~al.} 2006,
  \href{http://dx.doi.org/10.1086/497455}{\JournalTitle{\pasp}, 118, 129}

\bibitem[{{Kormendy} {et~al.}(2009){Kormendy}, {Fisher}, {Cornell}, \&
  {Bender}}]{2009ApJS..182..216K}
{Kormendy}, J., {Fisher}, D.~B., {Cornell}, M.~E., \& {Bender}, R. 2009,
  \href{http://dx.doi.org/10.1088/0067-0049/182/1/216}{\JournalTitle{\apjs},
  182, 216}

\bibitem[{{Kormendy} \& {Ho}(2013)}]{2013ARA&A..51..511K}
{Kormendy}, J., \& {Ho}, L.~C. 2013,
  \href{http://dx.doi.org/10.1146/annurev-astro-082708-101811}{\JournalTitle{\araa},
  51, 511}

\bibitem[{{Kormendy} {et~al.}(1997){Kormendy}, {Bender}, {Magorrian},
  {Tremaine}, {Gebhardt}, {Richstone}, {Dressler}, {Faber}, {Grillmair}, \&
  {Lauer}}]{1997ApJ...482L.139K}
{Kormendy}, J., {Bender}, R., {Magorrian}, J., {et~al.} 1997,
  \href{http://dx.doi.org/10.1086/310720}{\JournalTitle{\apjl}, 482, L139}

\bibitem[{{Krajnovi{\'c}} {et~al.}(2005){Krajnovi{\'c}}, {Cappellari},
  {Emsellem}, {McDermid}, \& {de Zeeuw}}]{2005MNRAS.357.1113K}
{Krajnovi{\'c}}, D., {Cappellari}, M., {Emsellem}, E., {McDermid}, R.~M., \&
  {de Zeeuw}, P.~T. 2005,
  \href{http://dx.doi.org/10.1111/j.1365-2966.2005.08715.x}{\JournalTitle{\mnras},
  357, 1113}

\bibitem[{{Kriek} {et~al.}(2009){Kriek}, {van Dokkum}, {Labb{\'e}}, {Franx},
  {Illingworth}, {Marchesini}, \& {Quadri}}]{2009ApJ...700..221K}
{Kriek}, M., {van Dokkum}, P.~G., {Labb{\'e}}, I., {et~al.} 2009,
  \href{http://dx.doi.org/10.1088/0004-637X/700/1/221}{\JournalTitle{\apj},
  700, 221}

\bibitem[{{Kriek} {et~al.}(2006){Kriek}, {van Dokkum}, {Franx}, {Quadri},
  {Gawiser}, {Herrera}, {Illingworth}, {Labb{\'e}}, {Lira}, {Marchesini},
  {Rix}, {Rudnick}, {Taylor}, {Toft}, {Urry}, \& {Wuyts}}]{2006ApJ...649L..71K}
{Kriek}, M., {van Dokkum}, P.~G., {Franx}, M., {et~al.} 2006,
  \href{http://dx.doi.org/10.1086/508371}{\JournalTitle{\apjl}, 649, L71}

\bibitem[{{Krist}(1995)}]{1995ASPC...77..349K}
{Krist}, J. 1995, in Astronomical Society of the Pacific Conference Series,
  Vol.~77, Astronomical Data Analysis Software and Systems IV, ed. R.~A.
  {Shaw}, H.~E. {Payne}, \& J.~J.~E. {Hayes}, 349

\bibitem[{{Kronawitter} {et~al.}(2000){Kronawitter}, {Saglia}, {Gerhard}, \&
  {Bender}}]{2000A&AS..144...53K}
{Kronawitter}, A., {Saglia}, R.~P., {Gerhard}, O., \& {Bender}, R. 2000,
  \href{http://dx.doi.org/10.1051/aas:2000199}{\JournalTitle{\aaps}, 144, 53}

\bibitem[{{Kuo} {et~al.}(2011){Kuo}, {Braatz}, {Condon}, {Impellizzeri}, {Lo},
  {Zaw}, {Schenker}, {Henkel}, {Reid}, \& {Greene}}]{2011ApJ...727...20K}
{Kuo}, C.~Y., {Braatz}, J.~A., {Condon}, J.~J., {et~al.} 2011,
  \href{http://dx.doi.org/10.1088/0004-637X/727/1/20}{\JournalTitle{\apj}, 727,
  20}

\bibitem[{{La Barbera} {et~al.}(2013){La Barbera}, {Ferreras}, {Vazdekis}, {de
  la Rosa}, {de Carvalho}, {Trevisan}, {Falc{\'o}n-Barroso}, \&
  {Ricciardelli}}]{2013MNRAS.433.3017L}
{La Barbera}, F., {Ferreras}, I., {Vazdekis}, A., {et~al.} 2013,
  \href{http://dx.doi.org/10.1093/mnras/stt943}{\JournalTitle{\mnras}, 433,
  3017}

\bibitem[{{L{\"a}sker} {et~al.}(2014){L{\"a}sker}, {Ferrarese}, {van de Ven},
  \& {Shankar}}]{2014ApJ...780...70L}
{L{\"a}sker}, R., {Ferrarese}, L., {van de Ven}, G., \& {Shankar}, F. 2014,
  \href{http://dx.doi.org/10.1088/0004-637X/780/1/70}{\JournalTitle{\apj}, 780,
  70}

\bibitem[{{L{\"a}sker} {et~al.}(2013){L{\"a}sker}, {van den Bosch}, {van de
  Ven}, {Ferreras}, {La Barbera}, {Vazdekis}, \&
  {Falc{\'o}n-Barroso}}]{2013MNRAS.434L..31L}
{L{\"a}sker}, R., {van den Bosch}, R.~C.~E., {van de Ven}, G., {et~al.} 2013,
  \href{http://dx.doi.org/10.1093/mnrasl/slt070}{\JournalTitle{\mnras}, 434,
  L31}

\bibitem[{{Lousto} {et~al.}(2010){Lousto}, {Nakano}, {Zlochower}, \&
  {Campanelli}}]{2010PhRvD..81h4023L}
{Lousto}, C.~O., {Nakano}, H., {Zlochower}, Y., \& {Campanelli}, M. 2010,
  \href{http://dx.doi.org/10.1103/PhysRevD.81.084023}{\JournalTitle{\prd}, 81,
  084023}

\bibitem[{{Lyubenova} {et~al.}(2013){Lyubenova}, {van den Bosch},
  {C{\^o}t{\'e}}, {Kuntschner}, {van de Ven}, {Ferrarese}, {Jord{\'a}n},
  {Infante}, \& {Peng}}]{2013MNRAS.431.3364L}
{Lyubenova}, M., {van den Bosch}, R.~C.~E., {C{\^o}t{\'e}}, P., {et~al.} 2013,
  \href{http://dx.doi.org/10.1093/mnras/stt414}{\JournalTitle{\mnras}, 431,
  3364}

\bibitem[{{Macci{\`o}} {et~al.}(2008){Macci{\`o}}, {Dutton}, \& {van den
  Bosch}}]{2008MNRAS.391.1940M}
{Macci{\`o}}, A.~V., {Dutton}, A.~A., \& {van den Bosch}, F.~C. 2008,
  \href{http://dx.doi.org/10.1111/j.1365-2966.2008.14029.x}{\JournalTitle{\mnras},
  391, 1940}

\bibitem[{{Marconi} \& {Hunt}(2003)}]{2003ApJ...589L..21M}
{Marconi}, A., \& {Hunt}, L.~K. 2003,
  \href{http://dx.doi.org/10.1086/375804}{\JournalTitle{\apjl}, 589, L21}

\bibitem[{{Marcum} {et~al.}(2001){Marcum}, {O'Connell}, {Fanelli}, {Cornett},
  {Waller}, {Bohlin}, {Neff}, {Roberts}, {Smith}, {Cheng}, {Collins},
  {Hennessy}, {Hill}, {Hill}, {Hintzen}, {Landsman}, {Ohl}, {Parise}, {Smith},
  {Freedman}, {Kuchinski}, {Madore}, {Angione}, {Palma}, {Talbert}, \&
  {Stecher}}]{2001ApJS..132..129M}
{Marcum}, P.~M., {O'Connell}, R.~W., {Fanelli}, M.~N., {et~al.} 2001,
  \href{http://dx.doi.org/10.1086/318953}{\JournalTitle{\apjs}, 132, 129}

\bibitem[{{Markarian}(1963)}]{1963SoByu..34....3M}
{Markarian}, B.~E. 1963, \JournalTitle{Soobshcheniya Byurakanskoj Observatorii
  Akademiya Nauk Armyanskoj SSR Erevan}, 34, 3

\bibitem[{{Mart{\'{\i}}n-Navarro}
  {et~al.}(2015{\natexlab{a}}){Mart{\'{\i}}n-Navarro}, {Barbera}, {Vazdekis},
  {Falc{\'o}n-Barroso}, \& {Ferreras}}]{2015MNRAS.447.1033M}
{Mart{\'{\i}}n-Navarro}, I., {Barbera}, F.~L., {Vazdekis}, A.,
  {Falc{\'o}n-Barroso}, J., \& {Ferreras}, I. 2015{\natexlab{a}},
  \href{http://dx.doi.org/10.1093/mnras/stu2480}{\JournalTitle{\mnras}, 447,
  1033}

\bibitem[{{Mart{\'{\i}}n-Navarro}
  {et~al.}(2015{\natexlab{b}}){Mart{\'{\i}}n-Navarro}, {La Barbera},
  {Vazdekis}, {Ferr{\'e}-Mateu}, {Trujillo}, \&
  {Beasley}}]{2015arXiv150501485M}
{Mart{\'{\i}}n-Navarro}, I., {La Barbera}, F., {Vazdekis}, A., {et~al.}
  2015{\natexlab{b}}, \JournalTitle{ArXiv e-prints: 1505.01485},
  \href{http://arxiv.org/abs/arXiv:1505.01485}{{\sffamily
  arXiv:arXiv:1505.01485}}

\bibitem[{{McConnell} {et~al.}(2013){McConnell}, {Chen}, {Ma}, {Greene},
  {Lauer}, \& {Gebhardt}}]{2013ApJ...768L..21M}
{McConnell}, N.~J., {Chen}, S.-F.~S., {Ma}, C.-P., {et~al.} 2013,
  \href{http://dx.doi.org/10.1088/2041-8205/768/1/L21}{\JournalTitle{\apjl},
  768, L21}

\bibitem[{{McConnell} \& {Ma}(2013)}]{2013ApJ...764..184M}
{McConnell}, N.~J., \& {Ma}, C.-P. 2013,
  \href{http://dx.doi.org/10.1088/0004-637X/764/2/184}{\JournalTitle{\apj},
  764, 184}

\bibitem[{{McConnell} {et~al.}(2011){McConnell}, {Ma}, {Gebhardt}, {Wright},
  {Murphy}, {Lauer}, {Graham}, \& {Richstone}}]{2011Natur.480..215M}
{McConnell}, N.~J., {Ma}, C.-P., {Gebhardt}, K., {et~al.} 2011,
  \href{http://dx.doi.org/10.1038/nature10636}{\JournalTitle{\nat}, 480, 215}

\bibitem[{{McDermid} {et~al.}(2015){McDermid}, {Alatalo}, {Blitz}, {Bournaud},
  {Bureau}, {Cappellari}, {Crocker}, {Davies}, {Davis}, {de Zeeuw}, {Duc},
  {Emsellem}, {Khochfar}, {Krajnovi{\'c}}, {Kuntschner}, {Morganti}, {Naab},
  {Oosterloo}, {Sarzi}, {Scott}, {Serra}, {Weijmans}, \&
  {Young}}]{2015MNRAS.448.3484M}
{McDermid}, R.~M., {Alatalo}, K., {Blitz}, L., {et~al.} 2015,
  \href{http://dx.doi.org/10.1093/mnras/stv105}{\JournalTitle{\mnras}, 448,
  3484}

\bibitem[{{Merritt} {et~al.}(2004){Merritt}, {Milosavljevi{\'c}}, {Favata},
  {Hughes}, \& {Holz}}]{2004ApJ...607L...9M}
{Merritt}, D., {Milosavljevi{\'c}}, M., {Favata}, M., {Hughes}, S.~A., \&
  {Holz}, D.~E. 2004,
  \href{http://dx.doi.org/10.1086/421551}{\JournalTitle{\apjl}, 607, L9}

\bibitem[{{Merritt} {et~al.}(2009){Merritt}, {Schnittman}, \&
  {Komossa}}]{2009ApJ...699.1690M}
{Merritt}, D., {Schnittman}, J.~D., \& {Komossa}, S. 2009,
  \href{http://dx.doi.org/10.1088/0004-637X/699/2/1690}{\JournalTitle{\apj},
  699, 1690}

\bibitem[{{Monnet} {et~al.}(1992){Monnet}, {Bacon}, \&
  {Emsellem}}]{1992A&A...253..366M}
{Monnet}, G., {Bacon}, R., \& {Emsellem}, E. 1992, \JournalTitle{\aap}, 253,
  366

\bibitem[{{Morganti} {et~al.}(2013){Morganti}, {Gerhard}, {Coccato},
  {Martinez-Valpuesta}, \& {Arnaboldi}}]{2013MNRAS.431.3570M}
{Morganti}, L., {Gerhard}, O., {Coccato}, L., {Martinez-Valpuesta}, I., \&
  {Arnaboldi}, M. 2013,
  \href{http://dx.doi.org/10.1093/mnras/stt442}{\JournalTitle{\mnras}, 431,
  3570}

\bibitem[{{Moster} {et~al.}(2010){Moster}, {Somerville}, {Maulbetsch}, {van den
  Bosch}, {Macci{\`o}}, {Naab}, \& {Oser}}]{2010ApJ...710..903M}
{Moster}, B.~P., {Somerville}, R.~S., {Maulbetsch}, C., {et~al.} 2010,
  \href{http://dx.doi.org/10.1088/0004-637X/710/2/903}{\JournalTitle{\apj},
  710, 903}

\bibitem[{{Naab} \& {Burkert}(2003)}]{2003ApJ...597..893N}
{Naab}, T., \& {Burkert}, A. 2003,
  \href{http://dx.doi.org/10.1086/378581}{\JournalTitle{\apj}, 597, 893}

\bibitem[{{Naab} {et~al.}(2006){Naab}, {Jesseit}, \&
  {Burkert}}]{2006MNRAS.372..839N}
{Naab}, T., {Jesseit}, R., \& {Burkert}, A. 2006,
  \href{http://dx.doi.org/10.1111/j.1365-2966.2006.10902.x}{\JournalTitle{\mnras},
  372, 839}

\bibitem[{{Naab} {et~al.}(2009){Naab}, {Johansson}, \&
  {Ostriker}}]{2009ApJ...699L.178N}
{Naab}, T., {Johansson}, P.~H., \& {Ostriker}, J.~P. 2009,
  \href{http://dx.doi.org/10.1088/0004-637X/699/2/L178}{\JournalTitle{\apjl},
  699, L178}

\bibitem[{{Naab} {et~al.}(2007){Naab}, {Johansson}, {Ostriker}, \&
  {Efstathiou}}]{2007ApJ...658..710N}
{Naab}, T., {Johansson}, P.~H., {Ostriker}, J.~P., \& {Efstathiou}, G. 2007,
  \href{http://dx.doi.org/10.1086/510841}{\JournalTitle{\apj}, 658, 710}

\bibitem[{{Naab} {et~al.}(2014){Naab}, {Oser}, {Emsellem}, {Cappellari},
  {Krajnovi{\'c}}, {McDermid}, {Alatalo}, {Bayet}, {Blitz}, {Bois}, {Bournaud},
  {Bureau}, {Crocker}, {Davies}, {Davis}, {de Zeeuw}, {Duc}, {Hirschmann},
  {Johansson}, {Khochfar}, {Kuntschner}, {Morganti}, {Oosterloo}, {Sarzi},
  {Scott}, {Serra}, {Ven}, {Weijmans}, \& {Young}}]{2014MNRAS.444.3357N}
{Naab}, T., {Oser}, L., {Emsellem}, E., {et~al.} 2014,
  \href{http://dx.doi.org/10.1093/mnras/stt1919}{\JournalTitle{\mnras}, 444,
  3357}

\bibitem[{{Navarro} {et~al.}(1996){Navarro}, {Frenk}, \&
  {White}}]{1996ApJ...462..563N}
{Navarro}, J.~F., {Frenk}, C.~S., \& {White}, S.~D.~M. 1996,
  \href{http://dx.doi.org/10.1086/177173}{\JournalTitle{\apj}, 462, 563}

\bibitem[{{Oser} {et~al.}(2010){Oser}, {Ostriker}, {Naab}, {Johansson}, \&
  {Burkert}}]{2010ApJ...725.2312O}
{Oser}, L., {Ostriker}, J.~P., {Naab}, T., {Johansson}, P.~H., \& {Burkert}, A.
  2010,
  \href{http://dx.doi.org/10.1088/0004-637X/725/2/2312}{\JournalTitle{\apj},
  725, 2312}

\bibitem[{{Patel} {et~al.}(2013){Patel}, {van Dokkum}, {Franx}, {Quadri},
  {Muzzin}, {Marchesini}, {Williams}, {Holden}, \&
  {Stefanon}}]{2013ApJ...766...15P}
{Patel}, S.~G., {van Dokkum}, P.~G., {Franx}, M., {et~al.} 2013,
  \href{http://dx.doi.org/10.1088/0004-637X/766/1/15}{\JournalTitle{\apj}, 766,
  15}

\bibitem[{{Peng} {et~al.}(2002){Peng}, {Ho}, {Impey}, \&
  {Rix}}]{2002AJ....124..266P}
{Peng}, C.~Y., {Ho}, L.~C., {Impey}, C.~D., \& {Rix}, H.-W. 2002,
  \href{http://dx.doi.org/10.1086/340952}{\JournalTitle{\aj}, 124, 266}

\bibitem[{{P{\'e}rez} {et~al.}(2013){P{\'e}rez}, {Cid Fernandes}, {Gonz{\'a}lez
  Delgado}, {Garc{\'{\i}}a-Benito}, {S{\'a}nchez}, {Husemann}, {Mast},
  {Rod{\'o}n}, {Kupko}, {Backsmann}, {de Amorim}, {van de Ven}, {Walcher},
  {Wisotzki}, {Cortijo-Ferrero}, \& {CALIFA
  Collaboration}}]{2013ApJ...764L...1P}
{P{\'e}rez}, E., {Cid Fernandes}, R., {Gonz{\'a}lez Delgado}, R.~M., {et~al.}
  2013,
  \href{http://dx.doi.org/10.1088/2041-8205/764/1/L1}{\JournalTitle{\apjl},
  764, L1}

\bibitem[{{Petrosian} {et~al.}(2007){Petrosian}, {McLean}, {Allen}, \&
  {MacKenty}}]{2007ApJS..170...33P}
{Petrosian}, A., {McLean}, B., {Allen}, R.~J., \& {MacKenty}, J.~W. 2007,
  \href{http://dx.doi.org/10.1086/511333}{\JournalTitle{\apjs}, 170, 33}

\bibitem[{{Ricciardelli} {et~al.}(2012){Ricciardelli}, {Vazdekis}, {Cenarro},
  \& {Falc{\'o}n-Barroso}}]{2012MNRAS.424..172R}
{Ricciardelli}, E., {Vazdekis}, A., {Cenarro}, A.~J., \& {Falc{\'o}n-Barroso},
  J. 2012,
  \href{http://dx.doi.org/10.1111/j.1365-2966.2012.21178.x}{\JournalTitle{\mnras},
  424, 172}

\bibitem[{{Rix} {et~al.}(1997){Rix}, {de Zeeuw}, {Cretton}, {van der Marel}, \&
  {Carollo}}]{1997ApJ...488..702R}
{Rix}, H.-W., {de Zeeuw}, P.~T., {Cretton}, N., {van der Marel}, R.~P., \&
  {Carollo}, C.~M. 1997,
  \href{http://dx.doi.org/10.1086/304733}{\JournalTitle{\apj}, 488, 702}

\bibitem[{{Robertson} {et~al.}(2006){Robertson}, {Cox}, {Hernquist}, {Franx},
  {Hopkins}, {Martini}, \& {Springel}}]{2006ApJ...641...21R}
{Robertson}, B., {Cox}, T.~J., {Hernquist}, L., {et~al.} 2006,
  \href{http://dx.doi.org/10.1086/500360}{\JournalTitle{\apj}, 641, 21}

\bibitem[{{Romanowsky} {et~al.}(2003){Romanowsky}, {Douglas}, {Arnaboldi},
  {Kuijken}, {Merrifield}, {Napolitano}, {Capaccioli}, \&
  {Freeman}}]{2003Sci...301.1696R}
{Romanowsky}, A.~J., {Douglas}, N.~G., {Arnaboldi}, M., {et~al.} 2003,
  \href{http://dx.doi.org/10.1126/science.1087441}{\JournalTitle{Science}, 301,
  1696}

\bibitem[{{Roth} {et~al.}(2005){Roth}, {Kelz}, {Fechner}, {Hahn}, {Bauer},
  {Becker}, {B{\"o}hm}, {Christensen}, {Dionies}, {Paschke}, {Popow}, {Wolter},
  {Schmoll}, {Laux}, \& {Altmann}}]{2005PASP..117..620R}
{Roth}, M.~M., {Kelz}, A., {Fechner}, T., {et~al.} 2005,
  \href{http://dx.doi.org/10.1086/429877}{\JournalTitle{\pasp}, 117, 620}

\bibitem[{{Rusli} {et~al.}(2011){Rusli}, {Thomas}, {Erwin}, {Saglia}, {Nowak},
  \& {Bender}}]{2011MNRAS.410.1223R}
{Rusli}, S.~P., {Thomas}, J., {Erwin}, P., {et~al.} 2011,
  \href{http://dx.doi.org/10.1111/j.1365-2966.2010.17610.x}{\JournalTitle{\mnras},
  410, 1223}

\bibitem[{{Rusli} {et~al.}(2013){Rusli}, {Thomas}, {Saglia}, {Fabricius},
  {Erwin}, {Bender}, {Nowak}, {Lee}, {Riffeser}, \&
  {Sharp}}]{2013AJ....146...45R}
{Rusli}, S.~P., {Thomas}, J., {Saglia}, R.~P., {et~al.} 2013,
  \href{http://dx.doi.org/10.1088/0004-6256/146/3/45}{\JournalTitle{\aj}, 146,
  45}

\bibitem[{{Rybicki}(1987)}]{1987IAUS..127..397R}
{Rybicki}, G.~B. 1987, in IAU Symposium, Vol. 127, Structure and Dynamics of
  Elliptical Galaxies, ed. P.~T. {de Zeeuw}, 397

\bibitem[{{S{\'a}nchez} {et~al.}(2012){S{\'a}nchez}, {Kennicutt}, {Gil de Paz},
  {van de Ven}, {V{\'{\i}}lchez}, {Wisotzki}, {Walcher}, {Mast}, {Aguerri},
  {Albiol-P{\'e}rez}, {Alonso-Herrero}, {Alves}, {Bakos}, {Bart{\'a}kov{\'a}},
  {Bland-Hawthorn}, {Boselli}, {Bomans}, {Castillo-Morales}, {Cortijo-Ferrero},
  {de Lorenzo-C{\'a}ceres}, {Del Olmo}, {Dettmar}, {D{\'{\i}}az}, {Ellis},
  {Falc{\'o}n-Barroso}, {Flores}, {Gallazzi}, {Garc{\'{\i}}a-Lorenzo},
  {Gonz{\'a}lez Delgado}, {Gruel}, {Haines}, {Hao}, {Husemann},
  {Igl{\'e}sias-P{\'a}ramo}, {Jahnke}, {Johnson}, {Jungwiert}, {Kalinova},
  {Kehrig}, {Kupko}, {L{\'o}pez-S{\'a}nchez}, {Lyubenova}, {Marino},
  {M{\'a}rmol-Queralt{\'o}}, {M{\'a}rquez}, {Masegosa}, {Meidt},
  {Mendez-Abreu}, {Monreal-Ibero}, {Montijo}, {Mour{\~a}o}, {Palacios-Navarro},
  {Papaderos}, {Pasquali}, {Peletier}, {P{\'e}rez}, {P{\'e}rez}, {Quirrenbach},
  {Rela{\~n}o}, {Rosales-Ortega}, {Roth}, {Ruiz-Lara},
  {S{\'a}nchez-Bl{\'a}zquez}, {Sengupta}, {Singh}, {Stanishev}, {Trager},
  {Vazdekis}, {Viironen}, {Wild}, {Zibetti}, \&
  {Ziegler}}]{2012A&A...538A...8S}
{S{\'a}nchez}, S.~F., {Kennicutt}, R.~C., {Gil de Paz}, A., {et~al.} 2012,
  \href{http://dx.doi.org/10.1051/0004-6361/201117353}{\JournalTitle{\aap},
  538, A8}

\bibitem[{{S{\'a}nchez-Bl{\'a}zquez} {et~al.}(2006){S{\'a}nchez-Bl{\'a}zquez},
  {Peletier}, {Jim{\'e}nez-Vicente}, {Cardiel}, {Cenarro},
  {Falc{\'o}n-Barroso}, {Gorgas}, {Selam}, \& {Vazdekis}}]{2006MNRAS.371..703S}
{S{\'a}nchez-Bl{\'a}zquez}, P., {Peletier}, R.~F., {Jim{\'e}nez-Vicente}, J.,
  {et~al.} 2006,
  \href{http://dx.doi.org/10.1111/j.1365-2966.2006.10699.x}{\JournalTitle{\mnras},
  371, 703}

\bibitem[{{Sani} {et~al.}(2011){Sani}, {Marconi}, {Hunt}, \&
  {Risaliti}}]{2011MNRAS.413.1479S}
{Sani}, E., {Marconi}, A., {Hunt}, L.~K., \& {Risaliti}, G. 2011,
  \href{http://dx.doi.org/10.1111/j.1365-2966.2011.18229.x}{\JournalTitle{\mnras},
  413, 1479}

\bibitem[{{Schlafly} \& {Finkbeiner}(2011)}]{2011ApJ...737..103S}
{Schlafly}, E.~F., \& {Finkbeiner}, D.~P. 2011,
  \href{http://dx.doi.org/10.1088/0004-637X/737/2/103}{\JournalTitle{\apj},
  737, 103}

\bibitem[{{Schwarzschild}(1979)}]{1979ApJ...232..236S}
{Schwarzschild}, M. 1979,
  \href{http://dx.doi.org/10.1086/157282}{\JournalTitle{\apj}, 232, 236}

\bibitem[{{Shields} \& {Bonning}(2013)}]{2013ApJ...772L...5S}
{Shields}, G.~A., \& {Bonning}, E.~W. 2013,
  \href{http://dx.doi.org/10.1088/2041-8205/772/1/L5}{\JournalTitle{\apjl},
  772, L5}

\bibitem[{{Silk} \& {Rees}(1998)}]{1998A&A...331L...1S}
{Silk}, J., \& {Rees}, M.~J. 1998, \JournalTitle{\aap}, 331, L1

\bibitem[{{Spiniello} {et~al.}(2014){Spiniello}, {Trager}, {Koopmans}, \&
  {Conroy}}]{2014MNRAS.438.1483S}
{Spiniello}, C., {Trager}, S., {Koopmans}, L.~V.~E., \& {Conroy}, C. 2014,
  \href{http://dx.doi.org/10.1093/mnras/stt2282}{\JournalTitle{\mnras}, 438,
  1483}

\bibitem[{{Spiniello} {et~al.}(2012){Spiniello}, {Trager}, {Koopmans}, \&
  {Chen}}]{2012ApJ...753L..32S}
{Spiniello}, C., {Trager}, S.~C., {Koopmans}, L.~V.~E., \& {Chen}, Y.~P. 2012,
  \href{http://dx.doi.org/10.1088/2041-8205/753/2/L32}{\JournalTitle{\apjl},
  753, L32}

\bibitem[{{Statler}(1987)}]{1987ApJ...321..113S}
{Statler}, T.~S. 1987,
  \href{http://dx.doi.org/10.1086/165619}{\JournalTitle{\apj}, 321, 113}

\bibitem[{{Tamura} \& {Ohta}(2003)}]{2003AJ....126..596T}
{Tamura}, N., \& {Ohta}, K. 2003,
  \href{http://dx.doi.org/10.1086/376469}{\JournalTitle{\aj}, 126, 596}

\bibitem[{{Thomas} {et~al.}(2005){Thomas}, {Saglia}, {Bender}, {Thomas},
  {Gebhardt}, {Magorrian}, {Corsini}, \& {Wegner}}]{2005MNRAS.360.1355T}
{Thomas}, J., {Saglia}, R.~P., {Bender}, R., {et~al.} 2005,
  \href{http://dx.doi.org/10.1111/j.1365-2966.2005.09139.x}{\JournalTitle{\mnras},
  360, 1355}

\bibitem[{{Thomas} {et~al.}(2007){Thomas}, {Saglia}, {Bender}, {Thomas},
  {Gebhardt}, {Magorrian}, {Corsini}, \& {Wegner}}]{2007MNRAS.382..657T}
---. 2007,
  \href{http://dx.doi.org/10.1111/j.1365-2966.2007.12434.x}{\JournalTitle{\mnras},
  382, 657}

\bibitem[{{Thomas} {et~al.}(2004){Thomas}, {Saglia}, {Bender}, {Thomas},
  {Gebhardt}, {Magorrian}, \& {Richstone}}]{2004MNRAS.353..391T}
---. 2004,
  \href{http://dx.doi.org/10.1111/j.1365-2966.2004.08072.x}{\JournalTitle{\mnras},
  353, 391}

\bibitem[{{Toft} {et~al.}(2012){Toft}, {Gallazzi}, {Zirm}, {Wold}, {Zibetti},
  {Grillo}, \& {Man}}]{2012ApJ...754....3T}
{Toft}, S., {Gallazzi}, A., {Zirm}, A., {et~al.} 2012,
  \href{http://dx.doi.org/10.1088/0004-637X/754/1/3}{\JournalTitle{\apj}, 754,
  3}

\bibitem[{{Toft} {et~al.}(2014){Toft}, {Smol{\v c}i{\'c}}, {Magnelli}, {Karim},
  {Zirm}, {Michalowski}, {Capak}, {Sheth}, {Schawinski}, {Krogager}, {Wuyts},
  {Sanders}, {Man}, {Lutz}, {Staguhn}, {Berta}, {Mccracken}, {Krpan}, \&
  {Riechers}}]{2014ApJ...782...68T}
{Toft}, S., {Smol{\v c}i{\'c}}, V., {Magnelli}, B., {et~al.} 2014,
  \href{http://dx.doi.org/10.1088/0004-637X/782/2/68}{\JournalTitle{\apj}, 782,
  68}

\bibitem[{{Tortora} {et~al.}(2011){Tortora}, {Napolitano}, {Romanowsky},
  {Jetzer}, {Cardone}, \& {Capaccioli}}]{2011MNRAS.418.1557T}
{Tortora}, C., {Napolitano}, N.~R., {Romanowsky}, A.~J., {et~al.} 2011,
  \href{http://dx.doi.org/10.1111/j.1365-2966.2011.19438.x}{\JournalTitle{\mnras},
  418, 1557}

\bibitem[{{Treu} {et~al.}(2010){Treu}, {Auger}, {Koopmans}, {Gavazzi},
  {Marshall}, \& {Bolton}}]{2010ApJ...709.1195T}
{Treu}, T., {Auger}, M.~W., {Koopmans}, L.~V.~E., {et~al.} 2010,
  \href{http://dx.doi.org/10.1088/0004-637X/709/2/1195}{\JournalTitle{\apj},
  709, 1195}

\bibitem[{{Trujillo} {et~al.}(2014){Trujillo}, {Ferr{\'e}-Mateu}, {Balcells},
  {Vazdekis}, \& {S{\'a}nchez-Bl{\'a}zquez}}]{2014ApJ...780L..20T}
{Trujillo}, I., {Ferr{\'e}-Mateu}, A., {Balcells}, M., {Vazdekis}, A., \&
  {S{\'a}nchez-Bl{\'a}zquez}, P. 2014,
  \href{http://dx.doi.org/10.1088/2041-8205/780/2/L20}{\JournalTitle{\apjl},
  780, L20}

\bibitem[{{Trujillo} {et~al.}(2006){Trujillo}, {F{\"o}rster Schreiber},
  {Rudnick}, {Barden}, {Franx}, {Rix}, {Caldwell}, {McIntosh}, {Toft},
  {H{\"a}ussler}, {Zirm}, {van Dokkum}, {Labb{\'e}}, {Moorwood},
  {R{\"o}ttgering}, {van der Wel}, {van der Werf}, \& {van
  Starkenburg}}]{2006ApJ...650...18T}
{Trujillo}, I., {F{\"o}rster Schreiber}, N.~M., {Rudnick}, G., {et~al.} 2006,
  \href{http://dx.doi.org/10.1086/506464}{\JournalTitle{\apj}, 650, 18}

\bibitem[{{Valdes} {et~al.}(2004){Valdes}, {Gupta}, {Rose}, {Singh}, \&
  {Bell}}]{2004ApJS..152..251V}
{Valdes}, F., {Gupta}, R., {Rose}, J.~A., {Singh}, H.~P., \& {Bell}, D.~J.
  2004, \href{http://dx.doi.org/10.1086/386343}{\JournalTitle{\apjs}, 152, 251}

\bibitem[{{Valluri} {et~al.}(2004){Valluri}, {Merritt}, \&
  {Emsellem}}]{2004ApJ...602...66V}
{Valluri}, M., {Merritt}, D., \& {Emsellem}, E. 2004,
  \href{http://dx.doi.org/10.1086/380896}{\JournalTitle{\apj}, 602, 66}

\bibitem[{{van de Sande} {et~al.}(2013){van de Sande}, {Kriek}, {Franx}, {van
  Dokkum}, {Bezanson}, {Bouwens}, {Quadri}, {Rix}, \&
  {Skelton}}]{2013ApJ...771...85V}
{van de Sande}, J., {Kriek}, M., {Franx}, M., {et~al.} 2013,
  \href{http://dx.doi.org/10.1088/0004-637X/771/2/85}{\JournalTitle{\apj}, 771,
  85}

\bibitem[{{van de Ven} {et~al.}(2008){van de Ven}, {de Zeeuw}, \& {van den
  Bosch}}]{2008MNRAS.385..614V}
{van de Ven}, G., {de Zeeuw}, P.~T., \& {van den Bosch}, R.~C.~E. 2008,
  \href{http://dx.doi.org/10.1111/j.1365-2966.2008.12873.x}{\JournalTitle{\mnras},
  385, 614}

\bibitem[{{van den Bosch}(1997)}]{1997MNRAS.287..543V}
{van den Bosch}, F.~C. 1997, \JournalTitle{\mnras}, 287, 543

\bibitem[{{van den Bosch} \& {de Zeeuw}(2010)}]{2010MNRAS.401.1770V}
{van den Bosch}, R.~C.~E., \& {de Zeeuw}, P.~T. 2010,
  \href{http://dx.doi.org/10.1111/j.1365-2966.2009.15832.x}{\JournalTitle{\mnras},
  401, 1770}

\bibitem[{{van den Bosch} {et~al.}(2012){van den Bosch}, {Gebhardt},
  {G{\"u}ltekin}, {van de Ven}, {van der Wel}, \&
  {Walsh}}]{2012Natur.491..729V}
{van den Bosch}, R.~C.~E., {Gebhardt}, K., {G{\"u}ltekin}, K., {et~al.} 2012,
  \href{http://dx.doi.org/10.1038/nature11592}{\JournalTitle{\nat}, 491, 729}

\bibitem[{{van den Bosch} {et~al.}(2015){van den Bosch}, {Gebhardt},
  {G{\"u}ltekin}, {Y{\i}ld{\i}r{\i}m}, \& {Walsh}}]{2015ApJS..218...10V}
{van den Bosch}, R.~C.~E., {Gebhardt}, K., {G{\"u}ltekin}, K.,
  {Y{\i}ld{\i}r{\i}m}, A., \& {Walsh}, J.~L. 2015,
  \href{http://dx.doi.org/10.1088/0067-0049/218/1/10}{\JournalTitle{\apjs},
  218, 10}

\bibitem[{{van den Bosch} \& {van de Ven}(2009)}]{2009MNRAS.398.1117V}
{van den Bosch}, R.~C.~E., \& {van de Ven}, G. 2009,
  \href{http://dx.doi.org/10.1111/j.1365-2966.2009.15177.x}{\JournalTitle{\mnras},
  398, 1117}

\bibitem[{{van den Bosch} {et~al.}(2008){van den Bosch}, {van de Ven},
  {Verolme}, {Cappellari}, \& {de Zeeuw}}]{2008MNRAS.385..647V}
{van den Bosch}, R.~C.~E., {van de Ven}, G., {Verolme}, E.~K., {Cappellari},
  M., \& {de Zeeuw}, P.~T. 2008,
  \href{http://dx.doi.org/10.1111/j.1365-2966.2008.12874.x}{\JournalTitle{\mnras},
  385, 647}

\bibitem[{{van der Marel} \& {Franx}(1993)}]{1993ApJ...407..525V}
{van der Marel}, R.~P., \& {Franx}, M. 1993,
  \href{http://dx.doi.org/10.1086/172534}{\JournalTitle{\apj}, 407, 525}

\bibitem[{{van der Wel} {et~al.}(2008){van der Wel}, {Holden}, {Zirm}, {Franx},
  {Rettura}, {Illingworth}, \& {Ford}}]{2008ApJ...688...48V}
{van der Wel}, A., {Holden}, B.~P., {Zirm}, A.~W., {et~al.} 2008,
  \href{http://dx.doi.org/10.1086/592267}{\JournalTitle{\apj}, 688, 48}

\bibitem[{{van der Wel} {et~al.}(2011){van der Wel}, {Rix}, {Wuyts}, {McGrath},
  {Koekemoer}, {Bell}, {Holden}, {Robaina}, \&
  {McIntosh}}]{2011ApJ...730...38V}
{van der Wel}, A., {Rix}, H.-W., {Wuyts}, S., {et~al.} 2011,
  \href{http://dx.doi.org/10.1088/0004-637X/730/1/38}{\JournalTitle{\apj}, 730,
  38}

\bibitem[{{van der Wel} {et~al.}(2012){van der Wel}, {Bell}, {H{\"a}ussler},
  {McGrath}, {Chang}, {Guo}, {McIntosh}, {Rix}, {Barden}, {Cheung}, {Faber},
  {Ferguson}, {Galametz}, {Grogin}, {Hartley}, {Kartaltepe}, {Kocevski},
  {Koekemoer}, {Lotz}, {Mozena}, {Peth}, \& {Peng}}]{2012ApJS..203...24V}
{van der Wel}, A., {Bell}, E.~F., {H{\"a}ussler}, B., {et~al.} 2012,
  \href{http://dx.doi.org/10.1088/0067-0049/203/2/24}{\JournalTitle{\apjs},
  203, 24}

\bibitem[{{van der Wel} {et~al.}(2014){van der Wel}, {Franx}, {van Dokkum},
  {Skelton}, {Momcheva}, {Whitaker}, {Brammer}, {Bell}, {Rix}, {Wuyts},
  {Ferguson}, {Holden}, {Barro}, {Koekemoer}, {Chang}, {McGrath},
  {H{\"a}ussler}, {Dekel}, {Behroozi}, {Fumagalli}, {Leja}, {Lundgren},
  {Maseda}, {Nelson}, {Wake}, {Patel}, {Labb{\'e}}, {Faber}, {Grogin}, \&
  {Kocevski}}]{2014ApJ...788...28V}
{van der Wel}, A., {Franx}, M., {van Dokkum}, P.~G., {et~al.} 2014,
  \href{http://dx.doi.org/10.1088/0004-637X/788/1/28}{\JournalTitle{\apj}, 788,
  28}

\bibitem[{{van Dokkum} \& {Conroy}(2012)}]{2012ApJ...760...70V}
{van Dokkum}, P.~G., \& {Conroy}, C. 2012,
  \href{http://dx.doi.org/10.1088/0004-637X/760/1/70}{\JournalTitle{\apj}, 760,
  70}

\bibitem[{{van Dokkum} {et~al.}(2009){van Dokkum}, {Kriek}, \&
  {Franx}}]{2009Natur.460..717V}
{van Dokkum}, P.~G., {Kriek}, M., \& {Franx}, M. 2009,
  \href{http://dx.doi.org/10.1038/nature08220}{\JournalTitle{\nat}, 460, 717}

\bibitem[{{van Dokkum} {et~al.}(2008){van Dokkum}, {Franx}, {Kriek}, {Holden},
  {Illingworth}, {Magee}, {Bouwens}, {Marchesini}, {Quadri}, {Rudnick},
  {Taylor}, \& {Toft}}]{2008ApJ...677L...5V}
{van Dokkum}, P.~G., {Franx}, M., {Kriek}, M., {et~al.} 2008,
  \href{http://dx.doi.org/10.1086/587874}{\JournalTitle{\apjl}, 677, L5}

\bibitem[{{van Dokkum} {et~al.}(2010){van Dokkum}, {Whitaker}, {Brammer},
  {Franx}, {Kriek}, {Labb{\'e}}, {Marchesini}, {Quadri}, {Bezanson},
  {Illingworth}, {Muzzin}, {Rudnick}, {Tal}, \& {Wake}}]{2010ApJ...709.1018V}
{van Dokkum}, P.~G., {Whitaker}, K.~E., {Brammer}, G., {et~al.} 2010,
  \href{http://dx.doi.org/10.1088/0004-637X/709/2/1018}{\JournalTitle{\apj},
  709, 1018}

\bibitem[{{Vazdekis} {et~al.}(1996){Vazdekis}, {Casuso}, {Peletier}, \&
  {Beckman}}]{1996ApJS..106..307V}
{Vazdekis}, A., {Casuso}, E., {Peletier}, R.~F., \& {Beckman}, J.~E. 1996,
  \href{http://dx.doi.org/10.1086/192340}{\JournalTitle{\apjs}, 106, 307}

\bibitem[{{Vazdekis} {et~al.}(2012){Vazdekis}, {Ricciardelli}, {Cenarro},
  {Rivero-Gonz{\'a}lez}, {D{\'{\i}}az-Garc{\'{\i}}a}, \&
  {Falc{\'o}n-Barroso}}]{2012MNRAS.424..157V}
{Vazdekis}, A., {Ricciardelli}, E., {Cenarro}, A.~J., {et~al.} 2012,
  \href{http://dx.doi.org/10.1111/j.1365-2966.2012.21179.x}{\JournalTitle{\mnras},
  424, 157}

\bibitem[{{Vazdekis} {et~al.}(2010){Vazdekis}, {S{\'a}nchez-Bl{\'a}zquez},
  {Falc{\'o}n-Barroso}, {Cenarro}, {Beasley}, {Cardiel}, {Gorgas}, \&
  {Peletier}}]{2010MNRAS.404.1639V}
{Vazdekis}, A., {S{\'a}nchez-Bl{\'a}zquez}, P., {Falc{\'o}n-Barroso}, J.,
  {et~al.} 2010,
  \href{http://dx.doi.org/10.1111/j.1365-2966.2010.16407.x}{\JournalTitle{\mnras},
  404, 1639}

\bibitem[{{Verheijen} {et~al.}(2004){Verheijen}, {Bershady}, {Andersen},
  {Swaters}, {Westfall}, {Kelz}, \& {Roth}}]{2004AN....325..151V}
{Verheijen}, M.~A.~W., {Bershady}, M.~A., {Andersen}, D.~R., {et~al.} 2004,
  \href{http://dx.doi.org/10.1002/asna.200310197}{\JournalTitle{Astronomische
  Nachrichten}, 325, 151}

\bibitem[{{Verolme} {et~al.}(2002){Verolme}, {Cappellari}, {Copin}, {van der
  Marel}, {Bacon}, {Bureau}, {Davies}, {Miller}, \& {de
  Zeeuw}}]{2002MNRAS.335..517V}
{Verolme}, E.~K., {Cappellari}, M., {Copin}, Y., {et~al.} 2002,
  \href{http://dx.doi.org/10.1046/j.1365-8711.2002.05664.x}{\JournalTitle{\mnras},
  335, 517}

\bibitem[{{Walsh} {et~al.}(2012){Walsh}, {van den Bosch}, {Barth}, \&
  {Sarzi}}]{2012ApJ...753...79W}
{Walsh}, J.~L., {van den Bosch}, R.~C.~E., {Barth}, A.~J., \& {Sarzi}, M. 2012,
  \href{http://dx.doi.org/10.1088/0004-637X/753/1/79}{\JournalTitle{\apj}, 753,
  79}

\bibitem[{{Weijmans} {et~al.}(2009){Weijmans}, {Cappellari}, {Bacon}, {de
  Zeeuw}, {Emsellem}, {Falc{\'o}n-Barroso}, {Kuntschner}, {McDermid}, {van den
  Bosch}, \& {van de Ven}}]{2009MNRAS.398..561W}
{Weijmans}, A.-M., {Cappellari}, M., {Bacon}, R., {et~al.} 2009,
  \href{http://dx.doi.org/10.1111/j.1365-2966.2009.15134.x}{\JournalTitle{\mnras},
  398, 561}

\bibitem[{{Weijmans} {et~al.}(2014){Weijmans}, {de Zeeuw}, {Emsellem},
  {Krajnovi{\'c}}, {Lablanche}, {Alatalo}, {Blitz}, {Bois}, {Bournaud},
  {Bureau}, {Cappellari}, {Crocker}, {Davies}, {Davis}, {Duc}, {Khochfar},
  {Kuntschner}, {McDermid}, {Morganti}, {Naab}, {Oosterloo}, {Sarzi}, {Scott},
  {Serra}, {Verdoes Kleijn}, \& {Young}}]{2014MNRAS.444.3340W}
{Weijmans}, A.-M., {de Zeeuw}, P.~T., {Emsellem}, E., {et~al.} 2014,
  \href{http://dx.doi.org/10.1093/mnras/stu1603}{\JournalTitle{\mnras}, 444,
  3340}

\bibitem[{{Williams} {et~al.}(2014){Williams}, {Giavalisco}, {Cassata},
  {Tundo}, {Wiklind}, {Guo}, {Lee}, {Barro}, {Wuyts}, {Bell}, {Conselice},
  {Dekel}, {Faber}, {Ferguson}, {Grogin}, {Hathi}, {Huang}, {Kocevski},
  {Koekemoer}, {Koo}, {Ravindranath}, \& {Salimbeni}}]{2014ApJ...780....1W}
{Williams}, C.~C., {Giavalisco}, M., {Cassata}, P., {et~al.} 2014,
  \href{http://dx.doi.org/10.1088/0004-637X/780/1/1}{\JournalTitle{\apj}, 780,
  1}

\bibitem[{{Wuyts} {et~al.}(2010){Wuyts}, {Cox}, {Hayward}, {Franx},
  {Hernquist}, {Hopkins}, {Jonsson}, \& {van Dokkum}}]{2010ApJ...722.1666W}
{Wuyts}, S., {Cox}, T.~J., {Hayward}, C.~C., {et~al.} 2010,
  \href{http://dx.doi.org/10.1088/0004-637X/722/2/1666}{\JournalTitle{\apj},
  722, 1666}

\bibitem[{{Zirm} {et~al.}(2007){Zirm}, {van der Wel}, {Franx}, {Labb{\'e}},
  {Trujillo}, {van Dokkum}, {Toft}, {Daddi}, {Rudnick}, {Rix},
  {R{\"o}ttgering}, \& {van der Werf}}]{2007ApJ...656...66Z}
{Zirm}, A.~W., {van der Wel}, A., {Franx}, M., {et~al.} 2007,
  \href{http://dx.doi.org/10.1086/510713}{\JournalTitle{\apj}, 656, 66}

\end{thebibliography}


\label{lastpage}

\end{document}